\documentclass[nofootinbib,superscriptaddress,eqsecnum,tightenlines,11pt,longbibliography,notitlepage]{revtex4-1}

\usepackage[draft]{showlabels}
\usepackage{hyperref}
\usepackage{graphicx}
\usepackage{amsmath,amssymb,amsfonts,amsthm,stmaryrd,mathtools,mathtools}
\usepackage{yfonts}
\usepackage{multirow,bigdelim}
\usepackage{dsfont}
\usepackage{color}
\usepackage{graphicx,color,epstopdf,epsfig,psfrag}
\usepackage[dvipsnames]{xcolor}
\usepackage{ mathrsfs }
\usepackage{footmisc}
\usepackage{enumerate}
\usepackage{tikz}
\usetikzlibrary{calc}
\usetikzlibrary{decorations.pathmorphing}
\usetikzlibrary{shapes.geometric}
\usetikzlibrary{arrows,decorations.markings}
\usepackage{bbold}

\hypersetup{
    colorlinks=true,       
    linkcolor=MidnightBlue,          
    citecolor=BrickRed,        
    filecolor=BrickRed,      
    urlcolor=BrickRed           
}
 
\makeatletter
\newcounter{subsubsubsection}[subsubsection]
\renewcommand\thesubsubsubsection{\thesubsubsection .\@alph\c@subsubsubsection}
\newcommand\subsubsubsection{\@startsection{subsubsubsection}{4}{\z@}%
                                     {-3.25ex\@plus -1ex \@minus -.2ex}%
                                     {1.5ex \@plus .2ex}%
                                     {\centering\normalfont\small\textit}}
\newcommand*\l@subsubsubsection{\@dottedtocline{3}{10.0em}{4.1em}}
\newcommand*{\subsubsubsectionmark}[1]{}
\makeatother

\numberwithin{equation}{section}
\numberwithin{paragraph}{subsection}

\newcommand{\E}{\mathrm{e}}

\renewcommand{\Re}{{\mathrm{Re}}}

\newcommand{\C}{{\mathbb C}}
\newcommand{\N}{{\mathbb N}}
\newcommand{\R}{{\mathbb R}}
\newcommand{\Q}{{\mathbb Q}}
\newcommand{\Z}{{\mathbb Z}}

\newcommand{\cJ}{{\mathcal J}}

\newcommand{\SU}{\mathrm{SU}}

\newcommand{\SO}{\mathrm{SO}}
\newcommand{\U}{\mathrm{U}}

\renewcommand{\d}{{\mathrm{d}}}

\newcommand{\be}{\begin{equation}}
\newcommand{\ee}{\end{equation}}
\newcommand{\beq}{\begin{eqnarray}}
\newcommand{\eeq}{\end{eqnarray}}
\newcommand{\bes}{\begin{eqnarray}}
\newcommand{\ees}{\end{eqnarray}}

\newcommand{\mat} [2] {\left ( \begin{array}{#1}#2\end{array} \right ) }

\newcommand{\su}{{\mathfrak{su}}}

\newcommand{\I}{{\mathrm{i}}}

\newcommand{\la}{\langle}
\newcommand{\ra}{\rangle}

\newcommand{\Tr}{{\mathrm{Tr}}}
\newcommand{\f}{\frac}
\newcommand{\tl}{\widetilde}
\def\nn{\nonumber}
\def\pp{\partial}

\newcommand{\id}{\mathbb{I}}

\newcommand{\G}{{\Gamma}}
\newcommand{\LS}{{\text{LS}}}

\newcommand{\up}{{\,\uparrow\,}}

\newcommand{\GG}{\mathbf{G}}

\newcommand{\DD}{\mathbf{D}}
\newcommand{\CC}{\mathbf{C}}
\newcommand{\bfa}{\mathbf{a}}

\renewcommand{\hat}{\widehat}

\renewcommand{\bar}{\overline}

\newcommand{\ellpl}{\ell_\text{Pl}}

\theoremstyle{definition}

\theoremstyle{remark}

\def\vsigma{\vec{\sigma}}
\def\vtau{\vec{\tau}}

\begin{document}

\title{\LARGE Quasi-local holographic dualities\\ in non-perturbative 3d quantum gravity
 \\\vspace{.5em} \large  II - From coherent quantum boundaries to BMS$_3$ characters}

\author{{\bf Bianca Dittrich}}\email{bdittrich@perimeterinstitute.ca}
\affiliation{Perimeter Institute, 31 Caroline St North, Waterloo ON, Canada N2L 2Y5}

\author{{\bf Christophe Goeller}}\email{christophe.goeller@ens-lyon.fr}
\affiliation{Laboratoire de Physique, ENS Lyon, CNRS-UMR 5672, 46 all\'ee d'Italie, Lyon 69007, France}
\affiliation{Perimeter Institute, 31 Caroline St North, Waterloo ON, Canada N2L 2Y5}

\author{{\bf Etera R. Livine}}\email{etera.livine@ens-lyon.fr}
\affiliation{Laboratoire de Physique, ENS Lyon, CNRS-UMR 5672, 46 all\'ee d'Italie, Lyon 69007, France}
\affiliation{Perimeter Institute, 31 Caroline St North, Waterloo ON, Canada N2L 2Y5}

\author{{\bf Aldo Riello}}\email{ariello@perimeterinstitute.ca}
\affiliation{Perimeter Institute, 31 Caroline St North, Waterloo ON, Canada N2L 2Y5}

\date{\today}

\begin{abstract}

We analyze the partition function of three-dimensional quantum gravity on the twisted solid tours and the ensuing dual field theory.
The setting is that of a non-perturbative model of three dimensional quantum gravity---the Ponzano--Regge model, that we briefly review in a self-contained manner---which can be used to compute quasi-local amplitudes for its boundary states. 
In this second paper of the series, we choose a particular class of boundary spin-network states which impose Gibbons--Hawking--York boundary conditions to the partition function.
The peculiarity of these states is to encode a two-dimensional quantum geometry peaked around a classical quadrangulation of the finite toroidal boundary. 
Thanks to the topological properties of three-dimensional gravity, the theory easily projects onto the boundary while crucially still keeping track of the topological properties of the bulk. 
This produces, at the non-perturbative level, a specific non-linear sigma-model on the boundary, akin to a Wess--Zumino--Novikov--Witten model, whose classical equations of motion can be used to reconstruct different bulk geometries: the expected classical one is accompanied by other ``quantum'' solutions. 
The classical regime of the sigma-model becomes reliable in the limit of large boundary spins, which coincides with the semiclassical limit of the boundary geometry.  
In a 1-loop approximation around the solutions to the classical equations of motion, we recover (with corrections due to the non-classical bulk geometries) results obtained in the past via perturbative quantum General Relativity and through the study of characters of the BMS$_3$ group. 
The exposition is meant to be completely self-contained.

\end{abstract}

\maketitle

\newpage
\small
\tableofcontents
\normalsize
\newpage
\section{Introduction}

\subsection{Motivations\footnote{For a (much) broader set of motivations, we invite the reader to consult the companion paper \cite{PART1}.}}

Three-dimensional gravity is an ideal testing ground for comparing different approaches to quantum gravity. One set of ideas,  grown out of the AdS/CFT correspondence \cite{Maldacena:1997re,Witten:1998qj,Horowitz:2006ct,Hubeny:2014bla,Giombi:2008vd}, consists in defining a theory of quantum gravity via a holographic dual on the asymptotic boundary of an AdS spacetime.  Large efforts have been made in order to generalize such boundary duals to other classes of asymptotic boundary conditions, in particular to the asymptotically flat case \cite{Ashtekar:1996cd,Arcioni:2003xx,Barnich:2006av,Barnich:2010eb,Barnich:2012aw,Barnich:2013axa,Carlip2016,BarnichEtAl2015}. The three dimensional case is where these attempts have been the most successful.

This gives us the opportunity to study the question whether and how holographic duals arise in local non-perturbative approaches to quantum gravity \cite{Freidel:2008sh,Gomes:2013qza,Dittrich:2013jxa,Bonzom:2015ova,BonzomDittrich2015,Smolin:2016edy,Han:2016xmb,Chirco:2017vhs,Livine:2017xww}. More specifically, we have in mind the spin foam framework, whose goal is to define a path integral for quantum gravity formulated using tools from topological quantum field theory  \cite{Rovelli:2004tv,Perez:2012wv,Reisenberger:1996pu,Baez:1997zt,Barrett:1997gw,Freidel:1998pt,Engle:2007wy,Freidel:2007py}.
 In fact, for spin foams, the case of a vanishing cosmological constant is particularly well understood and many techniques for the analyses of the spin foam partition functions are available, especially in three dimensions \cite{Livine:2007vk,ConradyFreidel2008,BarrettEtAl2009,DowdallGomesHellmann2010,HanZhang2011,HanZhang2011a,Freidel:2012ji,HellmannKaminski2013,HHKR2015}.
In the three--dimensional case, the spin foam model goes back to a clever observation by Ponzano and Regge \cite{PonzanoRegge1968} which led to what can be considered as the earliest proposal for a 3D quantum gravity theory.%
\footnote{The case with a cosmological constant can also be constructed \cite{TuraevViro1992,MizoguchiTada1992,TaylorWoodward2005,Noui:2011im,DupuisGirelli2014,Bonzom2014,Bonzom:2014wva,DittrichGeiller2017} through the use of quantum groups.
In this case the coherent states techniques largely used in this paper are less developed. 
}

Spin foam models are a non-perturbative---in the sense of manifestly background independent---approach to quantum gravity, constructed from local weights associated to some building blocks of the spacetime. For this reason, in this approach it is most natural to consider first  finite boundaries which are only eventually pushed asymptotically far away.
In fact, a first study of the one--loop partition function for 3D gravity with finite boundaries, using perturbative Regge calculus \cite{BonzomDittrich2015}, revealed that a holographic dual can actually be defined even before taking an asymptotic limit. The possibility to make use of holographic dualities also for finite boundaries offers the exciting perspective to get insight also on much more local properties of quantum gravity, than those that can be encoded on an asymptotic boundary. 
 Of course, this is possible in other contexts too, e.g. in the Chern--Simons formulation of 3D AdS gravity \cite{Witten1988}. In that case, the role of asymptotic condition is to automatically select a stricter set of boundary conditions, hence leading to a further specification of the boundary theory, from a Wess--Zumino--Novikov--Witten sigma model to scalar Liouville theory \cite{Carlip2005} (and references therein).\footnote{See also \cite{Carlip2005a} for a different derivation, and \cite{Carlip2016} for a tentative adaptation of the latter to the asymptotically flat case.
}
The approach we take here is to solve exactly the bulk part of the partition function for gravity---which is possible due to the topological nature of 3D gravity---and to finally study---in some approximation---the ``dual'' theory so induced onto the boundary. This allows us to compare the partition functions obtained in different approaches.  We will indeed find a match between the various one--loop partition functions.  More precisely, the match holds for contributions coming from the same background solution. 

As we stated, the evaluation of the partition function will reveal a  dual boundary field theory, from which we will show how to reconstruct a bulk geometry. On general grounds, the dual theory is expected to be akin to an $\SU(2)$ (or $\SO(3)$) sigma-model, whose details, however, shall strongly depend on the choice of a class of boundary state, i.e. on a class of boundary conditions for the path integral (see \cite{PART1}). In the case of spin-network boundary states considered here (as well as in the companion work \cite{PART1}), it is the induced metric on the boundary to be kept fixed. As a consequence, the coupling constants of the dual theory are found to be given by the very parameters encoding for such a metric.
Finally, the semiclassical bulk geometry reconstruction is achieved in the semiclassical limit of the boundary theory, which coincides with its large-spin limit.\footnote{This is, however, quite different from any sort of large $N$ limit taken at the level of the action functional.}

The evaluation of the partition function for general boundary states is still a daunting task.  The companion paper \cite{PART1}  found that the  partition function with boundary state based on the smallest possible discretization lengths --that is a spin network state with all edges carrying a spin $j=\tfrac{1}{2}$-- can be evaluated exactly, and, in the limit of a large boundary composed by arbitrarily many elementary cells, the main features of the continuum \cite{BarnichEtAl2015} and Regge calculus \cite{BonzomDittrich2015} results can be recovered. 

In this paper, we analyze the partition function in the ``opposite'' limit, that is in the large-spin regime where the discretization is large compared to the Planck length. The large-spin regime corresponds to the semi-classical limit of spin foam models, and can be studied via a saddle point analysis. This has been arguably the main tool so far for analyzing the dynamics encoded in spin foam models \cite{Livine:2006it,Christensen:2007rv,Barrett:1998gs, ConradyFreidel2008,BarrettEtAl2009,DowdallGomesHellmann2010,HanZhang2011,HanZhang2011a,HHKR2015,HHKR2016,SpezialeEtAl2017}. Most of the time, however, this approximation is used to analyze amplitudes for a single building block.\footnote{See, however, \cite{Riello2013} where saddle points techniques are applied to the study of divergences in the EPRL spinfoam model.}
Therefore, the key question is still largely open, which is that of understanding very fine discretizations with many building blocks.\footnote{See \cite{DittrichReview2014,DittrichMartinBenitoSchnetter,DittrichMartinBenitoSteinhaus,DittrichMizeraSteinhaus,DelcampDittrich2016,Bahr2014,BahrSteinhaus2015,BahrSteinhaus2016}
for recent work in this direction.} Fortunately, in the 3D case such a general understanding is not necessary, since the bulk theory is purely topological (i.e. discretization invariant) and  can be easily be solved for exactly. This fact, allows us to explore the large spin limit in a new regime: one where the only spins to be taken parametrically large are those on the boundary, while the bulk is treated exactly and non-perturbatively.   There remains however to understand how the choice of the discretization scale of the boundary influences the partition function           
\cite{DittrichCylCon2012,DittrichReview2014}.

This strategy has already been employed in the first study of the torus partition function by Dowdall, Gomes and Hellmann \cite{DowdallGomesHellmann2010}. However, their results are still on a very abstract level, and in particular the one-loop correction, which is of essential interest in this context of holographic dualities, has not been evaluated.  Here, we manage (possibly for the first time) the explicit geometric reconstruction via a saddle-point analysis of a complex boundary state, with an exact rather than semi-classical treatment of the bulk discretization. 

This allows us to shed some light on a number of important issues arising in spinfoams (i.e. covariant Loop Quantum Gravity), including:
\begin{enumerate}
\item[{\it i}\,)] How does the Ponzano--Regge model, which is based on a first order formulation of gravity, differ from other approaches based on a metric formulation \cite{Witten1988,Matschull1999,ChristodoulouEtAl2013}?
\item[{\it ii}\,)] The Ponzano--Regge model includes a sum over (space--time) orientations of each building block. Semi-classically, can a consistent orientation be imposed on the bulk geometry via the choice of an appropriate boundary state?
\item[{\it iii}\,)] The Ponzano--Regge model admits a holonomy formulation. Thus curvature defect angles which differ by $2\pi$ are not distinguishable. This is a key difference to Regge calculus, where the fundamental variables are the dihedral angles rather than holonomies. How will this feature show up in the geometric reconstruction?
\item[{\it iv}\,)] In which way does the partition function associated to a small-scale discretization with many building blocks differ from one associated to a large-scale discretization? This is the first time that an explicit comparison is possible and we will find agreement in some, rather subtle, features of the two partition functions.
\end{enumerate}

Thus beyond the  study of holographic properties of non--perturbative 3D quantum gravity we believe that this work has a number of lessons applicable to 4D quantum gravity, and in particular to 4D spin foam models.

Finally, we invite the reader to consult the companion paper \cite{PART1}  for a more exhaustive review of the motivations and general features of Ponzano--Regge holography. Henceforth, we will simply refer to it as {\it Part I}.

~\\
{\it Outline of the paper:} 
 In the remainder of the introduction we will give a short review on the Ponzano--Regge model (see Part I for a more detailed overview and more references). In section \ref{sec_bdrystate}, we will first review spin-network boundary states in general and then we will introduce a class of geometrical spin-network states specifically adapted to the two-dimensional boundaries to be studied in this paper. In doing this we address a couple of crucial technical subtleties: for instance, we clarify the geometry of coherent intertwiners as semi-classical polygons, with special attention to their orientation in the 3d space, as well as the local lift from $\SO(3)$ to $\SU(2)$.
 The following, section \ref{sec_PRamplLS}, is the core section of the paper: there, we first introduce the specific boundary state we use on the twisted torus, section \ref{sec_Lsstatedef}; thus set-up the amplitude calculation in a form suitable for a saddle point (or ``1-loop'') analysis, section \ref{sec_PRamplitude}; and, finally, we study the amplitude in that approximation, section \ref{sec_SemiclassicalLargespin}. The latter task is broken into a series of smaller steps: derivation of the saddle point equations, geometrical interpretation and analytic solution thereof, reconstruction of the whole geometry and analysis of the subtleties arising for non-geometrical saddles, calculation of the 1-loop determinant (or action's Hessian). Finally, the paper closes with an extended summary and discussion session \ref{sec_discussion} (for the reader experienced in spinfoam calculations, the summary section might be in fact be a good place to start). Particular attention will be dedicated to the comparison of our calculation with previous ones, which led to the same results from radically different perspectives. These include perturbative quantum Regge calculus, perturbative quantum General Relativity (at 1-loop), holographic methods, as well as the results for the Ponzano--Regge model obtained in \cite{PART1}. Four appendices detail some of the computations and discuss side-issues.

\subsection{The Ponzano--Regge partition function for 3D gravity\label{sec_PR}}

For the sake of completeness, and for fixing notations, we will now briefly review the group-variable formulation of the Ponzano--Regge model in presence of boundaries.
We warn the mathematically inclined reader that many of the manipulations performed in this section are purely formal.
For a more precise---and thorough---review of the Ponzano--Regge (PR) model, complemented by a wide bibliography, see Part I.

Euclidean first-order three-dimensional gravity is formulated in terms of an $\su(2)$-valued spin connection $\omega=\omega^a_\mu(x)\tau_a \d x^\mu$ and an $\su(2)\cong\mathbb R^3$-valued dreibein one-form $e=e^a_\mu(x) \tau_a \d x^\mu$, where $\tau^a = -\tfrac{\I}{2} \sigma^a$ is a basis of $\mathfrak{su}(2)$.
In absence of a cosmological constant, these fields are combined into a $BF$ action,
\be
Z(M)=\int \mathcal D e \mathcal D \omega\; \E^{-\I S_{BF}[e,\omega]} 
\qquad\text{where}\qquad
S_{BF} = \frac{1}{2\ellpl}\int_M e_a \wedge F^a[\omega],
\ee
where $\ellpl = 8\pi G_\text{N}$, and $M$ is (for now) a closed topological three-manifold. {\it No} sum over topologies is or will be implemented (see Part I for more extensive comments on this point).

For on-shell (i.e. torsionless) connections and everywhere invertible dreibeins, the $BF$ action equals the second-order Einstein--Hilbert action for $g_{\mu\nu}=\delta_{ab}e^a_\mu e^b_\nu$.
Notice the presence of the imaginary unit in front of the action, even for Euclidean geometries.

Integrating out the dreibein fields $e$, one obtains
\be
Z(M) = \int \mathcal D\omega\; \delta(F[\omega]),
\ee
which formally computes the volume of the moduli space of flat spin-connections on $M$. 

To start making sense of this formula, a discretization is introduced.
The most general discrete structure one can introduce is that of an (embedded) two-complex.
In the rest of the paper we will work with the two-complex induced by a cellular decomposition $\Delta$ of $M$. 
The two-complex of interest is actually the Poincar\'e dual $\Delta^*$ to $\Delta$: to the edges $e$, triangles $t$, and tetrahedra $\sigma$ of $\Delta$, there correspond the faces $f=e^*$, links $l=t^*$, and nodes $n=\sigma^*$ of $\Delta^*$.
In particular, the faces and links of $\Delta^*$ are arbitrarily oriented objects, and for every pair $(l,f)$ with $l\in f$, the function $\epsilon(l,f)=\pm1$ if their orientations are compatible or not, respectively.
In order to re-write the above partition function, group elements $g_l$ are associated to the links. 
Giving them the interpretation of parallel transports of the spin-connection, i.e. $g_\ell = \mathrm P\exp \int_l \omega$, the partition function discretized on $\Delta^*$ reads
\be
Z_\text{PR-group}(\Delta) = \left[\prod_{l}\int_{\SU(2)} \d g_{l}\right] \prod_{f} \delta\left( \overleftarrow{\prod_{l:l\ni f}} g_{l}^{\epsilon(l,f)} \right),
\label{eq_PRgroupdef}
\ee
where $\d g$ is the normalized Haar measure on $\SU(2)$ and $\delta$ the corresponding Dirac delta distribution, $\int \d g \delta(g^{-1} h) f(g) = f(h)$.

Being generally divergent, this formula is still merely formal.
We will deal with this issue later (see also Part I). For now, we invite the reader to interpret it simply as calculating the Haar volume of the space of (discrete) flat connections supported on the 1-skeleton $\Gamma$ of $\Delta^*$.

In presence of boundaries, $\pp M \neq \emptyset$, the above formula for $Z_\text{PR-group}$ is readily generalized to a function of the {\it boundary} discrete connection:
\be
Z_\text{PR-group}(\Delta|g_{l_\pp}) = \left[\;\prod_{l\notin \pp\Delta^*}\int_{\SU(2)} \d g_{l}\right] \prod_{f} \delta\left( \overleftarrow{\prod_{l:l\ni f}} g_{l}^{\epsilon(l,f)} \right).
\ee
Here, we supposed that the cellular decomposition of $M$ induces a cellular decomposition $\pp\Delta$ of $\pp M$, so that to every boundary edge $e_\pp \in\pp \Delta$ there correspond by duality a boundary link $l_\pp\in\pp\Delta^*$, with an obvious shift in dimensions with respect to the bulk duality relation.

This formula (of course still potentially divergent) can be interpreted as computing the Haar volume of the space of discrete flat connections on $M$ {\it with a fixed pull-back on $\pp M$}. 

Using standard loop quantum gravity techniques, the space of boundary connections can be endowed with a Hilbert-space structure \cite{Baez1994,Ashtekar1995,Ashtekar1995a,MarolfMurao1995}.
Restricted to a particular graph $\Gamma$, this space is simply the space of gauge-invariant square-integrable functions
\be
\Psi(g_{l_\pp}) = \Psi(G_{t(l)} g_{l_\pp} G_{s(l)}^{-1})\in L^2(\SU(2)^{|l_\pp|})\qquad \forall G_n\in \SU(2),
\ee
with $n=t(l)$ or $s(l)$ the target and source nodes of $l$ respectively, and inner product
\be
\la \Psi | \Phi \ra = \left[\prod_{l_\pp\in\Gamma}\int_{\SU(2)} \d g_{l_\pp}\right] \overline{\Psi(g_{l_\pp})} \; \Phi(g_{l_\pp}).
\label{eq_I6}
\ee
These states will be more carefully analyzed in the next section.

As a side note, we observe that $Z_\text{PR-group}(\Delta|g_{l_\pp})$ defines a (non-normalized) state in the Hilbert space associated to $\Gamma=\pp\Delta^*$, hence (morally) providing a realization of the Atiyah--Segal axioms of topological field theory \cite{Atiyah1989}. In the same spirit, it is immediate to see that gluing two discretized manifold along the common boundary one obtains
\be
Z_\text{PR-group}(\overline{\Delta_1}\sqcup_{\Gamma} \Delta_2) = \la Z_\text{PR-group}(\Delta_1) | Z_\text{PR-group}(\Delta_2) \ra,
\ee
where $\Gamma=\overline{\pp\Delta_1^*}=\pp\Delta_2^*$, and where the overline stands for orientation reversal.

More relevantly for what will follow, we also observe that this construction also provides the definition of the transition amplitude for a {\it boundary state} $\Psi$ supported on $\Gamma$ {\it given a three-dimensional discretized manifold} $\Delta$, with $\pp\Delta^*=\Gamma$:\footnote{The orientation-reversal on the right-hand side is conventional.}
\be
\la \text{PR}|\Psi\ra \equiv \la Z_\text{PR-group}(\overline{\Delta}) | \Psi \ra = \left[\prod_{l_\pp\in\Gamma }\int_{\SU(2)} \d g_{l_\pp}\right] \overline{Z_\text{PR-group}(\Delta|g_{l_\pp}) }\Psi(g_{l_\pp}).
\label{eq_SNgroup}
\ee
The left-most term of this equation is a short-handed notation which will be used in the following.

This amplitude is the integral of the value of the boundary state over the moduli space of flat boundary connections {\it induced by a flat connection in the bulk}. Clearly this amplitude ``knows'' about the topology of the {\it bulk} of the manifold, in particular, it keeps track of the contractible and non-contractible cycles of the {\it solid} torus.

The interpretation of this amplitude is as follows: if the boundary of $\Delta$ has two disconnected components corresponding to $\overline{\Gamma_1}$ and $\Gamma_2$, then the previous formula calculates the transition amplitude between two states across the ``history'' represented by $\Delta$. On the other hand, if $\Delta$ has a single boundary component, the amplitude can be interpreted as in the Hartle--Hawking no-boundary proposal: as the ``probability'' of nucleation of a given state from nothing.

Formally, the amplitude is clearly invariant under changes of the {\it bulk} discretization, as long as $\Delta$ is fine enough to captures all the non-contractible cycles of $M$, i.e. its first homotopy group $\pi_1(M)$. 
In practice, the expression above is generally divergent due to redundancies of Dirac distributions. 
We will address this fact later on (see section \ref{sec_PRamplitude}).

\section{Boundary states \label{sec_bdrystate}}

\subsection{Spin network states}

In the previous section, the Hilbert space of boundary states has been introduced for convenience in the group polarization, i.e. as gauge-invariant functions of group elements associated to the links of a graph dual to the discretization of $\pp M$. 
This group elements were introduced as the discretized analogue of the parallel transports, or holonomies, of the spin connection $\omega$ along the links of $\Gamma$. 
The product of such holonomies along a closed face of $\G$ encodes (three--dimensional) curvature.

From the viewpoint of general relativity, one is however often interested in calculating the transition amplitude between hypersurface (intrinsic) geometries (see the introduction to Part I).
Hence the holonomy representation introduced above is rather inconvenient for this purpose.  A convenient basis to represent the intrinsic geometry is the spin network basis \cite{RovelliSmolin1995}, which at the same time efficiently encodes the state's gauge invariance.\footnote{Gauge invariant states can be also described systematically in the recently introduced fusion basis, which diagonalizes curvature and additional torsion operators \cite{ABC2016}.}

Gauge invariant spin network states are constructed as follows. To each link $l$ of the graph $\Gamma$---we henceforth drop the subscript $\pp$ for boundary links---one associates a Wigner (representation) matrix $\sqrt{d_j} D^{j_l}(g_l)_{M_lN_l}$ in the spin $j_l$ representation.  Here $d_{j_l}$ is the dimension of the representation space $V_{j_l}$.  To contract the $(M_l,N_l)$ magnetic indices of these matrices we introduce at each node $n$ of $\Gamma$ an intertwiner $\iota_n$. Intertwiners are tensors invariant under the $\SU(2)$ gauge action. More precisely, in the case of $k$ outgoing and $(m-k)$  ingoing links at the node $n$, the intertwiner is a map between the outgoing and the incoming spins,
\be
\iota_n :\,\, V_{j_1} \otimes \dots \otimes V_{j_k} \,\, \longrightarrow\,\, V_{j_{k+1}} \otimes \dots \otimes V_{j_m},
\ee
which commutes with the $\SU(2)$ action
\be
\iota_n\,\Big{[}D^{j_1}(G)\otimes\dots \otimes D^{j_k}(G)\Big{]}
= 
\Big{[}D^{j_{k+1}}(G^{-1}) \otimes \dots \otimes D^{j_m}(G^{-1})\Big{]}\,\iota_n
\quad \forall G\in\SU(2).
\ee
Thus, a spin network basis state can be written as%
\footnote{{ It is possible to write intertwiners as if all links were incoming, by using the isomoprhism between a representation of spin $j$ and its conjugate. In that case, one needs to insert an orientation switch on each link, which is realized by the $\su(2)$ structure map $\cJ$. However, since $\cJ^{2}=-1$, we can not completely get rid of the choice of an orientation for the links and changing the orientation of a link carrying the spin $j$ will produce a factor $(-1)^{2j}$. For a planar graph, it is possible to choose a canonical choice of orientation i.e. a Kasteleyn orientation (see \cite{Bonzom:2015ova}), although it is not clear how this can generalize to arbitrary spin networks.}\label{footnoteorientation}}
\be
\Psi_{j,\iota}( g_l) \,=\left(\bigotimes_{n\in \Gamma} \iota_n  \right) \bullet_\Gamma \left(\bigotimes_{l \in \Gamma}   \sqrt{d_{j_l}}  D^{j_l}(g_l)\right),
\ee
where $\bullet_\Gamma$ stands for the contraction of all the magnetic indices are as prescribed by the graph $\Gamma$.
For an (ortho)normal spin-network basis one needs an (ortho)normal basis of intertwiners.

For a three--valent node, the intertwiner is uniquely given by the Clebsch--Gordan coefficient associated to the three adjacent representations $j_1,j_2,j_3$. 
They are non-vanishing only if the three representations satisfy the triangle inequalities
\be
j_1 \leq j_2+j_3 \quad \text{and cyclic permutations thereof},
\ee
together with the extra integrality condition 
\be
j_1+j_2+j_3 \in {\mathbb N}  .
\ee
This hints at the fact that the spins $j_l$ can be interpreted as the lengths of the edges $e$ dual to the links $l$. This fact is confirmed by the construction of length-measuring operators associated to the edges of $\partial \Delta$, which is indeed diagonalized by the spin network basis \cite{RovelliSmolin1995}. 

In the present work, we consider boundary states associated to quadrangulations rather than triangulations.  In this case the dual graph has four--valent nodes and the intertwiners are not unique anymore. While the spins encode the edge-lengths of these quadrangles, the intertwiners encode degrees of freedom describing their actual shape. Notice that the quadrangles need not be planar. Indeed, it can be shown that an $s$-channel recoupling basis for the intertwiners corresponds to resolving the quadrangle into two triangles via a fixed-length diagonal (given by the recoupling spin), while keeping the extrinsic angle between the two triangles totally undetermined\footnote{An even more detailed analysis shows that these two quantites are actually conjugated in the sense of symplectic geometry, as in \cite{KapovichMillson1996}. See also \cite{HaggardHanRiello2016} and references therein for the curved case.} (it would be encoded in the $t$-channel recoupling). In the following section we will explain a specific choice for these intertwiners, which keeps the geometry of the quadrangles ``maximally classical''.

\subsection{Two-dimensional coherent spin-network states (LS intertwiner states)}

In this work we will consider a spin network state, based on a four--valent graph embedded in a 2-dimensional surface.  We will prescribe the spin labels  associated to the edges and associate so--called coherent intertwiners to the nodes of the graph. Such a state describes a (quantum) quadrangulation of the 2-dimensional surface with fixed edge lengths.  But each quadrangle carries an `internal' degree of freedom, which is described by a coherent intertwiner.
The coherent 4--valent intertwiner describes a state that is semi--classical for the `internal' degree of freedom of the quadrangle. This internal degree of freedom can be described in different ways. E.g. by picking a decomposition into two triangles given by the choice of one of its two diagonals, this degree of freedom can be encoded into the following pair of conjugated variables:\footnote{This pair of variables is the one identified by Kapovich and Millson \cite{KapovichMillson1996}.} the length of the given diagonal (intrinsic geometry) and the dihedral angle between the two triangles (extrinsic curvature).  The change to the other diagonal constitutes a canonical transformation.
Of course, these descriptions of the ``internal'' degree of freedom correspond to resolving the 4-valent nodes of the spin-network states into three-valent ones (in the $s$ and $t$ channels; the $u$ channel has no straightforward geometrical interpretation).
Choosing coherent intertwiners allows us to peak the state on a specific point of the phase space of the quadrangle internal degree of freedom.
In other words, it allows us to obtain a quadrangle with ``maximally classical'' intrinsic and extrinsic geometry. 
In particular, we will consider states of quadrangles peaked on flat configurations (vanishing extrinsic curvature {\it within} the quadrangle). See also Sections IV.B and VI.B2 of Part I.

Coherent intertwiners with the required properties have been introduced by Speziale and one of the authors in \cite{LivineSpeziale2007} and are often referred to as LS (coherent) intertwiners. Here we will need to reinterpret these intertwiners for a 2-dimensional geometry.\footnote{{Indeed, in the 3-dimensional context in which they were originally developed, such intertwiners are interpreted as  quantum tetrahedra in the four-valent case $m=4$, and more generally as quantum polyhedra for higher valencies\cite{Barbieri:1997ks,Freidel:2009ck,BianchiDonaSpeziale2010,Livine:2013tsa}. See also \cite{HaggardHanRiello2016,Charles:2016xzi} for a discussion of polyhedra in homogeneously curved space. In the context of canonical 3d gravity, the interpretation of $\SU(2)$ intertwiners as polygons was never truly developed beyond the interpretation of 3-valent intertwiners as quantum triangles.}
}  
In the following we will review the construction of the LS intertwiners as well as provide a reconstruction of the associated geometric objects in the context of 3D gravity.

\subsubsection{Overview of the construction}

Before delving into the technical details, let  us outline the content of the following section.

The first step is to switch to a more flexible setting, where the magnetic indices (labels of an orthonormal basis in $V_j$) are replaced by spinors $\xi\in\mathbb C^2$ (labels of an overcomplete basis of $V_{\frac12}$ tensored $2j$ times with itself).
The upshot is that spinors are readily interpretable as encoding vectors, which in turn can be interpreted as representing the edges of a polygon (a quadrangle in our case).
This last interpretation is made possible by the fact that sets of vectors which sum to zero are enough to provide an (over-)complete basis of the intertwiner space\footnote{If non-closing configuration are excluded, their contribution is suppressed in the large-spin limit \cite{Livine:2007vk}, which admits in turn a semiclassical interpretation in terms of Regge geometries.} \cite{ConradyFreidel2009,Freidel:2009nu,Freidel:2010tt,Livine:2013tsa}.

More precisely, spinors do not encode only vectors, but a full dreibein, i.e. a full reference frame in $\mathbb R^3$, plus---predictably---an extra sign. 
This fact can be used to canonically fix the phase of the spinor (again, up to a sign) in the case of planar polygons: while one of the dreibein vectors naturally encodes the sides of the polygons, the others can be used to encode its oriented normal in $\mathbb R^3$.

Hence, in the following section we proceed as follows: first we discuss how to encode dreibeins via spinors in $V_\frac{1}{2}\cong \mathbb{C}^2$, and how to use these to build coherent states in $V_j$, where $j$ sets a length-scale for the encoded ``quantum geometry''.
Then, we discuss how to combine these coherent states in $V_j$ into coherent intertwiners representing flat polygons in $\mathbb R^3$. And finally, we show how to combine all these coherent intertwiners into a spin-network state representing a quantum quadrangulation of a toroidal two-surface.

\subsubsection{Spinors, vectors, and dreibeins}

In a given representation space $V_j$, we can choose a basis $\{|j,m\ra\,,\; m=-j,\ldots,+j\}$, which diagonalizes the angular momentum operator $J_z$, in addition to the $\SU(2)$ Casimir $J^2=J_x^2+J_y^2+J_z^2$. We can thus think of a  basis element $|j,m\ra \in V_j$  as a quantum vector of length $j$ and $z$-projection $m$ (the eigenvalue of $J_z$). On the other hand the azimuthal direction of these quantum vectors is totally uncertain.
 
The case of $|j,j\ra$ is, however, peculiar: it has maximal $z$-projection and must hence be peaked along this direction. Indeed, this turns out to be the case, and moreover with the minimal allowed uncertainty: $|j,j\ra$ is a coherent state in $V_j$ representing a vector of length $j$ pointing in the $z$-direction. 
The (relative) uncertainty about the direction in which the vector is pointing decreases with $j$.%
\footnote{Computing the expectation values of the $\su(2)$ generators on the state $|j,j\ra$, we get $\la \vec{J}\ra=(0,0,j)$. In turn, we can compute the variance $\la \vec{J}^{2}\ra=j(j+1)$, which is simply given by the $\su(2)$ Casimir. Thus the state $|j,j\ra$ corresponds to a semi-classical vector of length $j$ in the $z$-direction,  peaked on $(0,0,j)$ with spread $\f1{\la \vec{J}\ra}\sqrt{\la \vec{J}^{2}\ra-\la \vec{J}\ra^{2}}\,\sim\f1{\sqrt{j}}$.
The corresponding polar angle $\theta$ can be estimated to be $\theta\approx\arccos\left(\frac{j}{\sqrt{j(j+1)}}\right)\approx\frac{1}{\sqrt j}\to 0$.
}

To obtain coherent states representing unit three--vectors pointing in an arbitrary direction $\hat n\in S_2$, we rotate $|j,j\ra$ appropriately. 
Hence, we introduce a state of the form  
\be
|j, \hat n\ra = D^j(G_{\hat n})|j,j\ra
\label{eq_coherentvect_predef}
\ee
where $G_{\hat n}\in\SU(2)$ is some element of $\SU(2)$ which in the vectorial (spin 1) representation corresponds to a rotation $R_{\hat n}$, that takes the $z$-axis onto the $\hat n$ direction. 
Of course, there are many such $G_{\hat n}$ and they all differ by an initial rotation around $\hat z$. 
This ambiguity translates into a choice of phase  for the vectors $|j,\hat n\ra$.%
\footnote{Indeed, the $S_2$ in which $\hat n$ lives is better understood as $\SU(2)/\U(1)$. This corresponds precisely to the Hopf fibration $S_2 \cong S_3/S_1$.}
This phase encodes a completion of $\hat n$ to an orthonormal frame and an additional sign.

To clarify how this works, we focus at first on spins $j=\tfrac12$. 
Since $V_{\frac12} \cong \mathbb C^2$, we introduce spinors
\be
|w\ra = \mat{c}{w^0 \\ w^1} = w^0 |\uparrow\ra + w^1 |\downarrow\ra \in V_{\frac12},
\ee
where $|\uparrow\ra$ ($|\downarrow\ra$) is the $V_\frac{1}{2}$ basis element with $m=+\tfrac12$ ($m=-\tfrac12$, respectively),
as well as the notation
\be
\la w | = \mat{cc}{\bar w^0 & \bar w^1} = \mat{c}{w^0 \\ w^1}^\dagger
\qquad\text{and}\qquad
| w ] = \mathcal J |w\ra = \mat{c}{- \bar w^1 \\ \bar w^0}. \label{II.15}
\ee
Note that  the map ${\mathcal J}$ defined in (\ref{II.15}) is  {\it anti}linear with the properties
\be
\mathcal J^2 = -1 
\qquad\text{and}\qquad
G \mathcal J |w\ra = \mathcal J G|w\ra \quad \forall G\in\SU(2),\; w\in \mathbb C^2.
\label{eq_Jproperties}
\ee
where we used that $j=\tfrac12$ is the defining matrix representation, i.e. $D^{1/2}(G)$ is given by $G$ itself.

We will denote by Greek letters, e.g. $|\xi\ra$, normalized spinors, $|\xi|^2 = \la \xi | \xi \ra = 1$.

We now review how spinors can be used to not only encode three--vectors but a full three dimensional reference frame $(\hat e_i)_i\equiv(\hat e_x,\hat e_y, \hat e_z)$ together with a scale (the spinor's norm) and an extra sign. 

First, fix a standard orthonormal reference frame in $\mathbb{R}^3$, and name it $(\hat x_i)_i \equiv (\hat x, \hat y, \hat z)$. Then, declare that this frame is represented by the spinor $|\up\ra =  \left|\tfrac12,\tfrac12\right\ra = (1,0)^t$ representing this reference frame.
Such a frame can be sent to any other frame $(\hat e_x,\hat e_y, \hat e_z)$ by a rotation parametrized by the Euler angles $(\psi,\theta,\phi)\in[0,2\pi)\times[0,\pi)\times[0,2\pi)$:
\be
\hat e_i = R(\psi,\theta,\phi)\hat x_i =  R_{\hat z}(\phi) R_{\hat y}(\theta) R_{\hat z}(\psi) \hat x_i.
\ee
We lift the above rotation to $\SU(2)$ by 
\be
R_{\hat x_i}(\varphi) \to G_{\hat x_i}(\varphi) = \E^{\varphi \hat x_i . \vec \tau}
\qquad\text{where}\qquad
\tau^i = - \frac{\I}{2} \sigma^i\,,
\ee
where the $\tau^{i}$ are the anti-Hermitian generators of the $\su(2)$ Lie algebra, satisfying $[\tau^i,\tau^j]=\epsilon^{ij}{}_k\tau^k$, and $\sigma^i$  are the three Pauli matrices normalized such that $(\sigma^{i})^{2}=\id$ for all $i$'s.%
Applying this lifted rotation to $|\up\ra$ one obtains a norm one spinor
\be
|\xi\ra = |\xi(\psi,\theta,\phi)\ra = G(\psi,\theta,\phi)|\up\ra 
= \E^{-\I\frac{\psi}{2}}\mat{c}{\E^{-\I\frac{\phi}{2}}\cos\left(\frac\theta2\right) \\ \E^{\I\frac{\phi}{2}}\sin\left(\frac\theta2\right) }.
\label{eq_xiexplicit}
\ee

Thus orthonormal frames---with a fixed  orientation---can be mapped to spinors of unit norm. The map is injective; it fails, however, to be surjective. 
This is due to the lift of $\SO(3)=\SU(2)/\mathbb{Z}_2$ to $\SU(2)$ and thus the above map's image constitutes precisely half the space of normalized spinors: if $|\xi\ra$ belongs to it, then $-|\xi\ra$ does not.
Clearly, this is because a rotation of $2\pi$ fails to take a spinor back to its original phase, and rather multiplies it by $-1$. 
Therefore, in order to cover the whole of $\mathbb C^2$, we  enlarge the range of $\psi$ to be $[0,4\pi)$. 
Of course this extra minus sign represents the orientation of a spin frame ``on the top of the dreibein frame''.

\begin{figure}
\begin{center}
\includegraphics[width=.3\textwidth]{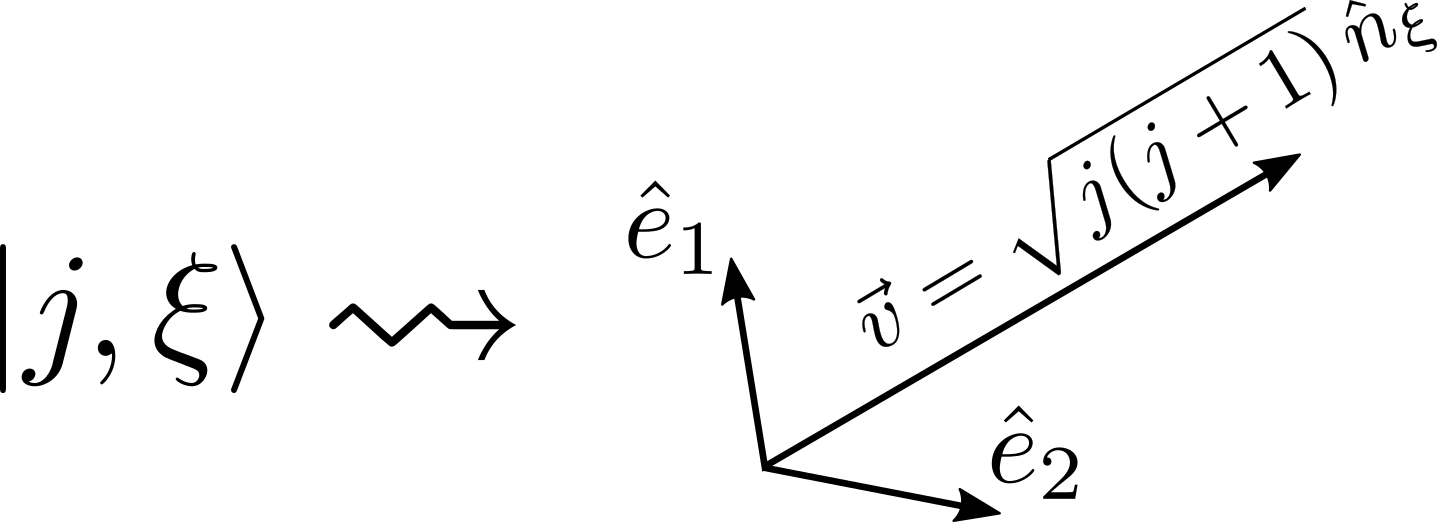}
\caption{The relation between a spinor $\xi$ with associated spin $j$, and the dreibein $(\hat e_x,\hat e_y, \hat e_z\equiv \hat n_\xi)$. The vector $\vec v = j \hat n_\xi$ has also been emphasized.
}
\label{fig_xi-3bein}
\end{center}
\end{figure}

The dreibein can be explicitly recovered from a spinor using (figure \ref{fig_xi-3bein}):%
\footnote{For more details see the very clear lecture notes by Samuel Gasster of the 1976's course ``Applied Geometric Algebra'' taught by Laszlo Tisza. They are available in \TeX format on the \href{{https://ocw.mit.edu/resources/res-8-001-applied-geometric-algebra-spring-2009/lecture-notes-contents/}}{\texttt{ocw.mit.edu}} website \cite{TiszaGasster2009}.}
\be
\hat n_\xi \equiv 
\hat e_z =  \la \xi | \vec \sigma |\xi\ra,
\qquad
\hat e_x + \I \hat e_y = [\xi|\vec \sigma |\xi\ra.
\qquad\text{and}\qquad
\hat e_x - \I \hat e_y = \la\xi|\vec \sigma |\xi].
\label{eq_dreibein}
\ee
Notice that $\xi \mapsto -\xi$ is the only norm-preserving operation on $\xi$ that leaves these formulas unchanged. 

To conclude the analysis of coherent dreibeins, we go back to elements of $V_j$ with arbitrary spin $j$.
Since $|j,j\ra = 
|\up\ra^{\otimes 2j}$, it is immediate that
\be
|j,\xi\ra = |\xi\ra^{\otimes 2j}
\label{eq_coherentvectdef}
\ee
can be taken as the proper definition of a coherent quantum vector $\vec v = j\, \hat n_\xi$, equipped with a local frame $\hat e_i$ defined as above.

Note that the states $|j,\xi\ra$ form an (overcomplete) basis of $V_j$:
\be
\mathbb 1_{V_j} = d_j \int_{\SU(2)} \d G \; G|j,j\ra\la j,j | G^{-1} = d_j \int_{\mathbb C^2\cong \mathbb R^4} \d^4 \xi\;\delta\left(|\xi|^2-1\right) \; |\xi\ra\la \xi | .
\label{eq_completeness}
\ee

\subsubsection{ LS intertwiners and coherent polygons}

With the notion of coherent vectors clarified, we can now proceed to construct coherent intertwiners which will encode polygons embedded in $\mathbb{R}^3$. 

These polygons can be described by a set of  $m\geq3$ three--vectors $\{\vec v_a\}_{a=1}^m$ which ``close",  i.e. sum up to zero (this set should be considered cyclically ordered).  The three--vectors can then  be taken as the edge vectors  of a (possibly degenerate) polygon, which in general will not be flat.  This polygon comes, however, also with a specified orientation in space.
If we want the polygon to be ``abstract'', i.e. loose the information about its orientation in $\mathbb{R}^3$, we need to take the set of  three-vectors modulo global rotations.

Quantum mechanically, the situation is very similar.
One takes a set of spins and normalized spinors such that the associated three--vectors have the properties above.

As we have seen in the previous section, however, spinors carry more information than simply the direction of these vectors, since they encode a full dreibein $(\hat e_x,\hat e_y, \hat e_z)$, as in equation \eqref{eq_dreibein}.
This additional information determines the phase of the states in $V_j$. To allow for a consistent reconstruction of the geometry we need to determine this phase consistently.

In the following, we will restrict to flat or planar polyhedra: in addition to satisfying the closure condition, the three--vectors are assumed to span a two--plane $\mathbb R^2\subset \mathbb R^3$.%
\footnote{The planarity condition is automatically satisfied for triangles.}
In this case we can demand that the dreibeins encoded in the spinors consistently describe the edge vectors of this polygon together with its plane and normal.
{ The phase specification for planar polygons is to our knowledge new.\footnote{ It is, however, interesting to compare it to the ``Regge phase'' condition of \cite{DowdallGomesHellmann2010}.}}

The definition up to rotations is finally provided by acting upon the polygon's state with an element of $\SU(2)$ and integrating over all such elements. As we will see, it is this operation that provides us in the end with an intertwiner which must be by definition rotation (or gauge) invariant.

Let us start by constructing a polygon in the $(xz)$-plane with normal pointing along the direction $+\hat y$. For this purpose, we have to specify a spinor variable  $\xi$, and thus the associated dreibein $(\hat e_x,\hat e_y,\hat e_z)$, for each edge of the polygon. This is done as follows: $\hat n_\xi = \hat e_z$ will provide the direction of the edge vector; $\hat e_x$ will be identified with a vector contained in the plane of the polygon and $\hat e_y$ with the normal to the polygon, i.e. $\hat e_y =\hat y$.

Implementing the latter condition in \eqref{eq_xiexplicit} it is easy to see that vectors $\vec v_a$ constituting such polygons in the $(xz)$ plane---if not parallel to $\hat z$---correspond to Euler angles $(\psi=0,\theta\neq 0,\phi=0)$ if $\hat x.\vec v_a>0$, or $(\psi=\pi,\theta\neq0,\phi=\pi)$ otherwise. 
Name these two classes of vectors ``positive'' and ``negative'' respectively, and label them accordingly with a $\pm$ sign.
In particular, two vectors pointing in opposite directions fall each in one of the two classes above, and moreover have $\theta^- = \pi - \theta^+$.
Therefore, their corresponding spinors are%
\footnote{The sign ambiguity is always present when going from vectors to spinors.}
\be\label{pmspinors}
\left| \xi^+ \right\ra= \pm \mat{c}{\cos\left(\tfrac\theta2\right) \\ \sin\left(\tfrac\theta2\right) } 
\qquad\text{and}\qquad
\left| \xi^- \right\ra= \pm\mat{c}{-\sin\left(\tfrac\theta2\right) \\ \cos\left(\tfrac\theta2\right) } =\pm | \xi^+].
\ee
These two classes of spinors are clearly distinguishable, since $\xi^+$ has two real components with the same sign, while $\xi^-$ has two real components with opposite signs.
Following this definition, we are led to associates to a polygon's edge pointing along $\pm\hat z$ the spinors $|\up\ra$ and $|\up]$, respectively.
For later use we also introduce the symbols $|+\ra$ and $|+]$ for the spinors characterized by $\left(\psi=0,\theta=\tfrac\pi2,\phi=0\right)$ and $\left(\psi=0,\theta=-\tfrac\pi2,\phi=0\right)$, respectively
\be
| + \ra = \tfrac{1}{\sqrt2}\mat{c}{1\\1}
\qquad\text{and}\qquad
| + ] =  \tfrac{1}{\sqrt2}\mat{c}{-1\\1}  .
\ee
They correspond to vectors pointing in the $+\hat x$  and $-\hat x$ direction, respectively. 

Note that the description above generalizes to arbitrarily rotated dreibeins: if $|\xi\ra$  represents the dreibein $(\hat e_x, \hat e_y,\hat e_z)$ then $|\xi] = {\mathcal J}| \xi\ra$ represents the dreibein $(-\hat e_x, +\hat e_y,-\hat e_z)$, that is ${\mathcal J}$  acts as rotation of $\pi$ radians in the plane $(\hat e_x,\hat e_z)$. 

We say that the polygon $\mathcal P$ is positively oriented if the orientation of its edge vectors agrees with that induced by its normal. (Notice that it is sufficient to invert the order of the spinors in the list above, $a\to m-a+1$, to obtain a polygon with opposite orientation---recall that $a=1,\dots,m$ labels the polygon's edges.)  

The $\mathbb Z_2$ ambiguity for the spinors, which  is reflected by the $\pm$ signs in (\ref{pmspinors}), can be fixed to a global one for $\mathcal P$ by requiring that the signs of the second component of all the spinors describing $\mathcal P$ agree. In this case we say that there exists a well-defined spinorial frame throughout $\mathcal P$.

Hence, we define a quantum polygon $\mathcal P$ on the $(xz)$-plane with normal pointing in direction $+\hat y$ and with side lengths $\{j_a\}_{a=1}^m$ to be given by a (cyclically) ordered set of representation vectors $\mathcal P = ( |j_a, \xi_a^{\pm_a}\ra)_{a=1}^m$ satisfying all the requirements discussed above. If a total ordering of the edge vectors is given we correspondingly define the based polygons $\mathcal P_\ast$ to be the polygon $\mathcal P$ with a distinguished vertex, given by the source vertex of the first edge vector.

\begin{figure}
\begin{center}
\includegraphics[width=.8\textwidth]{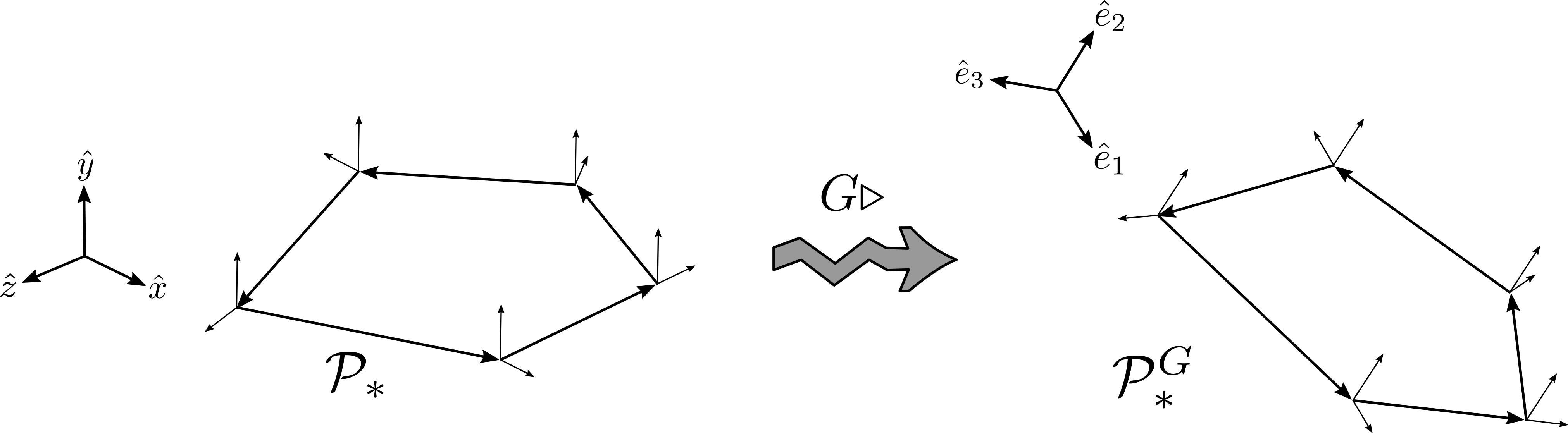}
\caption{A representation of the planar polygon $\mathcal P_\ast$ lying in the $(xz)$ plane, and of its arbitrarily oriented counterpart $\mathcal P^G_\ast$.}
\label{fig_polygon}
\end{center}
\end{figure}

A generic based polygon, arbitrarily oriented in space, is then obtained by diagonally acting with an element of $\SU(2)$, in the appropriate representation, on all the elements (edges) of a polygon $\mathcal P_\ast$.
 We will represent it as the state (figure \ref{fig_polygon})
\be
\left|\left|\,  \mathcal P^G_\ast(j_a, \xi_a) \,\right\ra\right. = G\triangleright |j_1, \xi_1\ra \otimes \dots \otimes |j_m,\xi_m\ra \in V_{j_1} \otimes \dots\otimes V_{j_m}.
\ee

As defined in \cite{LivineSpeziale2007}, a (coherent) LS intertwiner is finally defined as the rotationally invariant superposition of all rotated copies of a given based polygon:
\be\label{intII26}
\iota_{(j_a, \xi_a)} = \left|\left|\, (j_a, \xi_a)\, \right\ra\right. = \int_{\SU(2)} \d G \; \left|\left|\,  \mathcal P^G_\ast(j_a, \xi_a) \,\right\ra\right. \;\in \mathrm{Inv}\left(V_{j_1} \otimes \dots\otimes V_{j_m}\right).
\ee

As a side remark, let us notice that indeed any set $(|j_a,\xi_a\ra)_a$ (satisfying or not all the properties above) would define an intertwiner when fed into the previous two equations. The norm of such an intertwiner, $\bar \iota \cdot \iota =  \left\la\, (j_a, \xi_a)\,  \right|\left|\, (j_a, \xi_a)\, \right\ra$, would however be suppressed in the large-spin limit, if the associated vectors did not close \cite{LivineSpeziale2007}. Interestingly, it turns out that coherent intertwiners built out of a set of closing vectors are enough to provide an (over-)complete basis of the intertwiner space \cite{ConradyFreidel2009,Freidel:2009nu,Livine:2013tsa}, which means---in this two-dimensional context---that one can safely restrict the analysis to the space of polygons.

\subsubsection{LS spin-networks\label{sec_LSspin-networks}}

Having introduced coherent intertwiners, one can glue them together to form coherent spin network states  (e.g. \cite{Dupuis:2011fz}).  Here, we will call an LS spin-network is a spin-network whose intertwiners are all LS intertwiners.
These need to be slightly generalized in order to build a spin-network, to accommodate the fact that target nodes are in the contragradient (i.e. dual or complex conjugate) representation.
Unlike sometimes stated in the literature, the correct definition of this dualization involves not only a hermitian conjugation mapping ``kets'' into ``bras'', but also the antilinear map $\mathcal J$.
Geometrically, this corresponds  to the fact that the same edge appears with opposite orientations when seen from two neighboring polygons, provided these are consistently oriented.  

We thus need the following ingredients to define an LS spin-network:
For each node  $n$, with an ordered set of links $l$,  we construct the quantum polygon $({\cal P}_\ast)_n$, that is the associated ordered set of  $(|j_l,\xi^n_l \rangle)_{l \ni n}$.  This defines an intertwiner
\be
\iota_{(j_l, \xi^n_l)} = \left|\left|\, (j_l, \xi^n_l)\, \right\ra\right. = \int_{\SU(2)} \d G \; G\triangleright |j_1, \xi^n_1\ra \otimes \dots \otimes |j_m,\xi^n_m\ra .
\label{1intertw}
\ee

Assuming the spin-network graph is dual to a closed discretized surface (as in $\Gamma = \pp \Delta^*$), each of its links, $l$, carries two spinors $\xi^{s(l)}_l$ and $\xi^{t(l)}_l$ associated to the source $s(l)$ and the target node $t(l)$ of the link respectively. Of course, the spin $j_l$ has to agree for a consistent assignment. 

The definition of the intertwiner in (\ref{1intertw}) assumes all (half--)links to be outgoing, that is that the node is a source node for all adjacent links. Thus, we need to adjust this formula for the case that we have also ingoing links. 
Consequently, to any spinor  $\xi^{t(l)}_l$ in \eqref{1intertw} associated to an ingoing half-link we apply the replacement rule
\be
G | \xi_a^{t(l)} \rangle   \quad \mapsto    \quad  (G  {\cal J} |\xi_a^{t(l)} \rangle )^\dagger = [ \xi_a^{t(l)} | G^{-1}  .
\ee

Pairs of half--links meet at bi-valent vertices, and the replacement operation described above now allows us to contract neighboring intertwiners accordingly. In doing so, we have to insert the Wigner matrix $D^j(g_l)$ representing the holonomy between the two intertwiners, i.e. the (three-dimensional) parallel transport between the two polygons. The elements $g_l$ are precisely the argument of the spin-network state in the holonomy (or connection) polarization:
 \be
\Psi_{(j,\xi)} (g_l) = \left[ \prod_n \int_{\SU(2)} \d G_n\right] \prod_l [ \xi^{t(l)}_l| G_{t(l)}^{-1} g_l G_{s(l)} | \xi^{s(l)}_l\ra^{2j_l}  .
\label{eq_LSstate}
\ee
Notice that these states are holomorphic in  the components of the spinors entering the definition of the polygons, equation \eqref{1intertw}.

Evidently, these states are not normalized. This could be corrected by inserting factors of $\sqrt{d_j}$ for every link as well as a normalization factor for the LS intertwiners $\iota_{(j_a,\xi_a)}$. For the scope of this paper, there is, however, no particular interest in keeping track of these extra factors, which will therefore be neglected.

Finally, let us consider the behavior of the LS spin-network under the reversal of a link's orientation $l \mapsto l^{-1}$.
Spin network functions transform under a link reversal as
\be
\Psi(g_l, \dots) \xrightarrow{l\mapsto l^{-1}}   \Psi'(g_{l^{-1}}, \dots) \equiv \Psi(g_{l}^{-1}, \dots).
\label{eq_linkreversal}
\ee
This property can be enforced essentially by definition on our states, whose construction a priori depends on a chosen orientation of the graph links.
This dependence is actually mild as it can be shown by comparing the contribution given by a link $l$,
\be
[j_l,\xi_{t(l)}| G_{t(l)}^{-1} g_l G_{s(l)} | j_l,\xi_{s(l)} \ra
\ee
with that of its opposite oriented counterpart $l^{-1}$,
\begin{align}
[j_l,\xi_{t(l^{-1})}| G_{t(l^{-1})}^{-1} g_{l^{-1}}G_{s(l^{-1})} | j_l, \xi_{s(l^{-1})}
&=  [\xi_{s(l)}| G_{s(l)}^{-1} g_l G_{t(l)} | \xi_{t(l)} \ra^{2j_l} \notag\\
&  =  [G_{t(l)}^{-1} g^{-1}_l G_{s(l)}\xi_{s(l)}| \xi_{t(l)} \ra^{2j_l} \notag\\
& = (-1)^{2j} [j_l, \xi_{t(l)}| G_{t(l)}^{-1} g^{-1}_l G_{s(l)} | j_l, \xi_{s(l)} \ra ,
\end{align}
where in the second to last step we used the simple identity $[\xi|\eta\ra = - [\eta|\xi\ra$. 
From this simple computation it is clear that our construction automatically complies with \eqref{eq_linkreversal} possibly up to a sign (this sign becomes relevant only when one considers states with a superpositions of integer and half-integers spins).
For more on this point see footnote \ref{footnoteorientation} and, for a different more abstract perspective, also \cite[Section 2.5.1]{BarrettNaishGuzman2009}. Since these details are not relevant for the present paper, we will restrain from dealing with them in any greater detail.
~\\

Finally, let us recall that LS states play an important role in  four dimensional spinfoam models, where they are known to impose boundary conditions reproducing the GHY boundary term%
\footnote{More precisely, what is reproduced in the discrete spinfoam setting is the classical Regge--Hartle--Sorkin action. See the introduction to Part I.}
 on scales (much) larger than the Planck scale \cite{ConradyFreidel2008,BarrettEtAl2009,DowdallGomesHellmann2010,HanZhang2011,HanZhang2011a,HHKR2015,HHKR2016,SpezialeEtAl2017}. By ``much'' we mean in the formal large-spin limit. Nonetheless, the large-spin regime is often attained already for $j\sim10$ \cite{Livine:2006ab,Christensen:2007rv}.

\subsubsection{Geometric correspondence: from the single link amplitude to the Regge action
\label{sec_reconstr-overview}}

Before moving to the calculation of a LS spin-network amplitude for the twisted torus, let us elaborate on the geometric interpretation of the arguments of a link contribution:
\be
[j,\xi_{t}| G_{t}^{-1} g G_{s} |j, \xi_{s} \ra \equiv \la j,\mathcal J \xi_{t}| G_{t}^{-1} g G_{s} |j, \xi_{s} \ra,
\label{eq_linkamplitudedetailed}
\ee
where we have omitted the link label $l$ to avoid clutter.

Of course, we have discussed at length the meaning of the ``quantum edge vectors'' $|j, \xi_{s,t}\ra$, and in particular how they actually encode a full quantum dreibein. We recall that these vectors are written in a preferred, or standard, frame that we called $(\hat x, \hat y, \hat z)$. In particular, we will choose for our own construction in the next section, only spinors encoding dreibeins with $\hat e_y$ parallel to $+\hat y$, and hence $\hat n_\xi \equiv \hat e_z$ lying in the $(xz)$ plane. 

Now, the meaning of $G_n$ is to rotate these vectors in space. This rotations has two important features: it is common to all the spinors at one node---it corresponds indeed to a rotation to the polygon---and it is integrated over. The integration over $G_n$ has the algebraic role of implementing gauge invariance of $\Psi$ by geometrically removing any reference to the standard frame mentioned above. 

Finally, the geometrical meaning of $g$ is given by the very definition of the boundary states of the PR model: $g$ encodes the spin-connection holonomy between the two nodes of $\Gamma$ which are connected by the link $l$. In other words, it represent the parallel transport (lifted to $\SU(2)$) between the reference frames of the two polygons represented by $\iota_{t(l)}$ and $\iota_{s(l)}$. Also the spin-network arguments $g$ are integrated over---subject to flatness constraints---when calculating the state's amplitude $\la \text{PR} | \Psi \ra$, see \eqref{eq_SNgroup}.

Thus, we see that the link amplitude \eqref{eq_linkamplitudedetailed} calculates how much ``superposition'' there is between the frame at the edge $G_s|j,\xi_s\ra$, once parallel-transported by $g$, and the frame at the edge $G_t|j, \xi_t\ra$ with orientations appropriately changed by the map $\mathcal J$.  

At this point it is interesting to recall the fact that both $G_n$ and $g_l$ are eventually integrated over when calculating any physically relevant quantity.
This makes it meaningful to ask at which value these integrals happen to concentrate.
Intuitively---and slightly loosely---one expects these integrals to concentrate precisely where the superposition is maximal, which means at those value of the group variables which induce (provided it exists) a consistent gluing among {\it all} the edges of {\it all} the polygons in the discretization. Note how this is a {\it global} condition, which depends on the $G_n$'s and the $g_l$'s acting on different and interwoven partitions of the set of spinors.

Crucially, in the loose account above, ``maximal'' does not mean ``perfect'': the edge spinors will be allowed to coincide, after parallel transport, only up to a phase (this corresponds to maximizing the norm of \eqref{eq_linkamplitudedetailed}). This fact is crucial, because such a phase corresponds precisely to the dihedral angles $\psi$ between the polygons linked by $l$. In other words, the left-over phase encodes the extrinsic curvature between the two frames at the two nodes of $\Gamma$, hence contributing an edge-worth of the Regge--Hartle--Sorkin boundary action (think of this as a limiting version of the on-shell Gibbons--Hawking--York boundary term, for distributional extrinsic curvatures; see introduction to Part I):
\be
\text{for }j\gg 1, \qquad [j,\xi_{t}| G_{t}^{-1} g G_{s} |j, \xi_{s} \ra \sim \E^{-\I  j \psi} \qquad \text{where} \qquad j \approx {\ellpl}^{-1}\ell = (8\pi G_\text{N})^{-1}\ell,
\ee
with $\ell$ the length of the edge dual to link $l$. 

As we will see in great detail,  in the large spin limit, all of this can be made precise by a saddle point analysis.

\subsubsection{$\Z_{2}$-symmetry of LS intertwiners: from $\SU(2)$ to $\SO(3)$
\label{sec_SU2vsSO3}}

As we will sketch in the next section, and proved in detail in Part I, the Ponzano--Regge partition function in presence of boundaries amounts a certain evaluation of the boundary spin network states. Since we will define the boundary state in the present work as a semi-classical LS spin network, it is essential to understand the structure and symmetry of the LS spin network states. A key point, which has never been stressed before, is an enhanced $\Z_{2}$-gauge symmetry. This becomes crucial for the geometrical interpretation, since the integration group elements $G_{n}$ define the geometric angles and the $\Z_{2}$ allows to reduce the standard $4\pi$ periodicity of the phases of $\SU(2)$ group elements to the usual $2\pi$ periodicity of geometric angles.

Coming back to the explicit expression for the LS spin network wave-function given earlier in \eqref{eq_LSstate}:
\be
\Psi_{(j,\xi)} (g_l) = \left[ \prod_n \int_{\SU(2)} \d G_n\right] \prod_l [ \xi^{t(l)}_l| G_{t(l)}^{-1} g_l G_{s(l)} | \xi^{s(l)}_l\ra^{2j_l}
\,,
\nn
\ee
it is clear that the resulting function is invariant under local $\SU(2)$ transformation at each node, due to the group averaging over $G_{n}\in\SU(2)$.
On top of this, the integrand itself has an extra symmetry, where we can switch the sign of each $G_{n}$ group element individually, without changing the link amplitudes:
\be
G_{n}\mapsto -G_{n}\,.
\ee
The reason is that a necessary condition for a non-trivial intertwiner to exist is that the sum of the spins around each node must  be an integer.
This means, in turn, that the integrals $\int \d G_n$ can be meaningfully defined over $\SU(2)/\mathbb Z_2\cong \SO(3)$
\be
\int_{\SU(2)} \d G_n \leadsto \int_{\SU(2)/\mathbb Z_2} \d G_n =\int_{\SO(3)} \d G_{n}.
\label{eq_replacement}
\ee
More precisely, let us consider their parametrization as 2$\times$2 matrices:
\be
G(\psi,\hat{u})
=\E^{-\psi\hat{u}\cdot\vtau}
=\E^{\I\f\psi2\hat{u}\cdot\vsigma}
=\cos\f\psi2\,\id\,+\I\,\sin\f\psi2\,\hat{u}\cdot\vsigma
\,\quad\text{where}\quad
\psi\in[0,4\pi)
\,\quad\text{and}\quad
\hat{u}\in\mathbb{S}_{2}
\,,
\ee
and with the obvious redundancy $(\psi,\hat{u})\rightarrow (-\psi,-\hat{u})$. A sign switch $G\rightarrow -G $ corresponds to both the cosine and sine changing signs, i.e. to the mapping  $\psi\,\rightarrow \,\psi+2\pi$ while $\hat{u}$ remains unchanged. Therefore, quotienting by the extra $\Z_{2}$ symmetry exactly corresponds to modifying the periodicity condition for $\SU(2)$ group elements from the original $4\pi$ angle periodicity to the $2\pi$ angle periodicity of $\SO(3)$ group elements.

Quotienting by $\Z_{2}$ is by no means mandatory on any mathematical ground. Nevertheless, identifying $\Z_{2}$ as a spurious symmetry allows for a neater geometrical interpretation of the reconstructed geometry. Indeed it nicely resonates with the fact that the $G_n$'s are meant to represent the spatial orientation of the dreibeins encoded in the spinors.
Now working with $\SO(3)$ group elements, we actually need a lift to $\SU(2)$ group elements to write explicitly the integrand of the LS spin network wave-function. Nonetheless, the $\Z_{2}$ invariance makes the choice of  lift  completely irrelevant.%
\footnote{
A natural choice of section for the quotient $\SU(2)/\Z_{2}$ is to consider all group elements with $\cos\f\psi2\ge 0$, i.e. $\psi\in[0,2\pi)$, which corresponds to a projection $\SU(2)\rightarrow\SU(2)/\Z_{2}\sim\SO(3)$ defined by an absolute value map for $\SU(2)$ group elements:
\be
\big{|}G(\psi,\hat{u})\big{|}
=\,
\mathrm{sign}(\cos\tfrac\psi2)\, G\,.
\nn
\ee
}

\section{Twisted torus PR amplitude of an LS spin-network state \label{sec_PRamplLS}}

After having gone through a large amount of preliminary work, in this section we will finally be concerned with the actual study of the PR amplitude of the twisted (solid) torus spacetime%
\footnote{Recall, however, that we are in Euclidean signature.} 
with boundary conditions set by an LS spin-network state.
We will design this state so that it describes the intrinsic geometry of  {\it a homogenous (rectangular) quadrangulation} of the toroidal boundary.

Notice that LS states are precisely the boundary states which can be used to encode pure GHY-type\footnote{GHY stands for Gibbons--Hawking--York. See  the review section in Part I.}  boundary conditions on a quadrangulation, i.e. boundary conditions where only the intrinsic geometry is fixed.  The LS states will also allow us to apply a saddle point approximation to the partition function, that is, we will be able to evaluate the path integral to one--loop order.  

In fact, the rationale behind the use of a quadrangulation rather than a triangulation (as in \cite{DowdallGomesHellmann2010}) is that it allows us to perform calculations explicitly. In particular---in contrast to \cite{DowdallGomesHellmann2010}---we will be able to {\it explicitly} evaluate  the one--loop amplitude, and find agreement with the results of \cite{BonzomDittrich2015} in the context of perturbative quantum Regge calculus, and---in an appropriate sense---with those of \cite{BarnichEtAl2015} in the context of perturbative QFT of the metric perturbation.%
\footnote{The same result of \cite{BarnichEtAl2015} can be found as a particular limit of the same calculation in thermal AdS space \cite{GiombiEtAl2008}, which in turns coincides with the corresponding CFT one of \cite{MaloneyWitten2007}. Consistency with appropriate characters of the asymptotic symmetries (BMS$_3$ and Virasoro,  respectively) can also be checked \cite{BarnichOblak2014,Oblak2015}. See Part I for a review of all these results.}
The fact that any two of these three settings give compatible results is highly non-trivial, given the way the three calculations work.
For more comments on this fact, we refer the reader to the discussion section.

\subsection{LS state for homogeneous quadrangulations\label{sec_Lsstatedef}}

We start by introducing an LS intertwiner encoding a positively oriented quantum rectangle of ``horizontal'' (i.e. along $\hat x$) and ``vertical'' (along $\hat z$) side lengths $L, T\in\frac12\mathbb N$, respectively.

Hence, define $\mathcal Q_\ast$ to be the following based rectangle
\be
\mathcal Q_\ast = |L, +\ra \otimes  |T,\up ]\otimes | L , +] \otimes | T, \up\ra
\ee
It lies on the $(xz)$-plane, its normal points in the $\hat y$ direction, and is positively oriented (for orientation fixing purposes, the spinors are conventionally ordered right to left).
The corresponding LS intertwiner is 
\be
|| \, \mathcal Q \, \ra = \int_{\SU(2)}\d G\; G\triangleright \mathcal Q_\ast .
\ee

To quadrangulate the twisted torus consider first the homogeneous infinite rectangular lattice where each cell is labeled by discrete coordinates $(t,x)\in \mathbb N^2$, and then identify the cells according to the relation
\be
(t, x+N_x) \sim (t, x)\sim(t+N_t, x+N_\gamma), 
\ee
for some $0 \leq N_\gamma < N_x$. Notice that this implies a twisting  with angle
\be
\gamma = \frac{2\pi N_\gamma}{N_x}
\ee
before the cylinder with $\hat z$ axis is glued to a torus. 

\begin{figure}[t!]
\begin{center}
\includegraphics[width=.8\textwidth]{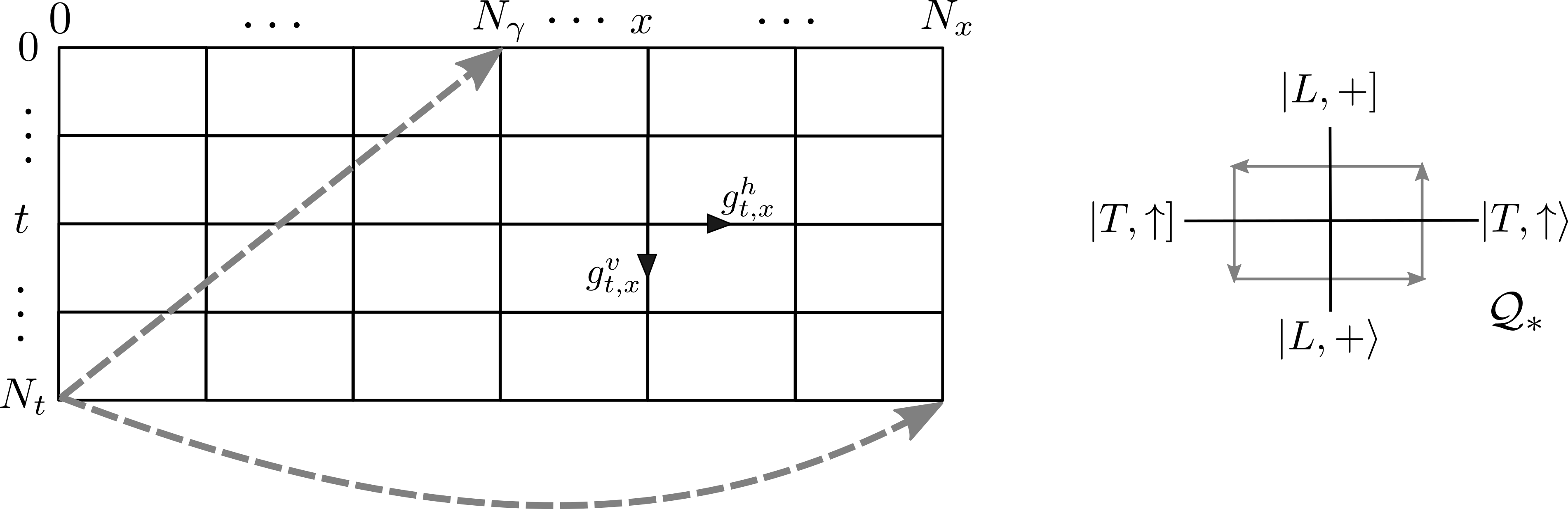}
\caption{{\it Left:} Graph and conventions for our LS spin-network state on the twisted torus. Dashed arrows emphasize the periodicity of the graph. {\it Right:} A representation of a node of the graph with its dual rectangle (gray arrows). }
\label{fig:lattice_Q}
\end{center}
\end{figure}

The quantum state describing the quantum twisted torus is then obtained by considering a spin-network graph dual  to the above quadrangulation, with LS intertwiner $|| \, \mathcal Q \, \ra$ at each of its nodes. The result is
\be
\Phi(g^h_{t,x},g_{t,x}^v) =  (-1)^{s} \left[ \prod_{(t,x)} \int_{\SU(2)/\mathbb Z_2} \d G_{t,x} \right] \, 
\prod_{t,x} \la \up| G_{t,x+1}^{-1} g^h_{t,x} G_{t,x} | \up\ra^{2T}  \la +| G_{t+1,x}^{-1} g^v_{t,x} G_{t,x} | +\ra^{2L},
\ee
where the replacement \eqref{eq_replacement} was implemented.
Here $h$ and $v$ refer to the direction of the spin-network links (either vertical or horizontal), see figure \ref{fig:lattice_Q}, and $s = 2(L+T)N_tN_x$.

The solid torus spacetime geometry is determined by choosing the ``vertical'' cycle to be non-contractible. As in a ``thermal Minkowski$_3$'', this choice identifies the $\hat z$ axis as the ``Wick-rotated'' time axis.\footnote{As well known, in quantum gravity the notion of Wick rotation is quite tricky. In particular we will use a model of quantum Euclidean GR, where each Riemannian geometry has a complex weigh $\E^{-\I S}$.}  
Correspondingly, the twist is along a spatial direction. 
This explains our notational conventions above---chosen for mnemonic reasons and in analogy with the Lorentzian calculations---where the letters $L$ and $T$ have been used.

Furthermore, we introduce 
\be
\beta = N_t T \ellpl.
\ee
as the inverse temperature, and 
\be
2\pi \tl a = N_x L \ellpl,
\ee
as the circumference of the contractible cycle of the torus.

\subsection{PR amplitude\label{sec_PRamplitude}}

\subsubsection{Gauge fixing and LS action}

To calculate the PR amplitude of the twisted solid tours with boundary state $\Phi$, we need to provide a cellular decomposition $\Delta$ of the manifold under investigation, $M=\mathbb D^2\times \mathbb S^1$, which is compatible with the boundary spin-network graph, $\Gamma=\pp\Delta^*$.

The PR being (formally) {\it bulk}-discretization independent, the choice of $\Delta$ can be performed out of mere convenience. 
In the companion paper, we detail the calculations for a ``wedding-cake'' discretization: the solid torus---think of it opened to a solid cylinder open in the ``time'' direction---is cut into horizontal layers with topology $\mathbb D^2\times[0,1]$, and each layer into vertical prism-like slices. 

Of course, as the formulas of section \ref{sec_PR} show, the formal amplitude is generally divergent. 
These divergences are indeed related to residual diffeomorphism symmetry of the internal vertices \cite{FreidelLouapre2003}  and need to be gauge-fixed.
After gauge-fixing, the amplitude turns into a well-defined expression, that can be shown to still be independent of the choice of bulk discretization.
The procedure of gauge-fixing is detailed in Part I, and in this simple case reduces to the removal of a few redundant Dirac distributions.

At this point, one is left with a well-defined expressions which integrates the boundary spin-network state over the moduli space of flat boundary connections {\it induced by a flat connection in the bulk}. Clearly this amplitude ``knows'' about the topology of the {\it bulk} of the manifold, in particular, it keeps track of the contractible and non-contractible cycles of the {\it solid} torus.

To write the amplitude in the form we will use in the following, it is now enough to use the $\SU(2)$-gauge invariance of the spin-network state as well as the invariance under translations (and the unit normalization) of the Haar measure:\footnote{In Part I it was more convenient to use $\theta_\text{there}:=\varphi_\text{here}/2$ as an integration variable. Beside the factor of 2, the two variables have the very same physical meaning.}
\begin{align}
\la \text{PR} | \Phi \ra 
&:= \left[\prod_{l_\pp}\int_{\SU(2)} \d g_{l_\pp}\right]   \overline{Z_\text{PR}(\Delta|g_{l_\pp}) } \, \Phi (g_{l_\pp}) \nn\\  
&=\left[ \prod_{t=0}^{N_t-1} \int_{\SU(2)} \d g_t \right] \, \Phi(g^h_{t,x} = \mathbb 1, g^v_{t,x} = g_t ) \nn\\
&=  \frac{1}{\pi}\int_0^{2\pi} \d \varphi \,\sin^2\left(\tfrac{\varphi}{2}\right) \,\, \, \Phi(g^h_{t,x} = \mathbb 1, g^v_{t,x} = \E^{\frac{\varphi}{N_t}\tau_z}),
\label{eq_PRtorus}
\end{align}
The first line is just the definition of the amplitude (see section \ref{sec_PR} for the notation, modulo the gauge-fixing of the redundant Dirac-distributions). The second line equality uses the fact that the ``spacelike'' cycle of the torus is contractible, plus the $\SU(2)$-gauge invariance to set all the horizontal holonomies $g^h_{t,x}$ equal to the identity. Indeed this gauge fixing, together with the flatness of the loops around each quadrangle implies that the holonomies in the time direction are homogeneous in space,  $g^v_{t,x}=g_t$.
Finally, the last line performs appropriate gauge transformations throughout time slices to distribute evenly the holonomies.
The meaningful group variable to evenly redistribute is $g=\prod_{t} g_{t}$.%
\footnote{Another way to proceed, as in Part I, is to perform suitable gauge transformations on all the time slices so as to gauge-fix to $g_{t}=\id$ on all the slices but the last one, leaving us with a single non-trivial holonomy $g^v_{N_t,x}\,=g=\overleftarrow\prod g_t$ on the last time slice. This group element $g=\prod g_t$ represents the holonomy along the only non-contractible cycle of the solid torus. Intuitively, one has cut open the torus around the time slice $t=N_t$, uses gauge invariance  to trivialize the flat connection throughout the resulting solid cylinder, thus pushing the non-triviality of the bundle to the transition functions between the two sides of the cut. Then, finally, one can redistribute this holonomy homogeneously throughout all the time slices.}
This group element was furthermore aligned along the $\hat z$-axis by use of a remaining global $\SU(2)$-symmetry, so that he integral can be expressed in terms of its class angle alone, i.e. $g=\E^{\varphi \tau_z}$ with $\varphi\in[0,2\pi)$. The choice of the $z$-axis is arbitrary and does not affect any of the results of the paper. This choice, however, does simplify some of the formulas.
Let us also emphasize that we integrate over $\varphi\in[0,2\pi)$, but that this overall angle is evenly redistributed throughout all the time slices, thus it is the angle $\f\varphi{N_{t}}$ that appears in the holonomies in the integrand. And the measure factor $\sin^{2}\left(\f\varphi2\right)$ comes from the Haar measure on $\SU(2)$  and ensures the equality in \eqref{eq_PRtorus}.

Choosing LS spin-network states---introduced in the previous section---as boundary states, we find:  
\begin{subequations} \label{2ampl}
\be
\la \text{PR} | \Phi \ra 
 =  (-1)^s\left[\frac{1}{\pi}\int_{0}^{2\pi} \d \varphi \,\sin^2\left(\tfrac{\varphi}{2}\right)  \prod_{(t,x)} \int_{\SU(2)} \d G_{t,x} \right]  \, \E^{-S (G_{t,x},\varphi)},
\ee 
where the newly defined {\it LS action} is
\be
 S  = - \sum_{t,x}
2T \ln \la \up| G_{t,x+1}^{-1}  G_{t,x} | \up\ra + 2L \ln  \la +| G_{t+1,x}^{-1} \E^{\frac{\varphi}{N_t}\tau_z} G_{t,x} | +\ra. 
\ee
\label{eq_PRtorusLS}
\end{subequations}

{Notice that the logarithm is just a mathematical shortcut and an abuse of notation, and in particular we do not require a choice of branch cut for the complex logarithm. Indeed, the exponential $\exp(-S)$ is well-defined and is all that actually matters: looking for stationary points of the action $S$ is exactly equivalent to looking for the stationary points of $\exp(-S)$. Nevertheless, the $\ln$-notation clarifies the role of the spins $T$ and $L$ as the parameters assumed to be large in the logic of a saddle point approximation of the integral.}

Although the whole amplitude seems entirely projected on the boundary state, information about the bulk appears in two places: first, it is encoded in the angle variable $\varphi$, which represent the holonomy around the torus non-contractible cycle, and, second, it appears in the presence of the twist, which is implicit in the boundary conditions we impose on the group elements $G_{t,x}$.
Finally, notice that there is a residual global symmetry in the LS action above, i.e. $G_{t,x} \mapsto \E^{\phi\tau_z} G_{t,x}$ for arbitrary phase shifts $\phi$. It can be resolved by fixing the initial group element $G_{0,0}$.

\subsubsection{$\SU(2)$ group elements vs. $\SO(3)$ group elements \label{sec_SU2vsSO3-PR}}

We obtained the Ponzano--Regge partition function in presence of boundary it terms of  the evaluation (and integration) of a boundary  LS spin-network state. Here the maths and geometry of the LS spin networks crucially enter into play, especially the $\Z_{2}$ symmetry and the effective projection $\SU(2)\rightarrow\SU(2)/\Z_{2}$ for the group elements $G_{t,x}$ that we discussed in section \eqref{sec_SU2vsSO3}.

Indeed, the group elements $G_{t,x}$ are defined up to a sign and can legitimately be considered as $\SO(3)$ group elements instead of $\SU(2)$ group elements, since the (exponential of the) action $S (G_{t,x},\varphi)$ is completely invariant under sign switches of the individual group elements $G_{t,x}$. The integral $\int \d G_{t,x}$ over $\SU(2)$ is truly an integral over $\SU(2)/\Z_{2}$:
\be
\int_{\SU(2)} \d G_{t,x}\; \E^{-S (G_{t,x},\varphi)}
\leadsto 
\int_{\SU(2)/\mathbb Z_2} \d G_{t,x}\; \E^{-S (G_{t,x},\varphi)}
\,.
\ee
In particular, the periodicity condition on the lattice becomes
\be
\pm^h_{t,x} G_{t,x+N_x} = G_{t,x} = \pm^v_{t,x} G_{t+N_t, x+N_\gamma}.
\label{eq_Gperiodicity}
\ee

At this point, we need to highlight that this modification is {\it not} about the Ponzano-Regge model being a gauge theory over $\SU(2)$ or $\SO(3)$. Indeed, the bulk and boundary holonomy of the Ponzano-Regge model are the $g_{t,x}^{h,v}$. They have all been gauge-fixed to $\id$ or to $\exp(\f\psi{N_{t}}\tau_{z})$. These remain legitimate $\SU(2)$ group elements, and the model still imposes  local flatness of the $\SU(2)$ bundle. The group elements $G_{t,x}$, on the other hand, are mere group-averaging variables introduced to define the LS intertwiner and boundary spin network state \eqref{eq_LSstate}. {\it The $\Z_{2}$-symmetry is  a property of the boundary state and does not change the definition of the bulk theory as a $\SU(2)$ gauge theory.}

The above up-to-a-sign periodicity  condition  will considerably simplify the geometrical interpretation of the stationary points of our action. They will correspond to quadrangulations of a ``cylinder'' whose section is characterized by a regular $N$-sided polygon with external dihedral angles  $\psi =2\pi/N_{x}$, instead of $\psi = 4\pi/N$ if we were to ignore this $\Z_{2}$ symmetry.

Let us nevertheless insist that the partition function $\la \text{PR} | \Phi \ra $ is not affected at all by integrating the $G_{t,x}$'s over $\SU(2)$ or $\SU(2)/\Z_{2}$. This switch in writing $\la \text{PR} | \Phi \ra $ as an integral is about clarifying the geometrical meaning of the variables appearing in the integral defining the partition function.

\subsection{Semi-classical or large-spin-limit\label{sec_SemiclassicalLargespin}}

Reinserting physical units for the spins, the LS action $S$ restricted to a link becomes schematically
\be
S_l \sim 2\frac{\ell_l}{\ellpl}\ln [\xi^{n'}_l | G_{n'}^{-1}g_l G_n|\xi^n_l \ra
\qquad\text{where}\qquad
\ell_l = j_l\ellpl.
\ee

Thus, committing to the discrete setting, that is to a {\it finite-resolution} boundary state, keeping the physical lengths $\ell_l$ fixed and sending $\hbar\to0$, provides a classical limit for the dual discrete theory.\footnote{Here we assume the Newton's constant to be held fixed, but the same formal limit could be obtained as a weak-gravity limit, $G_\text{N}\to0$, while keeping $\hbar$ and the physical lengths fixed. Of course, this holds only in the absence of matter.}
This limit is formally equivalent to a large-spin limit. 

We are going to study this limit, and the one-loop corrections, by means of a critical point approximation of the discrete path integral \eqref{eq_PRtorusLS}.

As is well known \cite{PonzanoRegge1968, Roberts:1998zka,KaminskiSteinhaus}  the PR amplitude for one tetrahedron does reduce to the one of quantum Regge action in the large-spin limit.,Here, however, we consider the PR amplitude for an entire triangulation and apply the large-spin-limit for the {\it boundary spins} only. In contrast, the bulk theory has been solved exactly, that is all bulk spin variables have been (morally\footnote{The actual sum would lead to the divergences we have regularized removing some Dirac distributions. This procedure corresponds to sum over only a subset of the spins while ``gauge-fixing'' the remaining one to zero.}) summed over.

\subsubsection{Critical point equations}

The dominant classical contribution is given by a critical configuration $o$, at which the real part of the action is an absolute minimum and its first derivative vanishes: \be
\Re(S)|_o \leq \Re(S) 
\qquad\text{and}\qquad
S' |_o = 0.
\label{eq_critical}
\ee

By the Cauchy--Schwarz inequality, for each link, one has schematically
\be
\Re( - \ln \la \xi | \xi' \ra ) = - \ln |\la \xi | \xi' \ra| \geq - \ln (|\xi| \cdot |\xi'|) = 0
\ee
and the equality sign holds if and only if $\xi \propto_{\mathbb C} \xi'$. 
Thus the first condition of \ref{eq_critical} leads to the following {\it gluing equations}%
\footnote{The choice of label $T$ or $L$ for the phases $\psi$ is dictated by the nature of the dual edge in the quadrangulation $\partial\Delta$: horizontal links are dual to ``time-like'' edges, and vertical links to ``space-like'' ones. Cf. the form of the LS action $S$, equation \eqref{eq_III20a}, and the next equation too.}
\begin{subequations}
\begin{align}
G_{t,x} |\up\ra & \,=\, \E^{\I\frac{\psi^T_{t,x}}{2}}G_{t,x+1} |\up\ra\\
G_{t,x} |+\ra & \,=\, \E^{\I\frac{\psi^L_{t,x}}{2}} \E^{-\frac{\varphi}{N_t} \tau_z}  G_{t+1,x} |+\ra
\end{align}
\label{eq_gluing}
\end{subequations}
which have to hold for some phases $\psi^{T,L}_{t,x}\in (-2\pi,2\pi]$.

On-shell of these equations, the LS action $S$ will take the form
\be
S|_o = - \I \sum_{t,x} T \psi^T_{t,x} + L \psi^L_{t,x}.
\label{eq_classicalaction}
\ee
We see that from the discrete gravitational action perspective it would be appealing to interpret these angles as the dihedral angles, so that the on-shell action above reproduces a discrete version of the GHY boundary term to the action, see section \ref{sec_reconstr-overview}. 
This expectation will indeed be confirmed by the geometrical analysis of the critical point equation. 

The unorthodox range of the angle variables $\psi$ is of course related to the $\SU(2)$ versus $\SO(3)$ discussion of section \ref{sec_SU2vsSO3}. We will come back to this point in the following.

Now, the stationarity condition $S'|_o=0$ is most easily studied by introducing right derivatives (left-invariant vector fields) of functions on $\SU(2)$. Schematically,\footnote{The unorthodox positioning of the indices is justified by later convenience.}
\be
\nabla^k f(G) = \frac{\pp}{\pp a^k}_{|\vec a =0 } f(G \E^{\vec a . \vec\tau}) =\frac{d}{d t}_{|t=0} f\left(G\E^{t\tau^k}\right) =  \frac{d}{d t}_{|t=0} f\Big(G(\mathbb{1} + t\tau^k)\Big).
\label{eq_graddef}
\ee
Thus, we obtain the first derivatives
\begin{subequations}
\begin{align}
\nabla^k_{t,x} S
= &   2T \frac{\la \up | \tau^k G_{t,x}^{-1}G_{t,x-1} |\up\ra }{\la \up | G_{t,x}^{-1}G_{t,x-1} |\up\ra} 
- 2T \frac{\la \up | G_{t,x+1}^{-1}G_{t,x} \tau^k |\up\ra }{\la \up | G_{t,x+1}^{-1}G_{t,x}  |\up\ra} +\notag\\
& + 2L \frac{\la + | \tau^k G_{t,x}^{-1}\E^{\frac{\varphi}{N_t} \tau_z} G_{t-1,x}  |+\ra }{\la + | G_{t,x}^{-1}\E^{\frac{\varphi}{N_t} \tau_z} G_{t-1,x}  |+ \ra}
- 2L \frac{\la + | G_{t+1,x}^{-1}\E^{\frac{\varphi}{N_t} \tau_z} G_{t,x}   \tau^k |+\ra }{\la + | G_{t+1,x}^{-1}\E^{\frac{\varphi}{N_t} \tau_z} G_{t,x}  |+ \ra},\\
\pp_\varphi S
= & -\frac{2L}{N_t} \sum_{t,x}  \frac{\la +| G_{t+1,x}^{-1}  \E^{\frac{\varphi}{N_t}\tau_z} \tau_z G_{t,x} | +\ra}{ \la +| G_{t+1,x}^{-1} \E^{\frac{\varphi}{N_t}\tau_z} G_{t,x} | +\ra}.
\end{align}
\label{eq_LSfirstvariations}
\end{subequations}
Evaluated on-shell of the gluing equations \eqref{eq_gluing}, they simplify to give the following stationarity conditions%
\footnote{The following identity for $G$ and $\hat n_\xi = \la\xi|\vec\sigma|\xi\ra$ was used,
$$
\hat n_{G\xi} = \la\xi | G^{-1} \vec \sigma G |\xi\ra =  \la\xi |\Big[G^{(1)}_{t,x} \triangleright  \vec \sigma \Big] |\xi\ra = G^{(1)}_{t,x} \triangleright \hat n_\xi.
$$
}
\begin{subequations}
\begin{align}
\nabla^k_{t,x} S |_o
= &   -\I\Big( T \la \up | \sigma^k |\up\ra 
- T \la \up | \sigma^k |\up\ra + L {\la + | \sigma^k  |+\ra }
- L \la + |   \sigma^k |+\ra\Big)
 \equiv 0, \label{eq_III20a}\\
\pp_\varphi  S|_o
= & \frac{\I L}{N_t} \sum_{t,x}  \la +| G_{t,x}^{-1}  \sigma_z G_{t,x} | +\ra %
= \frac{\I L}{N_t} \hat z. \sum_{t,x} G^{(1)}_{t,x}\triangleright \hat x = 0 \label{eq_stationarity}
\end{align} 
\end{subequations}
The first equation vanishes identically thanks to the closure condition, i.e. thanks to the fact that the intertwiners encode a semiclassical polygon.
The second equation, on the other hand, gives a global constraint the solution to the gluing equations must satisfy. Here, $G^{(1)}$ stands for the vectorial (spin 1) representation of $G$. 
We shall drop from now on the header $(1)$ if this does not cause confusion.

To ease the following analysis, define
\be
\tl G_{t,x} := \E^{- \frac{t}{N_t} \varphi \tau_z} G_{t,x}.
\ee
Notice that the periodicity condition \eqref{eq_Gperiodicity} now takes the form 
\be
 \pm_{t,x}^h \, \tl G_{t, x + N_x} = \tl G_{t,x} = \pm_{t,x}^v \, \E^{\varphi \tau_z} \tl G_{t+N_t,x+N_\gamma},
\label{eq_tlGperiodicity}
\ee
where the signs take into account the fact that $G_{t,x}$'s have been defined modulo $\mathbb Z_2$.
In term of these new variables, the gluing and saddle point equations read:
\begin{subequations}
\begin{align}
& \tl  G_{t,x} |\up\ra  = \E^{\I \frac{\psi^T_{t,x}}{2}} \tl G_{t,x+1} |\up\ra \label{eq_criticaleqs_a} ,\\
& \tl G_{t,x} |+\ra = \E^{\I \frac{\psi^L_{t,x}}{2}} \tl G_{t+1,x} |+\ra \label{eq_criticaleqs_b} , \\
& \hat z. \sum_{t,x} \tl G_{t,x} \triangleright \hat x = 0 \label{eq_criticaleqs_c}  .
\end{align}
\label{eq_criticaleqs}
\end{subequations}

\subsubsection{Geometrical interpretation of saddle point equations}

Using equation \eqref{eq_dreibein} to map spinors onto dreibeins, the gluing equations \eqref{eq_criticaleqs_a} and \eqref{eq_criticaleqs_b} can be translated into statements between any pair of reference frames $\hat e_i(t,x; l)$ associated to two adjacent $\tl G_{t,x}|\xi_l^{t,x}\ra$.
We are now going to show that these equations imply that, on the one hand, there is actually a single notion of what the boundary edge dual to the link $l$ is---a priori there is one from the perspective of vertex $s(l)$, and one from that of $t(l)$---hence the name ``gluing equations'', on the other, the normals to the cells of the quadrangulation can be twisted due to the presence of the phases $\psi_{t,x}$. As anticipated, these phases take the interpretation of extrinsic curvature.

To proceed, we explicitly write down the $\hat e_z$ component of \eqref{eq_criticaleqs_a} and \eqref{eq_criticaleqs_b} by sandwiching the Pauli matrices $\vec \sigma$ on both the left and right hand side of these equations, finding respectively
\begin{subequations}
\begin{align}
\tl G_{t,x} \triangleright \hat z & = \tl G_{t,x+1} \triangleright \hat z \label{eq_gluingvec3a} ,\\
\tl G_{t,x} \triangleright \hat x & =\tl  G_{t+1,x} \triangleright \hat x \label{eq_gluingvec3b} .
\end{align}
\label{eq_gluingvec3}
\end{subequations}
In words, these equations  state that---when brought to a common frame defined by the $\tl G_{t,x}$, rather than the standard frame $(\hat x, \hat y, \hat z)$ in which each rectangle $\mathcal Q$ has been originally defined---adjacent edges of the quadrangles must coincide.%
\footnote{A careful analysis of the origin of these equations shows that they are best understood as 
$$G|\xi\ra = \E^{\I\psi} G' \mathcal J|\xi'\ra,$$
except that we had already fixed $|\xi'\ra = |\xi]$ when we engineered $\mathcal Q_\ast$.
In the latter form, however, the gluing equations emphasize that each edge of the quadrangulation is identified with {\it minus} itself as seen from the neighbouring cells, as required by geometrical considerations (preservation of orientations). 
}

To understand the role of the phases $\psi_{t,x}$, the dreibein components $\hat e_x + \I \hat e_y$ have to be studied: 
\begin{subequations}
\begin{align}
\tl G_{t,x}\triangleright(\hat x + \I \hat y)  & = \E^{i{\psi^T_{t,x}}} \tl G_{t,x+1} \triangleright(\hat x + \I \hat y) \label{eq_gluingvec12a}\\
\tl G_{t,x}\triangleright(-\hat z + \I \hat y)  & = \E^{i{\psi^L_{t,x}}} \tl G_{t+1,x}  \triangleright(- \hat z + \I \hat y) \label{eq_gluingvec12b}
\end{align}
\label{eq_gluingvec12}
\end{subequations}
(Note that we have $\psi^{T,L}_{t,x}$  instead of $\psi^{T,L}_{t,x}/2$ appearing as we are using here the spin $1$ instead of the spin $1/2$ representation.)
These equations encode the rotation, that the plane orthogonal to the edge has to undergo in order to provide a matching between the frames.
Therefore, the phases $\psi^{T,L}_{t,x}$ encode precisely the dihedral angles between two neighbouring cells of the boundary quadrangulation. 

The geometrical meaning of the last saddle point equation \eqref{eq_criticaleqs_c}, the one coming from the variation of $\varphi$, is more subtle. Indeed, it constitutes a {\it global} constraint on the solution, and as such it has a quite different nature with respect to the gluing equations. 
This fact will become clear in the course of the next section.

\subsubsection{Solving the equations of motion}

{\it Preliminary note:} In this section we will be solving the equations of motion {\it as if} all phases $\psi^{T,L}_{t,x}$ and $\varphi$ are defined up to integer multiples of $2\pi$. This is not correct: all these phases are defined up to multiples of $4\pi$. Our ``mistake'' is a trick to look for all the solutions for the $\tl G_{t,x}$ which {\it are} defined up to a sign, since $\tl G_{t,x}\in\SU(2)/\mathbb Z_2$. For this reason we will rather work in $\SO(3)\cong\SU(2)/\mathbb Z_2$. In the next section, we shall consider a lift to $\SU(2)$ of the solutions we found in this way into the original equations, check their validity, and finally provide the actual values for $\psi^{T,L}_{t,x}$ and $\varphi$.\\

Equations \eqref{eq_criticaleqs_a} and \eqref{eq_criticaleqs_b} imply%
\footnote{$\{\vec J\}$ are the generators of three-dimensional rotation, i.e. of $\mathfrak{so}(3)$, i.e. $\vec J = \vec\tau^{(1)}$.  }
\be
\tl G^{-1}_{t,x}\tl G_{t,x+1} = \E^{\psi^T_{t,x}J_z}
\qquad\text{and}\qquad
\tl G^{-1}_{t,x}  \tl G_{t+1,x} = \E^{\psi^L_{t,x}J_x}.
\ee
Using these equations to go ``around'' a face in $\Gamma$, i.e. around four neighboring cells of $\Delta$, we obtain
\be
\E^{\psi^T_{t+1,x}J_z} = \tl G^{-1}_{t+1,x}\tl G_{t+1,x+1} = \E^{-\psi^L_{t,x}J_x}\tl G^{-1}_{t,x} \tl G_{t,x+1}\E^{\psi^L_{t,x+1}J_x} = \E^{-\psi^L_{t,x}J_x}  \E^{\psi^T_{t,x}J_z} \E^{\psi^L_{t,x+1}J_x}.
\label{eq_plaquette}
\ee
By uniqueness of the Euler decomposition, this equation has only two families of solutions, which we will name the $X$- and $Z$-family, respectively:
\be
X:\;\Big( \psi^L_{t,x} = \psi^L_t , \psi^T_{t,x} = 0\Big)
\qquad\text{and}\qquad
Z:\;\Big( \psi^L_{t,x} = 0 , \psi^T_{t,x} = \psi^T_x \Big)
\ee
(since the range of the $\psi_{t,x}^T$ is $(-\pi,\pi]$, there are actually a few other solutions. We discuss part of them later in this section, and part of them in section \ref{sec_pis}).

These two families of solutions lead to the Ansatz
\be
X:\;\tl G_{t,x} = \tl G_{0,0} \,\E^{\sum_{t'<t} \psi^L_{t'}J_x},
\qquad\text{and}\qquad
Z:\;\tl G_{t,x} = \tl G_{0,0}\, \E^{\sum_{x'<x} \psi^T_{x'}J_z}.
\ee

When expressed in terms of the variables $\tl G_{t,x}$, the gluing equations are totally symmetric in the directions $\hat x$ and  $\hat z$.
The asymmetry  between the two directions arises in the boundary conditions (\ref{eq_tlGperiodicity}) and is related to the presence of  the angle $\varphi$. 
This fact is natural considering the very origin of the variable $\varphi$ as encoding the holonomy around the non-trivial cycle of the solid torus.

We will now analyze one family of candidate solutions at a time. \\

\paragraph*{\bf $X$-family}
In terms of the $\tl G_{t,x}$, the role of $\varphi$ is encoded in the boundary conditions \eqref{eq_tlGperiodicity}.
The solution being constant in the horizontal direction, the corresponding periodicity is trivially satisfied.
One is left with the time periodicity condition
\be
\tl G_{0,0} \E^{\sum_{t'=1}^{t} \psi^L_{t'}J_x}
 = \E^{\varphi J_z} \tl G_{0,0} \,\E^{\sum_{t'=1}^{t+ N_t} \psi^L_{t'}J_x}.
\ee
i.e.%
\footnote{Notice, for $R\in\SO(3)$: $R^{-1}(\hat n. \vec J ) R = \hat n. (R\triangleright \vec J) = (R^{-1} \triangleright\hat n).\vec J $.}
\be
 \E^{-\sum_{t'=t}^{t+N_t} \psi^L_{t'}J_x} 
 =\E^{\varphi (\tl G_{0,0}\triangleright \hat z). \vec J}.
\ee
From this,
\be
\varphi  = - \sum_{t'=1}^{N_t} \psi^L_{t'} \; 
\mathrm{mod}\,2\pi
\qquad\text{and}\qquad
\tl G_{0,0}^{-1}\triangleright\hat z = \hat x
\qquad\text{with}\qquad
\psi^L_{t+N_t} = \psi^L_t
\mathrm{mod}\,2\pi,
\ee
where the second equation needs to hold unless $\varphi = 0$, in which case $\tl G_{0,0}$ is unconstrained. We will treat this case in Appendix \ref{app_n=0}. If $\varphi\neq0$, on the other hand, we can plug the ensuing solution into the global constraint \eqref{eq_criticaleqs_c} obtained from the stationarity condition on $\varphi$, to find a contradiction.%
\footnote{ 
The reasoning fails, however, when the sum in \eqref{eq_criticaleqs_c} vanishes on its own.
For the family of solutions above (with $N_x$ odd), however, this happens only if the timelike sequence of edges from $t=1$ to $t=N_t$ sums to zero (independently of $x$). This, in turns, gives back the condition $\varphi=0$ mod $2\pi$. Therefore, the above analysis, although apparently fallacious, covers all cases.
}
\\

\paragraph*{\bf $Z$-family}
In this case, conversely, one finds that both periodic boundary conditions provide non-trivial constraints:
\be
\tl G_{0,0}\, \E^{\sum_{x'=1}^{x+N_x} \psi^T_{x'}J_z}
 =  \tl G_{0,0} \,\E^{\sum_{x'=1}^x \psi^T_{x'}J_z}
 = \E^{\varphi J_z} \tl G_{0,0} \, \E^{\sum_{x'=1}^{x+ N_\gamma} \psi^T_{x'}J_z} .
\ee
From these we derive
\begin{subequations}
\be\label{subeq33a}
\sum_{x'=1}^{N_x} \psi^T_{x'}  =0 \;
\mathrm{mod}\,2\pi
\qquad
\varphi  = \sum_{x'=1}^{N_\gamma} \psi^T_{x'} \;
\mathrm{mod}\,2\pi
\qquad\text{and}\qquad
\tl G_{0,0}\triangleright\hat z = \hat z
\ee
with
\be
\psi^T_{x+N_x} = \psi^T_x\;
\mathrm{mod}\,2\pi
\qquad\text{and}\qquad
\psi^T_{x+N_\gamma} = \psi^T_x\;
\mathrm{mod}\,2\pi.
\label{eq_alphaperiodicity}
\ee
\end{subequations}

Notice that $\tl G_{0,0}$ is determined up to a rotation around $\hat z$. Name the corresponding angle $\bar\varphi$. This could have been expected from the fact that this rotation corresponds to the residual (global) symmetry left in the LS action. 
Hence,
\be
\tl G_{t,x} = \E^{\left(\bar\varphi + \sum_{x'=1}^x \psi^T_{x'}   \right)J_z}
\qquad\text{or equivalently}\qquad
G_{t,x} = \E^{\left(\bar\varphi + \frac{t}{N_t}\varphi + \sum_{x'=1}^x \psi^T_{x'}   \right)J_z}
\ee
which trivially satisfies equation \eqref{eq_criticaleqs_c}.

Now, using equation \eqref{eq_alphaperiodicity}, one can go even further.
In particular,
\be
\text{if}\quad K:=\mathrm{GCD}(N_\gamma, N_x) =1
\quad\text{then}\quad
\psi^T_x = \psi^T\;
\mathrm{mod}\,2\pi,
\ee
and hence from \eqref{subeq33a}
\be
\psi^T = \frac{2\pi}{N_x} n \;
\mathrm{mod}\,2\pi
\quad\text{and}\quad
\varphi = -\gamma n  \;
\mathrm{mod}\,2\pi
\label{eq_SOLUTION}
\ee
for some $n\in\mathbb Z $, $|n| \leq \lfloor\frac{N_x}{2}\rfloor$. \\

There are two cases which stand out, i.e. $n=0$ and---if $N_x$ is even---also $n=\frac{N_x}{2}$.

It is not complicated to see that the status of the $n=0$ solution is somewhat different, since it superposes to the allowed $\varphi=0$ case of the $X$-family solution. In particular it is part of a continuum set of solutions to the saddle point equations. In appendix \ref{app_n=0}, we will argue that the contribution associated to this solution by the saddle point approximation, is suppressed. For this reason, we will henceforth discard this solution altogether. 

The case of $n=\frac{N_x}{2}$ corresponds to $\psi^T_{t,x} = \pi$, and for such a value of $\psi^T_{t,x}$ equation \eqref{eq_plaquette} also admits a continuum set of solutions for which $\psi^L_{t,x+1}=-\psi^L_{t,x}$ and $\sum_{t'}\psi^L_{t',x}=0\,\text{mod}\,2\pi$. The origin of the continuum set of solutions is similar to the $n=0$ case, and for this reason it will also be discussed in appendix \ref{app_n=0}. Nonetheless, this contribution is not suppressed, and its full treatment is consequently much more subtle. Hence, beside where explicitly stated otherwise, we will restrict from now on to the case where $N_x$ is odd.

{
As a final remark, let us notice that the role of equation \eqref{eq_criticaleqs_c} is to select along which direction the embedded torus is bent. This is compatible with the fact that this is the equation of motion for $\varphi$, which is in turn the monodromy variable keeping track of which cycle of the torus is contractible in the bulk.\\
}

In summary, {\it if $N_x$ is odd and $K:=\mathrm{GCD}(N_\gamma, N_x)=1$, there is a finite number of (relevant) solutions labeled by an integer parameter $n$,  $1\leq|n|\leq \frac{N_x-1}{2}$. }
{\it If $K>1$, on the other hand, each of the solutions above is part of a continuum $(K-1)$-dimensional family of solutions.}

This last statement will be proven shortly.
To ease this task, and to gain insight into the solutions to the equations of motion, we will have to analyze the geometry they encode.

Before doing this, however, we have to go back to reconsider the interval of definition of the phases $\psi^{T,L}_{t,x}$ and $\varphi$.

\subsubsection{Lift to $\SU(2)$}\label{sec_4pi}

We will restrict our attention to the case $K=1$.
In this case, using the results of the previous section, we see that the most general candidate solution we have is given by 
\be
\text{for }K=1,\quad
\tl G_{t,x} = \E^{\left(\bar\varphi + \frac{2\pi}{N_x}n x  + 2\pi \epsilon_{t,x} \right)\tau_z}
\quad\text{or equivalently}\quad
G_{t,x} = \E^{\left(\bar\varphi - \frac{\gamma +2\pi n'}{N_t}t + \frac{2\pi}{N_x}n x  + 2\pi \epsilon_{t,x} \right)\tau_z},
\label{eq_SOLUTION_k=1}
\ee
where two solutions with different $\epsilon_{t,x}\in\{0,1\}$ have to be identified, since $\tl G_{t,x}$ and $G_{t,x}$ are elements of $\SU(2)/\mathbb Z_2$.

Evaluating these solutions at $(t,x+N_x)$ and $(t+N_t,x+N_\gamma)$, we find
\be
\tl G_{t,x+N_x} = (-1)^{n + \epsilon_{t,x+N_x} -\epsilon_{t,x}}  \tl G_{t,x}
\ee
and
\be
\tl G_{t+N_t,x+N_\gamma} = (-1)^{\epsilon_{t+N_t,x+N_\gamma} -\epsilon_{t,x}}\E^{\gamma n \tau_z} \tl G_{t,x},
\ee
respectively.
At the light of the $\tl G_n$ being in $\SU(2)/\mathbb Z_2$, the first equation is readily compatible with the space periodicity condition of \eqref{eq_tlGperiodicity}  for any value of $n$ and $\epsilon_{t,x}$.
A similar consideration applies to the second equation and the time periodicity condition as well. The only difference being that this equation also constraints the value of $\varphi$ (and it is the only one doing so). In particular it fixes
\be
\varphi = -\gamma n + 2\pi n',
\label{eq_varphiSOLUTION}
\ee
where we recall $1\leq|n|\leq \lfloor\frac{N_x}{2}\rfloor$, and $n'\in\mathbb Z$ is uniquely fixed by $\varphi\in[0,2\pi)$. 

Now, reinserting the above candidate solutions in the saddle point equations \eqref{eq_criticaleqs}, and using them to {\it find} the values of $\psi^{T,L}_{t,x}\in(-2\pi,2\pi]$, we get
\begin{subequations}
\begin{align}
& \psi^T_{t,x} =  \frac{2\pi n}{N_x} + 2\pi \left(\epsilon_{t,x+1} - \epsilon_{t,x} \right) \label{eq_criticaleqsBIS_a} \\
& \psi^L_{t,x} =  2\pi \left( \epsilon_{t+1,x} - \epsilon_{t,x} \right)  \label{eq_criticaleqsBIS_b} .
\end{align}
\label{eq_criticaleqsBIS}
\end{subequations}
Therefore we see that the phases $\psi^{T,L}_{t,x}$ take slightly different value for any choice of lift $\epsilon_{t,x}$ of the $\tl G_{t,x}$ to $\SU(2)$.
Although this might look troublesome, it is not so: recall that the phases $\psi$ where auxiliary objects useful to determine the value of the on-shell LS action, while the actual variables where the $\tl G_{t,x}$, themselves. 
Thus, the only thing that needs to be checked, is that the on-shell LS action does not depend on the lift to $\SU(2)$.
Of course, this must be so from the general arguments of section \ref{sec_SU2vsSO3}, but can also be readily verified in an explicitly manner from equations \eqref{eq_criticaleqsBIS} and \eqref{eq_classicalaction}.

We can now go back to the geometric interpretation of the solutions we found.

\subsubsection{Geometry reconstruction}\label{sec:reconstr}

The solution to the saddle point equations encode a twisted torus {\it locally} embedded in $\mathbb R^3$ as a quadrangulated cylinder of height $\beta = N_t T \ellpl$ and ``circumference'' $2\pi \tilde a = N_x L \ellpl$.
Recall that we refer to the horizontal direction of the torus as its ``spatial'' direction, and to the vertical one as its ``time'' direction (cf. section \ref{sec_Lsstatedef}).

The details of the geometry can be read from the data above by juxtaposing neighboring quadrilateral cells identifying their respective sides according to the gluing equations \eqref{eq_gluing} and orienting them in the embedding space according to the action of the $\tl G^{(1)}_{t,x}$. 

For $n=1$, this allows to build a right prism whose base is an $N_x$-sided polygon embedded in $\mathbb R^3$. The twisted torus is finally obtained by identifying the first and the last time slice after application of the twist encoded in the periodicity condition \eqref{eq_tlGperiodicity}.
The resulting spatial cycle is contractible in the bulk, while the time cycle is not due to the topological identification.
This is in agreement with the non-triviality of the holonomy $g=\E^{\varphi\tau_z}$ along the time cycle.
Between two spatially neighboring rectangular cells, there is a dihedral angle equal to $\psi^T=2\pi/N_x$, while the dihedral angle $\psi^L$ between two temporally neighboring cells vanishes (figure \ref{fig_cyl}).
\begin{figure}
\begin{center}
\includegraphics[width=.5\textwidth]{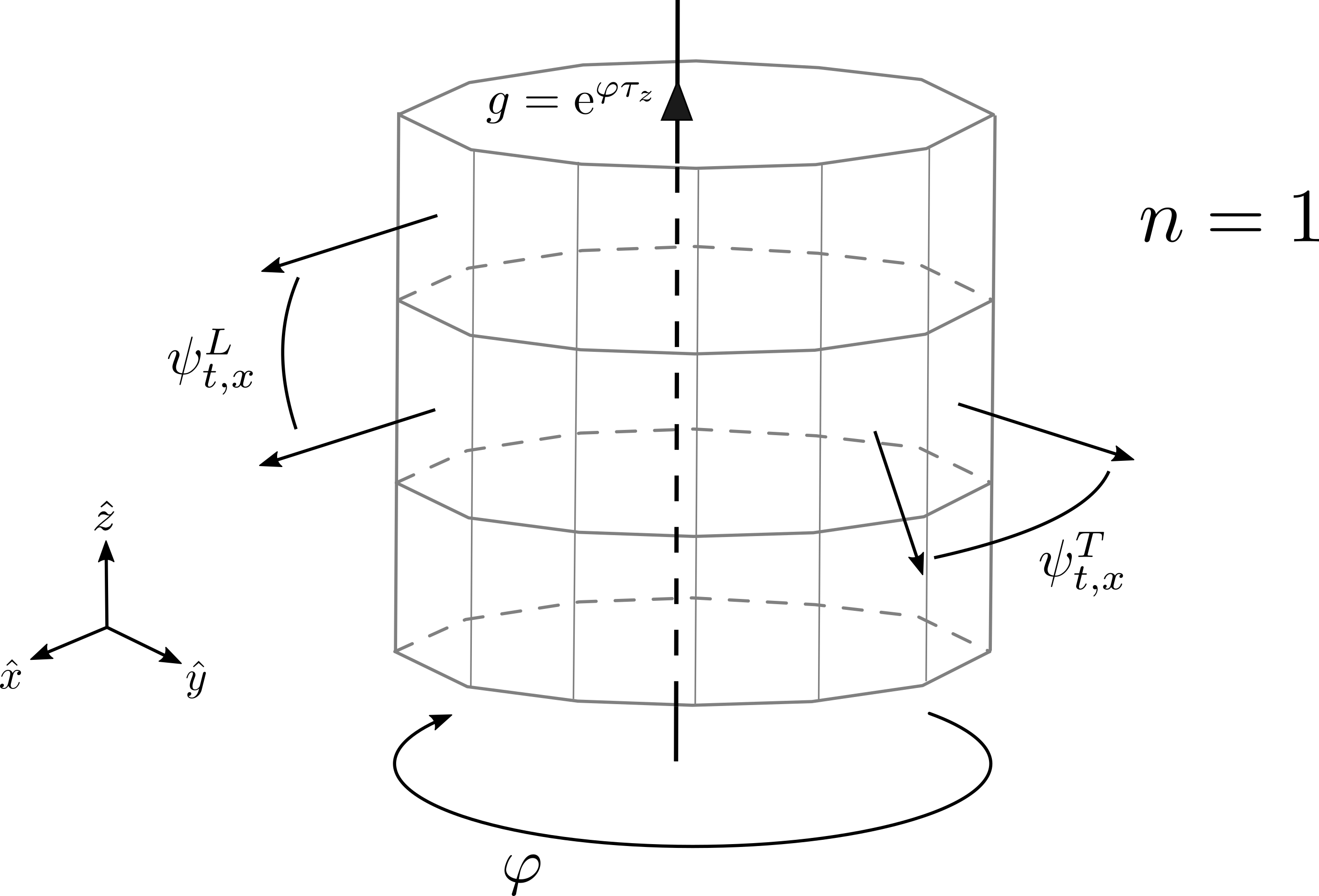}
\caption{The reconstructed toroidal geometry, for $n=1$, represented as a cylinder with ends identified up to a twist of an angle $\varphi$. The definition of the dihedral angles $\psi^{T,L}_{t,x}$ has also been highlighted. }
\label{fig_cyl}
\vspace{1cm}
\hspace{1.7cm}\includegraphics[width=.3\textwidth]{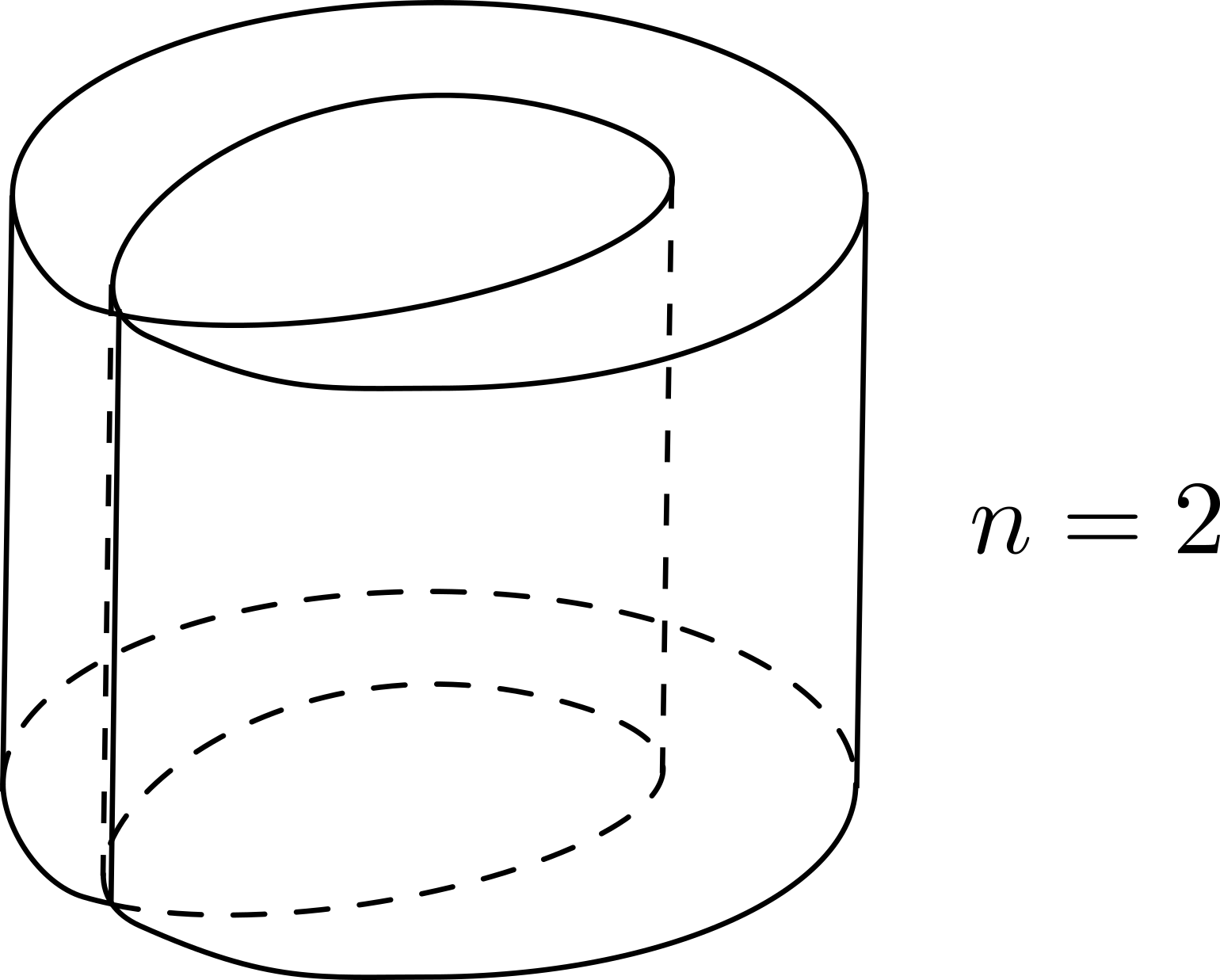}
\caption{A sketch of the surface reconstructed for $n=2$.}
\label{fig_cyl}
\vspace{1.5cm}
\includegraphics[width=.6\textwidth]{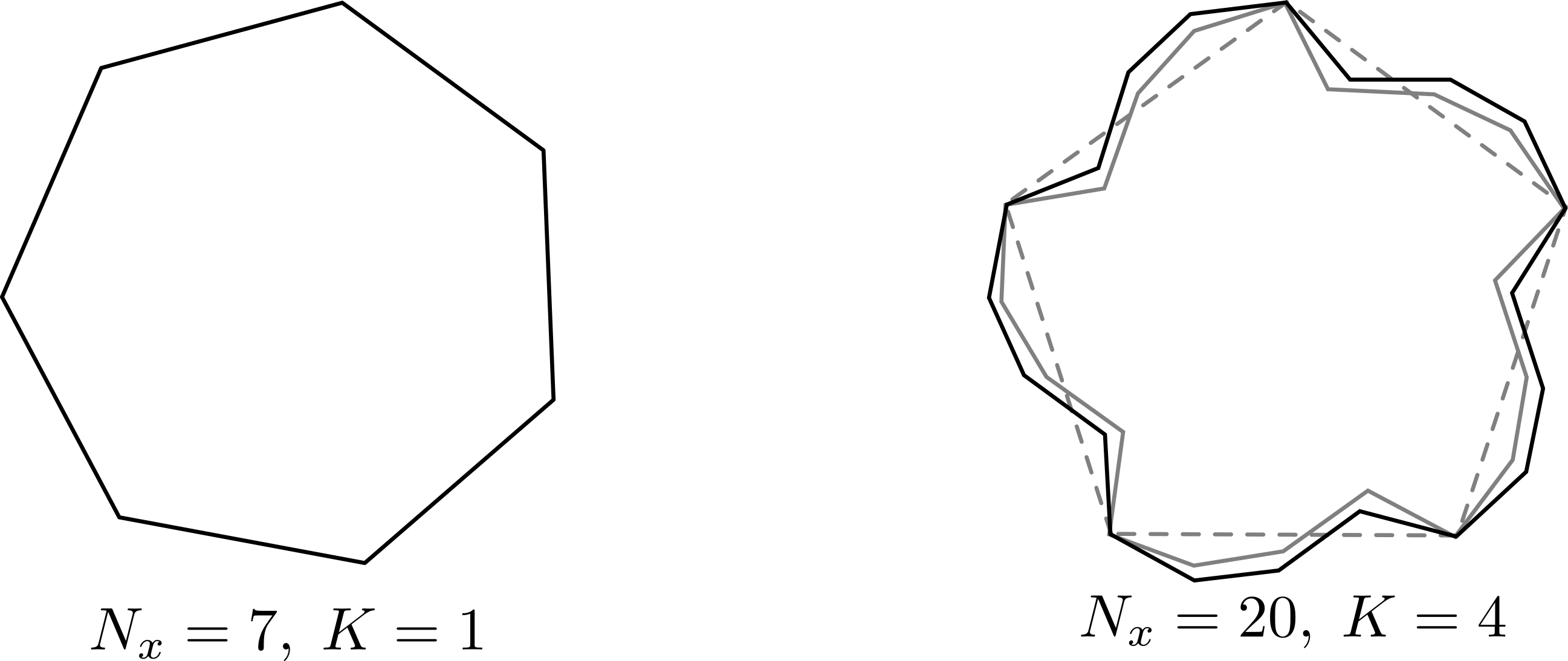}
\caption{The reconstruction of a single time slice for $K=1$ and $K>1$. In the second case, two infinitesimally close solutions of the saddle point equations are shown.}
\label{fig:constant_time_slice}
\end{center}
\end{figure}

For a generic $n\neq 0$, the surface of the cylinder wraps around itself exactly $n$ times before closing.
This surface cannot be embedded in $\mathbb R^3$ (it can, however, be immersed, see \cite{DowdallGomesHellmann2010}). 
As we anticipated, the case $n=0$ is peculiar and is discussed separately in appendix \ref{app_n=0}.

If $K:=\mathrm{GCD}(N_x,N_\gamma)=1$, the operation of hopping from one cell to the temporally following one takes the cell-hopper to visit {\it all} the cells before coming back to the initial one. This fact is what gives ``rigidity'' to the structure, and forces all the $\psi^T_x$ to be constant. Hence, for $K=1$, the prism described above has a regular polygon for a basis.

If $K>1$, on the other hand, the hopping procedure produces exactly $K$ independent closed cycles of cells.
The extrinsic geometry structure needs to be periodic only modulo $K$.
Considering groups of $K$ spatially consecutive cells as a single unit, we find again the same regular structure as the one discussed above for the regularly quadrangulated torus, the only difference being that the fundamental cells are now not-necessarily-planar polygons. 
As a consequence, one expects that a regular solution, $\tl G_{t,x} = \E^{\alpha_{t,x} \tau_z}$ with $\alpha_{t,x} = ( \bar\varphi + \tfrac{\varphi}{N_t} t + \psi^T x)$, can be deformed to another neighboring solution by adding first-order perturbations of the type 
\be
\alpha_{t,x} \mapsto \alpha_{t,x} +  \epsilon\sum_{m=1}^{K-1}\alpha_m\sin\left( \frac{2\pi }{K} m x \right).
\ee
where $\epsilon\ll1$.
It is easy to explicitly check that these are---at first order in $\epsilon$---still solutions of the equations of motion, at least if $N_x$ is even and $K$ is odd. In full generality, however, this fact is imprinted in the zeros of the 1-loop determinant.%
\footnote{Similar redundancies arise in the Regge calculus treatment of \cite{BonzomDittrich2015}, as discussed in Part I.}
 A constant-time section is sketched in figure \ref{fig:constant_time_slice}.

Finally, we comment on the interpretation of negative values of $n$.
Under the change $n\mapsto-n$, nothing major changes in the geometric interpretation apart from $\psi^T\mapsto-\psi^T$, globally.
The presence of two sectors of solutions for the dihedral angles is well-known in the PR model, and is in general attributed to the contribution of two oppositely oriented geometries. 
Indeed, fixing the boundary metric of a manifold, the saddle point analysis is supposed to determine the corresponding classical conjugated momentum (provided the chosen intrinsic metric admits one). The sign of the momentum cannot, however, be determined by this analysis due to time-reversal invariance. 
In gravity, such momentum is precisely the extrinsic curvature here encoded in the $\psi^T$.

\subsubsection{Foldings\label{sec_pis}}

In this section, we go back to the solutions of equation \eqref{eq_plaquette}.
The argument that led us to consider the $X$- and $Z$-families consisted made use of the uniqueness of the Euler decomposition of rotations. This, however, does not strictly apply to the present context, because the range of all the angles is $(-\pi,\pi]$ (recall we were in the setting where $\psi$ were ``artificially'' treated modulo $2\pi$).

We have already seen that for $\psi^T_{t,x}=\pi$ (which is only possible if $N_x$ is even) there is a continuum of solutions, which falls outside the $X$- and $Z$-family classification.

Similarly, the other solutions to equation \eqref{eq_plaquette} we have been missing are (the following equation is written for the moment for {\it one} value of $(t,x)$)
\be
\Big(\psi^T_{t,x} = \psi^T, \psi^T_{t+1,x} = - \psi^T,  \psi^L_{t,x} = \psi^L_{t,x+1} = \pi \Big).
\ee

These equations imply
\be
\tl G_{t,x+1} = \tl G_{t,x} \E^{\psi^T J_z}
\qquad\text{and}\qquad
\tl G_{t+1,x} = \tl G_{t,x} \E^{\pi J_x},
\ee
as well as 
\be
 \tl G_{t,x+1}\E^{\pi J_x} = \tl G_{t+1,x+1} = \tl G_{t+1,x} \E^{-\psi^TJ_z}  .
\ee
Extending these solutions homogeneously on a spacial slice, we see that the geometry encoded is that of a folding along a line of equal-time spatial edges of the quadrangulation. 
In particular, the difference in sign of $\psi^T_{t,x}$ from one time-slice to the next across the folding, means that the ``inside'' and the ``outside'' of the cylinder get swapped across the folding itself.

\begin{figure}[t]
\begin{center}
\includegraphics[width=.25\textwidth]{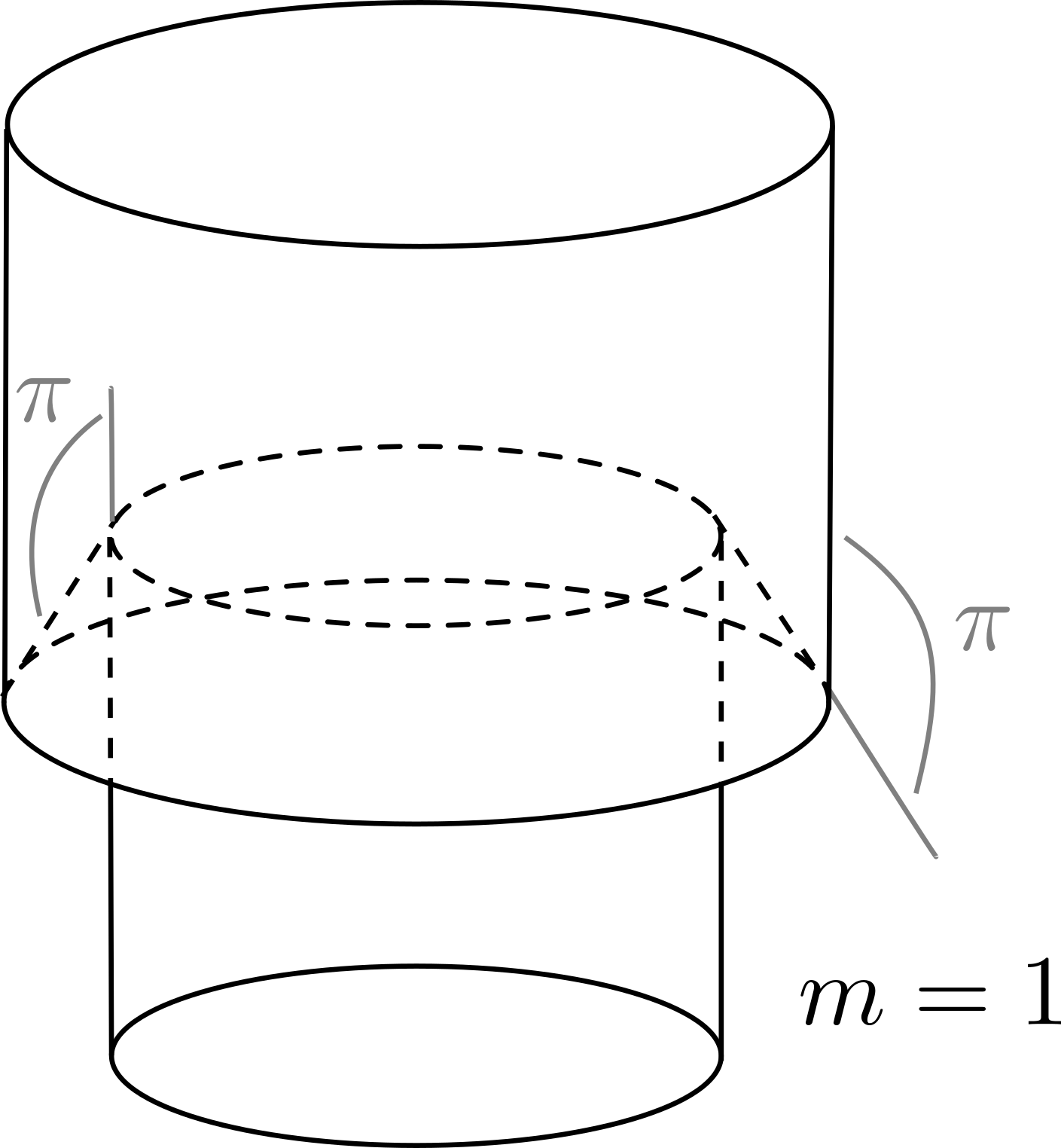}
\caption{A schematic representaiton of the folding for $m=1$.}
\label{fig_folding}
\end{center}
\end{figure}

Of course, periodicity in time enforces an even number $2m<N_t$ of such foldings. The case $m=1$ is depicted in figure \ref{fig_folding}.

The on-shell value of the LS action of one such configurations, for $\psi^T=\tfrac{2\pi}{N_x}n$, is given by
\be
S|_o = - 2\I \pi  T(N_t^+ - N_t^-)n   - 2\I  \pi L N_x m,
\ee
where $N_t^\pm$ are the number of time slices with positive and negative values of $\psi^T_{t,x}$, respectively.
E.g., if $m=0$, $N_t^+=N_t$ and $N_t^-=0$.
Thus, we find that the on-shell action effectively ``sees'' a shorter cylinder of inverse temperature
\be
\beta = (N_t^+ - N_t^-) T \ellpl.
\ee
The second term in the action simply counts the number of foldings.

\subsubsection{One-loop determinant}

In order to calculate the one-loop determinant, we need to develop the action to quadratic order around the solution of interest and calculate the ensuing Hessian.

First, fix a solution $o$ of the saddle point equations  $(G^o_{t,x}, \varphi^o)$
\be
G_{t,x}^o = \E^{\left(\bar\varphi+\frac{\varphi^o}{N_t} t + \psi^T_o x  \right)\tau_z},
\qquad\text{with}\qquad
\psi^T_o = \frac{2\pi}{N_x}n
\qquad\text{and}\qquad
\varphi^o = - N_\gamma \psi^T_o  + 2\pi n',
\ee
and where $\bar\varphi$ is an arbitrary global rotation parameter. 
This solution is valid for any $n$ and $K$. If $N_x$ is odd, $n\neq0$, and $K=1$, this solution is isolated---modulo the innocuous parameter $\bar\varphi$ which we will essentially ignore---and the Hessian of the LS action at this solution will be non-degenerate. 
As we will prove, confirming the claims of the previous section, this is not the case for $n=0$ or $K>1$.

Introduce, then, a parametrization of the linear perturbations around $(G^o_{t,x}, \varphi^o)$ by $\vec a_{t,x}\in\mathbb R^3$ and $\phi \in \mathbb R$ as follows:
\be
G_{t,x} = G^o_{t,x} \E^{\vec a_{t,x} . \vec \tau} 
\qquad\text{and}\qquad
\varphi = \varphi^o + \phi.
\ee

Hence, to second order
\begin{align}
S & = S_o
+ \left(%
\frac{1}{2}\left.\frac{\pp^2  S}{\pp{a^j_{s,y} }\pp{a^k_{t,x} }}\right|_o  a^j_{s,y} a^k_{t,x}%
+ \left.\frac{\pp^2  S}{\pp\phi \pp{a^k_{t,x} } }\right|_o       a^k_{t,x} \phi%
+ \frac{1}{2}\left.\frac{\pp^2  S}{\pp\phi^2} \right|_o  \phi^2 %
\right) + \mathrm o(a^2, \phi^2, a \phi)\notag\\
& = S_o
+\frac{1}{2} \left(%
 H^{j;k}_{s,y;t,x} a^j_{s,y} a^k_{t,x} %
+ (H^{k}_{\phi; t,x} + H^{k}_{t,x; \phi})    a^k_{t,x} \phi%
+  H_{\phi;\phi} \phi^2 %
\right) + \mathrm o(a^2, \phi^2, a \phi),
\end{align}
where in the last line we have introduced the following notation $H_{\alpha,\beta}$ for the Hessian matrix:
\begin{subequations}\label{Hessiandef1}
\begin{align}
H^{j;k}_{s,y;t,x} & = \frac{1}{2}\left.(\nabla^j_{s,y} \nabla^k_{t,x} + \nabla^k_{t,x} \nabla^j_{s,y} ) S \right|_o \\
 H^{k}_{t,x; \phi} & =  \left.  \nabla^k_{t,x} \pp_\phi S \right|_o  \\
  H^{k}_{\phi; t,x} & = \left. \pp_\phi \nabla^k_{t,x}  S \right|_o  =  H^{k}_{t,x; \phi}  \\
H_{\phi;\phi} & = \left. \pp^2_\phi S \right|_o 
\end{align}
\end{subequations}
where $\nabla^k_{t,x}|_o:=\pp/\pp a^k_{t,x}$ and $j,k=1,\ldots, 3$ are indices for the $\su(2)$ Lie algebra components.

The explicit form of the Hessian can be worked out by further deriving the first order variation of equation \eqref{eq_LSfirstvariations}. Details can be found in appendix \ref{appHess1}. Here, we proceed by giving the result.

~\\
\paragraph*{\bf Hessian matrix}
Organizing the $(1 + N_t\times N_x )$-dimensional perturbation vector as
\be
\bfa^T = \Big(\phi, (\vec a_{t=1,x=1}, \vec a_{t=1, x=2} , \cdots) , \cdots, (\vec a_{t=N_t,x=1}, \vec a_{t=N_t, x=2}, \cdots)\Big)^T,
\ee
the Hessian matrix of second derivatives of the action can be put into the form (empty entries are vanishing entries)
\be
\mathbf{H} = 
\mat{c|ccccc}%
{%
F & \DD & \DD & \DD &\cdots& \DD  \\\hline%
\DD^T &\GG & \CC & &   &\CC^T_\gamma \\%
\DD^T&\CC^T& \GG &\CC&&\\%
\DD^T&& \CC^T&\GG&\ddots&\\%
\vdots&&&\ddots&\ddots&\CC\\%
\DD^T&\CC_\gamma&&&\CC^T&\GG\\%
}.
\ee
This is a matrix made of $(1+ N_t)\times(1+N_t)$ blocks built as follows.

In the top left corner there is a $1\times1$-dimensional block (remember that $L$ and $T$ denote the values of the spins associated to the edges along the space and time direction, respectively),
\be
F\,=\,H_{\phi\phi}\,=\,\frac{L N_x}{2}.
\ee
The first row and the first columns are occupied by $\DD$ and its transpose, respectively, where $\DD$ is the $(3N_x)$-dimensional covector
\be
\DD=\overbrace{D\otimes \cdots \otimes D}^{N_x-\text{times}},
\ee
where $D$'s three components $k=1,2,3$ are
\be
 D_k\,=\, H^{k}_{t,x; \phi}
    \,=\, \frac{\I L}{N_t} \delta^k_2 .
\ee

Similarly, the blocks $\GG$ are ${3N_x\times 3N_x}$ blocks. They encode the spatial coupling of perturbations on a given time-slice. They are defined by 
\be
\GG =  \mat{ccccc}%
{%
A &B&&& B^T\\
B^T&A&B&&\\
&B^T&A&\ddots&\\
&&\ddots&\ddots&B\\
B&&&B^T&A
}
\ee
with $A$ and $B$ given by the following $3\times3$ matrices 
\begin{align}
 A^{jk}&\,=\, H^{j;k}_{t,x;t,x}
    \,=\,  T \left( \delta^{jk} - \delta^k_3\delta^j_3 \right) + L \left( \delta^{jk} - \delta^k_1\delta^j_1 \right) ,\\
    B^{jk} &\,=\, H^{j;k}_{t,x-1;t,x}\,=\,
     -\frac{T}{2}\left(  R_z(-\psi^T_o)^k{}_j  + \I R_z(-\psi^T_o)^k{}_i \epsilon^{ij3} -   \delta^k_3\delta^j_3    \right) .
\end{align}
($R_z(\alpha)$ denotes the matrix describing the rotation around the $z$--axis by an angle $\alpha$.)

The blocks $\CC$ are also ${3N_x\times 3N_x}$ blocks. They encode the coupling between subsequent time slices, 
\be
\CC =  \mat{cccc}%
{%
\,C\, &&&\\
&\,C\,&&\\
&&\,C\,&\\
&&&\,\ddots
},
\ee
with $C$ given by the $3\times3$ matrix 
\be
C^{jk}\,=\,H^{j;k}_{t-1,x;t,x} \, =\, -\frac{L}{2} \Big(  \delta^{kj} + \I\epsilon^{kj1} - \delta^k_1\delta^j_1 \Big) .
\ee
The first and last time-slices of the cylinder, however, couple under a twist of $N_\gamma$ units. This is encode in the shifted $\CC$ block we named $\CC_\gamma$:
\be
\CC_\gamma = 
\begin{array}{c} 1\\2\\\vdots\\N_\gamma-1\\ N_\gamma\\ N_\gamma+1\\\vdots\\N_x \end{array}
\mat{cccc|cccc}%
{%
&&&&\,C\,&&&\\
&&&&&\,C\,&&\\
&&&&&&\,\ddots&\\
&&&&&&&\,C\,\\
\hline
\,C\,&&&&&&&\\
&\,C\,&&&&&&\\
&&\,\ddots&&&&&\\
&&&\,C\,&&&&\\
}.
\ee

Developed at second order around the solution, the action can now be written as
\be
S = S_o + \frac12 \bfa. \mathbf{H} \bfa + \mathrm O(\bfa^2).
\ee
The one-loop determinant is therefore simply given by the determinant $\det(\mathbf H)$.

Notice that $\det(\mathbf H)$ is essentially a band matrix, but that its entries (almost) do not depend on the $(t,x)$ labels.  Such matrices can be diagonalized via a Fourier transform in $(t,x)$. We have however to introduce a `twist' due to the shifts appearing in the blocks $\CC_\gamma$. This will enable us to compute $\det(\mathbf H)$.\\

\paragraph*{\bf Twisted Fourier transform}
For the Fourier transform to respect the boundary conditions we imposed on the perturbations, we have to twist it as follows (indices in $\su(2)$ have been omitted in the following formulas)
\be
\hat a_{E,p} = \frac{1}{\sqrt{N_t N_x}} \sum_{t,x} \E^{\I\frac{2\pi}{N_x}p\left(x - \frac{N_\gamma}{N_t}t\right) + \I \frac{2\pi}{N_t} E t} a_{t,x},
\label{eq_TwFT}
\ee
Consequently, the Brillouin zone is reciprocally twisted:
\be
\hat a_{E+N_\gamma, p+N_x} = \hat a_{E,p} = \hat a_{E + N_t, p}.
\ee
Also, since $a_{t,x}\in \mathbb R$, one has
\be
\hat a_{-E,-p} = \hat a^*_{E, p} .
\ee
Or, in terms of momenta solely in the first Brillouin zone
\be
\hat a_{N_\gamma - E + N_t \Theta(E-N_\gamma)  , N_x -p}=\hat a^*_{E, p}
\ee
where $\Theta(s) =0$ if $s\leq 0$ and $\Theta(s)=1$ if $s>0$.
Excluded the case $(E,p)=(0,0)$, the above equation always relates modes at two different momenta, unless
\begin{subequations}
\be
\Big(\;p = N_x/2 \qquad\text{and}\qquad E = N_\gamma/2\;\Big),
\ee
or
\be
\qquad\quad\Big(\;p=N_x/2\qquad\text{and}\qquad E=(N_\gamma+N_t)/2\;\Big).
\ee
\end{subequations}\label{eq_oddeven}%
However, notice that the above $E$ and $p$ have to be integers to be admissible parameters in the Fourier transform. 
Therefore, the odd-$N_x$ requirement we introduced earlier on, turns out to automatically exclude this possibility.

Therefore, by considering the $\hat a_{E,p}$ as independent complex variables, one is exactly doubling each degree of freedom, exception made for $\hat a_{0,0}$ which is real. This remark will be useful when computing the 1-loop determinant.

Thus, in the Fourier-transformed basis, the Hessian matrix becomes a block-diagonal matrix with all its blocks $(3\times3)$-dimensional, but one. The latter is associated to the spacetime independent perturbations. In one line,
\begin{align}\label{HessianFT1}
\mathbf{b}. \mathbf H \bfa 
= & F \phi' \phi 
+ \Big( \phi'  {} D.\hat{{} a}_{0,0} + \phi\, \hat{{} b}_{0,0}. {} D \Big) + \sum_{E,p} \hat{{} b}{}_{-E,-p}. \hat{H}_{E, p}  \hat{{} a}_{E,p}
\end{align}
where 
\be
 \hat{H}_{E,p} =
 \mat{ccc}{ %
 T\left[1- \E^{-\I n \psi }\cos(\psi  p) \right] &- T\E^{-\I n \psi }\sin(\psi  p)   & 0 \\%
 T\E^{-\I n \psi }\sin(\psi p)   & T\left[1 - \E^{-\I n  \psi }\cos(\psi p)\right] + L\left[1- \cos\left (\chi_{E,p}\right)\right] & -L\sin\left (\chi_{E,p}\right)\\%
 0  & L\sin\left ( \chi_{E,p}\right)  & L\left [1 -\cos\left ( \chi_{E,p}\right)\right] }.
 \label{eq_hatHEp}
\ee
and
\be
\psi  = \frac{2\pi }{N_x},
\qquad
\chi_{E,p} = \frac{2\pi}{N_t} \left( E - \frac{ \gamma}{2\pi}p \right)
\qquad\text{and}\qquad
\gamma = \frac{2\pi N_\gamma}{N_x}.
\label{eq_psiANDchi}
\ee

~\\

\paragraph*{\bf One-loop amplitude}
Finally back to the PR amplitude 
\be
\la \text{PR} | \Phi \ra 
 =  (-1)^s\left[\frac{1}{\pi}\int_{0}^{2\pi} \d \varphi \,\sin^2\left(\tfrac{\varphi}{2}\right)  \prod_{(t,x)} \int_{\SU(2)} \d G_{t,x} \right]  \, \E^{-S (G_{t,x},\varphi)} ,
\ee
we evaluate the one-loop contribution of the $n$-th presented studied above.
This is given by the following Gaussian integral 
\begin{align}\label{AmplF1}
\la \text{PR} | \Phi \ra^\text{1-loop}_{o,n} 
& =2\pi\times \frac{1}{\pi}\sin^2\left( \frac{\gamma n}{2} \right) \times (-1)^s\E^{-S_o} \left( \int_{\mathbb R^+}\d \phi \int_{\mathbb R^3} \d\hat a_{0,0} \;  \E^{\frac12 F \phi^2 + \phi {} D. \hat{{} a}_{0,0} + \frac12 \hat{ a}{}^*_{0,0} \hat H'_{0,0} \hat a_{0,0}   } \right)\times\notag\\
&\hspace{5cm}\times\prod_{(E,p)\neq(0,0)}\left( \int_{\mathbb C^3} \d \hat a_{E,p}  \,  \E^{ \frac12  \hat{{} a}{}^*_{E,p} \hat H_{E,p} \hat a_{E,p}   } \right)^{1/2} ,
\end{align}
where we have introduced the notation $\hat H'_{0,0}$ for the matrix $\hat H_{0,0}$ from which one has removed the 3rd column and row. Indeed, these are identically zero as a consequence of the residual global symmetry of $S$, $G_{t,x} \mapsto \E^{\alpha \tau_z} G_{t,x}$.
This exact symmetry of the action and hence of the critical locus produces the $2\pi$ volume factor (essentially, an integral over $\bar\varphi$), i,e. the first factor in the expression above.

The second factor in the above expression, on the other hand, comes from evaluating the volume of the relevant conjugacy class of $\varphi=\varphi_o$, equation \eqref{eq_SOLUTION}.

The third factor is, of course, the value of the on-shell action at the given solution.

Finally, there appear the Gaussian integrals on the linear perturbations, which have been approximated---as usual---to be integrals over the full $\mathbb R^+$ or $\mathbb R^3$, rather than their original compact spaces.
Notice that we have used the trick mentioned at the end of the previous paragraph of doubling the degrees of freedom in the first Brillouin zone: for this reason we have taken a square root in the second series of integrals.

To evaluate the on-shell value of the LS action, we observe that from equation \eqref{eq_classicalaction}, and the values of the dihedral angles $\psi^T_o = \psi n$ and $\psi^L_o = 0$, one obtains%
\footnote{Recall that $L,T\in\frac12 \mathbb N$ is some (large) spin.}
\be
\E^{-S_o} = \E^{ \I 2 \pi T N_t n } = \E^{ \I \frac{2 \pi \beta}{\ellpl} n } = (-1)^{2T N_t n}  .
\label{eq_SoSL}
\ee
Thus, we see that, {\it although the on-shell action takes formally the expected form of an on-shell Regge--Hartle--Sorkin action (at least for the geometrically most natural case $n=1$), as a consequence of the discreteness of the lengths, this is just a sign factor.}

The Gaussian integrals appearing in (\ref{AmplF1}) are evaluated in appendix \ref{App_Gauss}, and are well defined whenever $N_x$ and $N_\gamma$ are such that $K:=\text{GCD}(N_x,N_\gamma)=1$.%
\footnote{ See the discussion sessions for comments on the case $K>1$.} 
Here, we give directly the result of these integrals.  We restrict the attention to the saddles with $m=0$ foldings, and $N_x$ odd (see appendix \ref{App_Gauss} and especially the second part of appendix \ref{app_n=0} for the $N_x$ even case).
We write
\begin{align}\label{AmplF2}
\la \text{PR} | \Phi \ra^\text{1-loop}_{o,n} 
& = {\cal D}(\gamma, n)  \times {\cal A}_\text{LS}(n) \times  (-1)^{s+2T N_t n} ,
\end{align}
where the label 1-loop means that the amplitude is evaluated in the  saddle point approximation, around the critical point $o$ labeled by $n$.

In this expression, the last factor is the contribution of the on-shell LS action $S|_o$ together with a sign $(-1)^s$ appearing in the definition of the state $\Phi$.
On the other hand, the first two factors come directly from the integration measure and from the evaluation of the Hessian at the saddle, and as such they are the 1-loop contribution to the amplitude. 

Let us analyze these two terms.
The first one,
${\cal A}_{\text{LS}}(n)$, does {\it not} depend on the twisting angle $\gamma$, and is explicitly given by 
\begin{align}
{\cal A}_\text{LS}(n)  =&\frac12 \left(\frac{(2\pi)^3}{LT(1-\E^{-i\psi n})\left( \frac{L}{4 N_t^2} + \frac{TN_t}{2}(1- \E^{-\I\psi n}) \right) }\right)^{1/2} 
 \times\notag\\
 &\times
 \left(\frac{(2\pi)^3}{LT \E^{-\I \psi n}}\right)^{\frac{N_x N_t -1}{2}}%
\left(\frac{1}{(L+T)(\cos(n\psi)-1) + \I L \sin(n\psi)}\right)^{\frac{N_t-1}{2}} \frac{1}{N_t} 
\times\notag\\%
&\times%
\prod_{p=1}^{\frac{N_x-1}{2}} \left(   \frac{1}{(L+T)(\cos(n\psi)-\cos(p\psi)) + \I L \sin(n\psi)}  %
\right)^{N_t},
\label{eqn:ALS}
\end{align}
(recall that $\psi=2\pi/N_x$ is the dihedral angle).
Notice that for $n\mapsto -n$, $\mathcal A_\LS$ goes into its complex conjugate:
\be
\mathcal A_\LS(-n) = \overline{\mathcal A_\LS(n)}
\ee
The other terms are invariant under this replacement. Thus, in the 1-loop approximation, the amplitude is real as a consequence of the combination of equal solution with opposite orientations.

As a function of $n$, ${\cal A}_\LS$ is peaked around $n=1$, which is the ``geometric'' configuration, and around $n= (N_x-1)/2$, which corresponds to a very ``crampled'' saddle---the latter turns out to be the true maximum (this peakedness is greatly enhanced at large $(N_t, N_x)$, especially so around the true maximum).
As discussed in appendix \ref{app_n=0}, this is  a consequence of the fact that at (boundary-) flat%
\footnote{We mean configurations for which the reconstructed 2D boundary is planar.}
configurations new ``gauge'' symmetries emerge, leading to vanishing Hessians. Although the exactly planar configurations, where the symmetry is actually present, must be treated separately and might even be suppressed (as it happens for the $n=0$ case), the configurations which are {\it close} to flat turn out to be enhanced. In \cite{Dittrich2008,BahrDittrich2009a,Rovelli2011a} a qualitative similar mechanism was discussed. This phenomenon appears, however, to be strictly related to the specific regular quadrangulation we choose. A possible strategy to select the geometrical stationary point is to make use of coherent boundary states obtained as superpositions of different values of the spins as well, in such a way to peak amplitude on a specific value of the extrinsic geoemetry too \cite{ThiemannCohStates,BahrThiemannCohStates1,Rovelli2006,LivineSpeziale2006,BianchiMagliaroPerini2009}.

The second term is certainly the most interesting contribution to the amplitude (beside $S|_o$ which however reduces to a mere sign contribution), since it encodes the amplitude's dependence on the twisting angle $\gamma$:
\be\label{Dete1}
{\cal D} (\gamma,n)
=  4 \sin^2 \left( \frac{\gamma n}{2} \right )  \times  \prod_{p=1}^{\frac{N_x-1}{2}}       \frac{1}{ 2-2\cos(\gamma p) } 
= \big(2-2\cos(\gamma n)\big)  \times  \prod_{p=1}^{\frac{N_x-1}{2}}       \frac{1}{ 2-2\cos(\gamma p) } 
\,,
\qquad
\gamma=\f{2\pi N_{\gamma}}{N_{x}}
\,. 
\ee
We  distinguish the front factor, which  comes directly from the integration measure over $\varphi$, from the product over $p$, which gives the contribution from  (part of) the Hessian.

The measure factor cancels exactly the contribution of the Fourier mode $p=n$. For the first winding mode $n=1$ we get a truncated product starting at $n=2$,
\be
{\cal D} (\gamma,n=1)
=   \prod_{p=2}^{\frac{N_x-1}{2}}       \frac{1}{ 2-2\cos(\gamma p) } 
\,.
\ee
This reproduces the results derived in \cite{BonzomDittrich2015} obtained from the path integral of Regge calculus, and fitting the 1-loop quantum General Relativity and BMS$_3$-character calculations of \cite{BarnichEtAl2015,BarnichOblak2014,Oblak2015}. This first winding mode $n=1$ allows for an embedding of the boundary geometry in $\R^{3}$ (modulo the identification in Euclidean time), while the higher winding modes $n\ge 2$ only allow for local embeddings (immersions) and seem to represent non-perturbative modes (instantons).

This beautiful interpretation needs to be put in balance against the fact that the product over $p=1,\dots,\frac12(N_x-1)$ is actually computable, and gives%
\footnote{It comes from evaluating the polynomial $(X^{N_{x}}-1)/(X-1)$ at $X=1$.}
\be
 \prod_{p=1}^{\frac{N_x-1}{2}} \big(2-2\cos(\gamma p) \big)=  \prod_{p=1}^{\frac{N_x-1}{2}}   4\sin^{2}\left(\f{\pi N_{\gamma} p}{N_{x}} \right)
 =
\begin{cases}
N_x \quad\textrm{if}\,\, K:=\mathrm{GCD}(N_{\gamma},N_{x})=1 \\
0 \quad\,\,\,\,\textrm{if}\,\, K:=\mathrm{GCD}(N_{\gamma},N_{x})>1
\end{cases}
\,.
\ee
Note, however, that the formula \eqref{Dete1} given above for  ${\cal D} (\gamma,n)$ is anyway only valid for $K=1$.\\

At this point two remarks are necessary: 

\begin{itemize}

\item[{\it i})]  It seems that the closed formula for the product over the Fourier modes kills the dependence of the Ponzano--Regge partition function in the twist angle $\gamma$. However, one key point is the tremendous difference in behavior between the case $K:=\mathrm{GCD}(N_{\gamma},N_{x})=1$ and the case $K>1$. Taking $K=1$ gives a constant finite  result (simply $N_{x}$) for the product over Fourier modes $p$ thus leaving us simply with the measure factor $\sin^2\left(\f{\gamma n}2\right)$, while having a larger gcd $K>1$ leads formally to  a divergent amplitude. This divergence  is  actually due to a continuum of stationary points, which actually requires a finer analysis. In the asymptotic limit $(N_{\gamma},N_{x})\rightarrow \infty$, which is the continuum limit for the angle $\gamma$, the case $K=1$ corresponds to an irrational value $\gamma\in2\pi(\R\setminus \Q)$, while $K>1$ corresponds to rational values $\gamma\in2\pi\Q$ (see Part I). Having poles for all rational values of the twist angle is the same key feature that is obtained in the companion paper \cite{PART1} using simpler boundary spin network  states with 0-spin intertwiners. 

\item[{\it ii})]   Even if the product over modes $p$ simplifies, giving the final result for the Ponzano--Regge partition function as a simple number is not revealing as providing its explicit Fourier mode decomposition. Indeed the product formula \eqref{Dete1}  for  ${\cal D} (\gamma,n)$ promises an interesting limit  $N_{x}\to\infty$ in terms of the inverse squared Dedekind $\eta$ function, which would establish explicitly the bridge between the Ponzano--Regge model for 3d quantum gravity and the AdS${}_{3}$/CFT${}_{2}$ correspondence (the interested reader will find a detailed discussion of this point in the review part of  \cite{PART1}). However, the Dedekind function is only well-defined for $\gamma$ on the upper  complex half-plane.\footnote{Coming from the AdS case, one obtains $\gamma$ as the zero cosmological constant limit of the torus modular parameter $\tau:=\frac{1}{2\pi}(\gamma + \I \frac{\beta}{\ell_\text{c}})\to \frac{1}{2\pi}(\gamma + \I \epsilon^+)$, where $\Lambda=-1/\ell_\text{c}^2$.}
Actually, the perturbative one-loop calculations also need to introduce an ad hoc  regularization  $\gamma\mapsto\gamma + \I \epsilon^+$ to obtain meaningful amplitudes \cite{GiombiEtAl2008,BarnichEtAl2015}. In the present framework, the angle $\gamma$ is hardcoded into the calculation as a geometrical property of the lattice, as a ratio $\gamma=\f{2\pi N_{\gamma}}{N_{x}}$, which seems to make it unfeasible to extend to complex values.
Strategies to solve these issues include the introduction of a Laplace transform in time, or of a modified Wick rotation. Overall, this issue of understanding how to complexify the parameter $\gamma$ highlights the necessity to test the flexibility of our framework and  explore how to generalize our rigid rectangular lattice to more general boundary geometries. For instance, we can consider LS intertwiners corresponding to parallellograms, thus producing tilted lattices. Not only would this allow us to consider modular transformations of our boundary lattice, but a preliminary analysis suggests that our stationary points become saddle points in the complex plane (actually in the complexified $\SU(2)_{\C}$), thus requiring a deeper analysis of the amplitudes. Such an extension of the  Ponzano--Regge amplitude as a function of complexified holonomies promises the possibility of a rigorous analytical continuation of the partition function to complexified values of the twist angle $\gamma$.

\end{itemize}
%

\section{Discussion and comparison with previous results\label{sec_discussion}}

We have succeeded in computing the semi-classical limit, or equivalently the one-loop approximation, of the Ponzano--Regge (PR) amplitude for a boundary state encoding the intrinsic geometry of a discretized torus.
The result is compatible with previous calculations done in various framework, and in particular in perturbative quantum Regge calculus \cite{BonzomDittrich2015} and in perturbative quantum field theory of the Einstein--Hilbert action \cite{BarnichEtAl2015}. Now we first summarize the main ingredients that went into our computation, and then compare it to other approaches.

\begin{itemize}
\item[{\it bulk 1} )] The bulk theory is given by the PR model. This model can be (formally) defined as a local state sum model which is invariant under changes of the (bulk-)discretization. The PR model is a quantization of first-order gravity. 
\item[{\it bulk 2} )] During the whole calculation the bulk topology has been kept fixed and equal to that of a solid torus with a twist $\gamma$. This information percolated through the calculation via the (global) variable $\varphi$ which encoded (even on the  boundary) the holonomy along the only non-contractible cycle.
\item[{\it bdry 1} )] The boundary state is chosen in the class of states that diagonalize the geometry of the {\it intrinsic} metric on the boundary (spin-network states). The state was built in such a way to encode the intrinsic metric proper of a regular quadrangulation of the toroidal boundary (with a $\gamma$-twist). The only ambiguities (due to the fact that we used a quadrangulation rather than a triangulation) were solved by using the LS coherent states, which allows to build the closest analogue to ``flat'' rectangular cells, which were here characterized quantum mechanically for the first time. Also,  to solve the $\SU(2)$ vs $\SO(3)$ conundrum of the PR model (see section \ref{sec_SU2vsSO3}) we noticed that a particular $\mathbb Z_2$ symmetry could be gauged without spoiling the $\SU(2)$ nature of the flatness condition, but ``saving'' the geometry reconstruction. 
\item[{\it bdry 2} )] In the sense of the item above, the boundary state is supposed to be the closest analogue to the Gibbons--Hawking--York boundary action term. In the discrete context used here, the right analogue is the Regge--Hartle--Sorkin boundary term, whose on-shell value will emerge automatically in the semiclassical limit.
\item[{\it hologr 1} )] The formalism of PR boundary states and amplitudes, of loop quantum gravitational origin, explicitly shows the equivalence between ({\it i}) boundary conditions, ({\it ii}) boundary states, ({\it iii}) dual boundary theories. Given the discrete nature of the spin-network states, the dual boundary theory is a 2d lattice theory. Different examples of such theories were discussed in the companion paper \cite{PART1}.
\item[{\it hologr 2} )] The boundary theory for the class of states we analyzed is akin to a $\SO(3)$ sigma-model. Its form is not very illuminating (the search of boundary states corresponding to more interesting boundary theories is already undergoing).
\item[{\it ampl 1} )] The amplitude was evaluated by {\it solving exactly} the bulk theory, and being hence left with a purely boundary theory which was in turn solved at ``1-loop''. That is, we  first identified all the critical points of the boundary action, we then expanded to second order in the perturbation around these configurations, and finally we evaluated the ensuing Gaussian integrals.
\item[{\it ampl 2} )] Every critical configuration of the boundary action was used to reconstruct a 3d geometry. If the twist variable $N_\gamma$ (the twist angle is $\gamma=\frac{2\pi N_\gamma}{N_x}$, $N_x$ the number of ``horizontal'' cells) is such that $K:=\mathrm{GCD}(N_\gamma,N_x)=1$, the geometry is that of a straight cylinder with basis a regular polygon of exterior dihedral angle $\psi=\tfrac{2\pi}{N_x}n$, with $1\leq|n|\leq \frac{N_x-1}{2}$ an integer (the cylinder can showcase foldings too, see section \ref{sec_pis}). The case $n=1$ correspond to the intuitive classical background, and the case $n=-1$ to its orientation reversal. The cases $|n|>1$ can be interpreted as non-classical or ``quantum'' backgrounds which wind multiple times on themselves. No reasonable bulk can be reconstructed in these cases. For a tentative interpretation see the second part of this discussion. 
\item[{\it ampl 3} )] Opposite values of $n$ can be interpreted as solutions with opposite orientations (see the end of section \ref{sec:reconstr} for an explanation of why these contributions are expected in any theory of gravity---and not only in a first order theory\footnote{For an analysis of the divergence structure of a first order theory of quantum gravity like the PR and its relations to orientation changes, see \cite{ChristodoulouEtAl2013}.}). Their contributions come with complex conjugate amplitudes, and combine to give a real amplitude. This phenomenon is well-known for the fundamental building blocks of the PR model. Here, we show explicitly that in the saddle point approximation there are still exactly two contributions per solution, corresponding to {\it global} orientation changes.
\item[{\it ampl 4} )] The twist, i.e. $\gamma$, dependence of the amplitude is contained in a very simple term, ${\cal D}(\gamma , n)$ which partly comes from the evaluation of the measure on $\varphi$ at the saddle and partly from the evaluation of the 1-loop determinant. Formally, this factor explodes for $N_\gamma$ such that $K\neq1$, but we know this case has to be treated differently, and also gives rise (by construction) to a finite amplitude. We did not explore the amplitude of these cases in detail. In the continuum, the condition analogue to $K=1$ would be $\gamma \in 2\pi(\mathbb R\setminus\mathbb Q)$. 
\end{itemize}

We can now compare our result to a series of previous calculations, and provide further comments in relations to these.\\

\subsection{PR vs. perturbative quantum Regge calculus}
As we stated multiple time now, the 1-loop results of our PR calculation around the geometrical $|n|=1$ saddle and that of perturbatively quantum 
Regge calculus \cite{BonzomDittrich2015} match perfectly: the on-shell LS action (although it eventually evaluates to a sign) takes locally the same form of the on-shell Regge--Hartle--Sorkin action of Regge calculus, and the 1-loop determinants coincide in their dependence on $\gamma$, too (the rest are normalization-dependent results, and therefore not so useful to compare). 
This result might (wrongly!) not look so surprising at first sight, since it is well known that the PR model reduces to Regge calculus in the large-spin approximation (e.g. this is one method to obtain the ``path integral'' measure of \cite{BonzomDittrich2015}, although this can be characterized independently too---see also \cite{BaratinFreidel3D,DittrichSteinhausMeas}). However, this intuition is misleading, because in the present paper we integrated out the bulk exactly, and performed the saddle point approximation on the boundary spin only.\footnote{This is also different form the  "standard" spinfoam asymptotic analysis, which either consider the single weight asymptotics \cite{PonzanoRegge1968, BarrettEtAl2009,BarrettEtAl2010,HnybidaFreidel,HHKR2015}, or require a choice of bulk background as in the lineraized Regge calculation \cite{HanPRD2013}, or yet uses the large-spin approximation for some specific bulk sub-complexes \cite{PeriniRovelliSpeziale2009,Riello2013}. It matches, however, the earlier treatment by \cite{DowdallGomesHellmann2010} of the torus partition function, where the bulk variables were also solved exactly. See also \cite{HellmannKaminski2013} for a general framework to employ only a large-spin limit on the boundary.}
Therefore no trace of the bulk degrees of freedom is left in out calculation, while the Regge calculus computation relies on the choice of fixed bulk background. As a consequence, the dynamical degrees of freedom (dof's) and hence the ``1-loop'' expansions are extremely different in the two cases: in the Regge calculus case the dof's are the bulk edge-lengths, in the PR one with LS states the dof's are the boundary's $G_{t,x}\in\SO(3)$ plus a residual global bulk dof $\varphi\in[0,2\pi)$.

\smallskip
{\bf Symmetries and $p\geq2$.}\quad At the light of the previous remarks, it is quite surprising that the results actually match (around the geometrical $|n|=1$ saddle). 
Focusing on the $\gamma$ dependence, it is particularly interesting to compare the mechanisms of cancellation of the $p\in\{0,\pm1\}$ factors from the product appearing in 
\be
\mathcal D(\gamma,n=1) = \prod_{p\geq2} \frac{1}{2-2\cos(\gamma p)}.
\label{eq_Dn=1}
\ee
In the case of Regge calculus, the $p=0$ mode corresponds to angle-independent perturbations of the bulk edge-lengths. These perturbations correspond to translations of the bulk vertices in a direction parallel to the cylinder's axis. Similarly, the $p=\pm1$ modes correspond to the two translations of the bulk vertices on a plane orthogonal to the above mentioned axis. Therefore, one sees that the $p\in\{0,\pm1\}$ mode correspond precisely to the residual diffeomorphism symmetry of (flat-space) Regge calculus \cite{RocekWilliams1,RocekWilliams2,DittrichFreidelSpeziale2007,Dittrich2008}. Being related to gauge directions, it is natural that these modes do not appear in the 1-loop determinant (in \cite{BonzomDittrich2015} these modes have been gauge-fxed altogether). In the PR case studied here, on the other hand, only the $p=0$ mode does not appear because it corresponds to a symmetry (the global rotations of the boundary frames around the cylinder axis). Indeed, the $p=\pm1$ modes {\it are} present in the 1-loop determinant, but get canceled by the contribution of the measure on $\varphi$! Therefore, the mechanisms at play which lead to the final result for the 1-loop amplitudes are extremely different in the two cases.

\smallskip
{\bf Diffeomorphisms in PR.}\quad Before moving to the next point, let us comment on the fate of diffeomorphism symmetry in the PR model. 
The vertex translation symmetry of Regge calculus corresponds (on-shell) to the shift (or ``translation'') symmetry of $BF$ theory  \cite{FreidelLouapre2003}, which acts on the fields canonically conjugated to the connection. These fields are already integrated out in the holonomy formulation of the PR partition function. Divergences, however, result from integrating over non-compact gauge orbits and give rise to redundancies in the Dirac distributions that appear in the ``naive'' formulation of the (group-representation) PR amplitude.\footnote{In the spin representation of the model, on the other hand, the symmetries has the exact same form as in the Regge calculus case: it corresponds to (discrete) bulk-vertex translations.} Gauge fixing the shift symmetry amounts to removing these redundancies and hence regularizing the PR partition function. 
In the spin representation, the PR model displays an exact bulk-vertex translation symmetry. This was shown to be directly related to the shift (or translation) symmetry of $BF$-theory  \cite{FreidelLouapre2003,BaratinGirelliOriti2011,ChristodoulouEtAl2013}, of which the PR model is a quantization. In the  group representation, the presence of this symmetry leads to the redundancy of the Dirac distributions present in the ``naive'' PR amplitude. In our context, the redundant Dirac distributions have been removed, which precisely corresponds to gauge fixing the diffeomorphism symmetry. 

\smallskip
{\bf Poles vs. Finiteness.}\quad A distinctive feature of the result is that the factor ${\cal D}(\gamma, n=1)$ showcases a series of poles in the parameter $\gamma$, which become relevant whenever $K:=\mathrm{GCD}(N_\gamma,N_x)>1$. 
As discussed in more detail in \cite{PART1}, in the case of perturbative Regge calculus, this pole structure is the result of a peculiar property of the {\it linearized} Regge equations: for homogeneous (intrinsic) boundary data the solutions for the bulk edges are not anymore unique if $K>1$. 
The very same feature emerged in studying the solutions of the saddle point equations for the variables $G_{t,x}$ (see section \ref{sec:reconstr}). Thus, despite using very different variables we have found the same degeneracy structure in the solutions. 

Note that this degeneracy in the solutions is most likely an artifact of the approximation: In the Regge case we discussed in \cite{PART1} that the same feature that leads to the degeneracy of solutions for homogeneous boundary data, prevents the existence of solutions for certain types of boundary inhomogeneities. This  indicates that the degeneracy in the solutions does {\it not} result from some kind of residual gauge symmetry. 
Similarly for  the PR partition function, the pole structure in the saddle point approximation means that it features divergencies. 
On the other hand the PR partition function itself, after gauge fixing the translation symmetry, is at least for finite boundary, finite. This latter fact makes the matching to the Regge results with its peculiar divergence structure even more surprising.

\smallskip
{\bf Winding number $n$.}\quad
A key difference between the PR amplitude studied here and the Regge calculus one is the appearance of solutions labeled by a ``winding number'' $1\leq|n|\leq \frac{N_x-1}{2}$.
This interpretation of $n$ comes from the following fact.
For simplicity consider the geometry of the saddle points before the identification of the first and last ``time slices'', at $t=0$ and $t=N_t$.
Then, as we said multiple times, the solution with $n=1$ corresponds to a straight prism with basis a regular (if $K=1$) $N_x$-polygon, with external dihedral angles $\psi=\frac{2\pi}{N_x}$.
However, the generic solution displays $\psi=\frac{2\pi}{N_x}n$ which means that the polygon, and hence the prism, winds around $n$ times before closing back on itself.
The total parallel transport around the polygon cumulates a total rotation angle of $2\pi n$, which corresponds to a deficit angle
\be
\epsilon = 2\pi(1-|n|). \label{eq_deficitangle_n}
\ee
Clearly, the appearance of these solutions is rooted in the use of a holonomy formulation, where the angular variable is automatically compactified. Flatness can therefore be expressed only modulo $2\pi$, i.e. in the form $\epsilon=0\,\text{mod}\,2\pi$ compatibly with the equation above.
In contrast, in the Regge calculus formulation, the fundamental variable is the dihedral angle itself and therefore the only available solution is $\epsilon$ strictly equal to zero. 

The appearance of the winding number hints at the possibility that the large scale limit of spin foams might include rather unwanted contributions (see \cite{ZapataDiazMarin,ZapataMeneses} for proposals to potentially control these winding numbers).
On the other hand, the appearance of the winding number $n$ can be easily controlled by using more general boundary states.
In fact, the LS boundary states we used control the intrinsic geometry of the boundary only. 
It is not hard, however, to generalize their construction to obtain states peaked on both the intrinsic and extrinsic geometries.
Analogously to the construction of quantum mechanical semi-classical wave packets, this states can be obtained most simply by considering Gaussian superpositions of spins times a $\E^{\I j \psi}$ factor.
More sophisticated coherent states have also been proposed in the literature, see e.g. \cite{ThiemannCohStates,BahrThiemannCohStates1,BahrThiemannCohStates2}.
Such coherent states  now correspond to  a different type of boundary condition and are therefore expected to lead to different boundary action.

Finally, for a tentative interpretation of the winding number in physical terms, see below (section \ref{sec_PRvsholo}).

\smallskip
{\bf Sum over orientations, aka $n\mapsto -n$.}\quad
This topic has been discussed already in some detail at the end of section \ref{sec:reconstr}. 
Briefly, the idea is that this change in sign of the reconstructed dihedral angles correspond to a swapping of the ``inside'' and the ``outside'' of the solid torus. From a ``mechanical'' viewpoint, it corresponds to a swap in the sign of the canonical momenta conjugated to the boundary's intrinsic metric, which is of its extrinsic curvature. This had to be expected, since the intrinsic geometry of the boundary cannot know anything about the direction of time. 
Moreover, this effect is not specific to the present situation at all, the same happens in the WKB analysis of the harmonic oscillator, where to a fixed energy and position, there correspond always two opposite momenta. Finally, this same effect is also known to appear in the local asymptotic of each $\{6j\}$ symbol, which is the local amplitude for a single bulk tetrahedron in the spin-representation of the PR model. The {\it local} presence of both orientations has been related to the structure of symmetries and divergences of the model (cf. the discussion in the paragraph ``Diffeomorphisms in PR'' here above). 

\smallskip
{\bf Foldings.}\quad For this class of PR saddles, for which we did not compute the 1-loop contribution, we refer the reader to section \ref{sec_pis}.
%

\subsection{PR vs. 1-loop General Relativity} 
In \cite{BarnichEtAl2015} the partition function for thermal Minkowski space at temperature $\beta$ was calculated in presence of a twist $\gamma$.\footnote{See also \cite{GiombiEtAl2008} for the same calculation performed in AdS$_3$. }
The setting of the computation was that of perturbative QFT applied to three-dimensional General Relativity. The QFT computation is in spirit closer to that of quantum perturbative Regge calculus, in as far as the choice of a background structure is chosen. However, on the one hand it deals with an infinite number of local degrees of freedom, and on the other it avoids to deal with space-like boundaries in an explicit manner, since it sets them infinitely far away.
The formal result is nonetheless the same as in the computation above, in particular the one loop determinant (formally) equals $ {\cal D}(\gamma,n=1)$ in the obvious limit $N_x\to \infty$ (corresponding both to a continuum setting and an infinitely large space-like boundary).
In the continuum, the infinite product turns out to be more suitably regularized by the replacement $\gamma \mapsto \gamma + \I\epsilon^+$, as it results from a heat-kernel regularization of the functional determinants.
Again, it is quite interesting to compare the reason why the product  $ {\cal D}(\gamma,n=1)$ starts at $p=2$. 
In the perturbative QFT computation, the modes $p\in\{0,\pm1\}$ are effectively ``killed'' by inverse contributions coming from the needed ghost degrees of freedom for the spin 0 and spin 1 modes of the metric. Therefore, it is again ``gauge'' at the origin of the disappearance of these modes from the final result.  

It is often stated that the 3D gravity partition function (e.g. \cite{BarnichEtAl2015}) is perturbatively one-loop exact. This can however only hold for asymptotic boundary conditions. Here we find divergencies (for $K:=\text{GCD}(N_\gamma,N_x)>1$) in the one-loop approximation to the PR partition function. On the other hand the (gauge fixed) partition function for finite boundaries is finite. We thus conclude that the divergencies are an artifact of the approximation and that the saddle--point or one--loop approximation is {\it not} exact.

On the other hand \cite{PART1} shows, that the same divergence structure can be recovered even for the non-perturbative PR partition function, in a limit where the spatial boundary goes to infinity. (This limit is different from the large spin limit, as all boundary spins are equal to $1/2$.) 
Thus we find another surprising convergence between results based on a large-spin limit and a result involving a very large number of small spins, describing an infinite boundary. This is---to our knowledge---the first computation to explicitly obtain such a convergence of results.

\subsection{PR vs. holography\label{sec_PRvsholo}}
From an holographic perspective, one expects the partition function of thermal Minkowski with temperature $\beta$ and twist $\gamma$ to be given by 
\be
Z(\beta, \gamma) = \Tr(\E^{-\beta H} \E^{-\I\gamma P}),
\label{eq_Zholo}
\ee
where $H$ and $P$ are the Hamiltonian and momentum operators of the dual theory, respectively, and $\Tr$ is the trace in the Hilbert space of the dual theory.
This is nothing else than the character---in an appropriate representation---of the element $g=\E^{-\beta H} \E^{-\gamma P}$ of the boundary symmetry group.
In Ads$_3$/CFT$_2$ \cite{MaloneyWitten2007}, this symmetry group is the Virasoro group, and the trace above gives an expression analogous to $ {\cal D}(\gamma,n=1)$, with the product starting again at $p=2$ (see also \cite{GiombiEtAl2008}, and the introduction of the companion paper \cite{PART1} for a more detailed summary and references). There, the reason for $p\geq2$ is that the CFT vacuum is annihilated by the $L_{-1}$ and $\tl L_{-1}$ generators of the Virasoro group (and as such these operators do not engender any descendant states of the CFT vacuum).
Taking the limit $\Lambda\to0$ appropriately in the Virasoro characters, leads directly to analogous BMS$_3$ characters \cite{BarnichOblak2014,Oblak2015}. 
The latter being the group of asymptotic symmetries of 3D Minkowski space, it is precisely the group whose characters are expected to appear in the computation summarized in the previous paragraph. And so it is.
A direct analysis shows that for a {\it massless} BMS$_3$ representation \cite{BarnichOblak2014,Oblak2015}, the character of a generic group element $g=(f,\alpha)$, with $f\in\mathrm{Diff}(\mathbb S_1)$ the ``super-rotation'' and $\alpha\in\mathbb C^\infty(\mathbb S_1)$ the ``super-translation'' component respectively, depends only on the zero-mode of $\alpha$, $\alpha^0$ and on the ``Poincar\'e rotation'' angle $\mathrm{Rot}(f)$ encoded in $f$, in the following way:
\be
\chi_\text{vac}[(f,\alpha)] = \E^{-\frac{\I  \pi \beta}{\ellpl} } \prod_{p\geq2} \frac{1}{2-2\cos(\gamma p)} \qquad \text{where} \quad\alpha^0=\frac{\beta}{\ellpl}\quad\text{and}\quad \mathrm{Rot}(f)=\gamma.
\ee
where $\ellpl=8\pi G$. We thus recognize the consistency with both equation \eqref{eq_Zholo} (up to a Wick rotation) and with our result \eqref{AmplF2} (up to normalizations). The latter, however, comes modulo a mismatch of a factor of 2 in the exponent, since we find $S|_o = \frac{2\pi \beta}{\ellpl}$, with $\beta=TN_t \ellpl$. For a discussion of this factor of 2, see the introduction to the companion paper \cite{PART1}.

Interestingly, in \cite{Oblak2015} also the characters of {\it massive} BMS$_3$ representations were computed. The result for a mass $M=\ellpl^{-1}m$ and an angular momentum $j$ is
\be
\chi_{m,j}[(f,\alpha)] = \E^{\I j \gamma} \E^{-\frac{\I (\pi - m)  \beta}{\ellpl} } \prod_{p\geq1} \frac{1}{2-2\cos(\gamma p)}.
\ee
Modulo the above mentioned factor of 2 (i.e. $\pi$ vs. $2\pi$ in the exponent), this result is beautifully consistent with the geometry of a cylinder in Minkowski space characterized by a deficit angle $m$. Importantly, however, the product starts now at $p=1$. This fact makes it tempting to interpret the saddles with $|n|>1$ in terms of these characters, possibly with an angular momentum and/or a mass which are integers in Planck units. However, this is possible only by appealing to linear superpositions of these possibilities.

Despite its difficulties, this interpretation is made compelling by its analogy to the sum implementing modular invariance in \cite{MaloneyWitten2007}.
Indeed, from a purely boundary CFT perspective, one expects the quantum gravitational amplitude to be invariant under the full modular group of symmetries of the torus, even if this fact implies a sum over topologies from the bulk perspective. A standard interpretation of this fact is that a ``good'' quantum gravity theory has to implement a sum over all possible geometries and topologies compatible with the asymptotic boundary conditions, here included non-trivial (BTZ) black hole solutions. 
Thus, at the light of the above considerations, the sum over winding numbers can be given an analogous interpretation in terms of non-perturbative Planck-scale excitations of the bulk theory. 

Notice also, that in both cases these sums break the holomorphic/antiholomorphic factorizability of the amplitude as a function%
\footnote{See also \cite{Witten3dRevisited}.} of $q=\E^{2\pi \I \tau}=\E^{\I \gamma - \beta/\ell_c}$ ($\Lambda \equiv -\ell_c^{-2}$ vanishes in the PR case):
\be
 Z_\text{AdS/CFT}\sim  \sum_{\Gamma\in\text{modular}}\frac{ (1- q_\Gamma)}{\prod_{p} (1-q^p_\Gamma)}\frac{ (1-\overline{q_\Gamma})}{\prod_{p} (1-\overline{q_\Gamma}^p)}
\quad
\text{vs.}
\quad
 Z_\text{PR}\sim  \sum_{n}\frac{ (1-q^n)}{\prod_{p} (1-q^p)}\frac{ (1-\overline q^n)}{\prod_{p} (1-\overline q^p)},
\ee
where $q_\Gamma := \exp(2\pi\I \Gamma\triangleright\tau)$ is the action of the modular group on the modular parameter $\tau = \gamma + \I \frac{\beta}{\ell_c}$.

\subsection{Remarks on the dual boundary theory and holographic interpretation}

As already remarked on, in computing the Ponzano--Regge amplitude for our chosen boundary state we did solve first for the bulk degrees of freedom and where left with an integral over boundary degrees of freedom and one global holonomy variable. The resulting form of the partition function in   (\ref{2ampl}) defines on a fully non-perturbative level a boundary field theory dual to three dimensional quantum gravity.  This is analogous to the way Wess--Zumino--Novikov--Witten (WZNW) theory can be extracted from the Chern--Simons (CS) path integral with appropriate boundary conditions---i.e. boundary action---see e.g. \cite{Carlip2005}.

For LS boundary states, the boundary theory degrees of freedom are a group valued field $G(t,x)\in \SU(2)$ plus a global variable $\phi$ of topological origin. Although the latter seems coupled to the $G$-field non isotropically, this is an artifact of gauge-fixing. Crucially, the coupling to this variable breaks the symmetry between the two cycles of the torus. This is in stark contrast with the requirement that the dual theory be modular invariant \cite{MaloneyWitten2007}. The action is non analytic (contains logarithms) and depends on boundary data $\{j_l \xi^n_l\}$ which parametrize the LS boundary state. These encode nothing but the discrete boundary metric. The path integral is a function(al) of these boundary state data, which enter as classical (or background) sources in the dual theory. All of this is parallels what happens in the CS-WZNW duality. Finally, it is clear that the same results would have been formally found for a LS state based on a triangulation: these qualitative features would not have changed.

The derivation of the Hessian (\ref{HessianFT1})  (also shown in appendix \ref{DBF}) around the saddle points allows us some insight into the dynamics described by the dual boundary field theory.   Let us in particular consider the variables $G(t,x) \in SU(2)$ assigned to each node $(t,x)$ in the boundary.  In the Hessian (\ref{Hess01})  we can recognize lattice Laplacians $(2-2\cos(\psi p))$  and $(2- 2\cos\chi_{E,p})$  in spatial and time direction respectively.  The lattice laplacian in spatial direction is  modified by a phase $(2-2\cos(\psi p)) \rightarrow (2-2 \exp({\rm{i}}n \psi ) \cos(\psi p))$, whereas the lattice laplacian in time direction is `twisted' and via $\chi_{E,p}= \frac{2\pi}{N_t}(E- \frac{\gamma}{2\pi} p)$ it is also the only place, where a dependence on the twisting angle $\gamma$ occurs.

The three components describing the perturbations around the saddle points come each with a different kinetic term: the $x$-component with a (modified) spatial Laplacian, the $z$-component with a Laplacian in the time direction, and the $y$-component has a `full' kinematic term, that is the Laplacian in both directions is appearing.  Furthermore the $x$- and $z$-fields are only coupled to the $y$-field, but not to each other.

Interestingly, if we integrate out only the $z$-field---which has a Laplacian in time direction only---we are left with a {\it local} theory, that is without inverse Laplacians (see equation \eqref{Hess02} in appendix \ref{DBF}). Furthermore the resulting Hessian does not depend anymore on $\chi_{E,p}$, that is neither on the energy label $E$ nor on the twisting angle $\gamma$.
Thus the integration of the $z$-field is solely responsible for the $\gamma$--dependent contribution to the one-loop correction. This $z$-field has only a (twisted) kinematic term in time direction. This is analogous to the dual field theory identified in \cite{BonzomDittrich2015} which had also a kinematic term in time direction only.

We can also first integrate out the $x$-field. Here however, due to the modification of the Laplacian in spatial direction we obtain one term which involves an inverse spatial Laplacian. Apart from this term the situation is analogous to integrating out the $z$-field: all other terms are local and do involve the frequency $p$ only via $\chi_{E,p}$.

Integrating out two fields leads to effective Hessians for only one variable. These do however involve inverse Laplacians, and thus encode for non-local theories.

\subsection{Outlook} 

The topological nature of three dimensional gravity allowed us to define dual boundary field theories at a non-perturbative level.
Different classes of boundary states lead in principle to different types of dual boundary theories, and it would be valuable to investigate this relationship more systematically. 
Among other theories, we know for example that interesting connections arise to two dimensional integrable models, such as the 6-vertex model discussed in the first paper of this series \cite{PART1} and the Ising model of \cite{Dittrich:2013jxa,Bonzom:2015ova}. Thus the natural question arises of which class of integrable models can be encoded in (superpositions of) spin-network evaluations.

Another appealing direction of investigation is a generalization of this work to four spacetime dimension. 
To still make use of  topological invariance one has to consider a $BF$ theory rather than gravity. 
Nonetheless, $BF$ theory is the starting point for the construction of the four dimensional gravitational spin-foam models and therefore we expect that at least certain features of the boundary theories dual to $BF$ should also hold in the gravitational context. 
It is at this point relevant to notice that the techniques developed in the present work can be straightforwardly applied to the four dimensional  $BF$ case. 
In particular, the same kind of non-linear sigma model studied here will appear in one dimension more, and---in light of the relation of the Ponzano--Regge model to integrable two dimensional theories---we also expect connections to one-dimension-higher statistical models.

Finally, to better understand the relation of the presently calculated amplitude with the characters of the BMS3 group, it seems unavoidable to perform a deep study of the analytical continuation of the boundary amplitudes over complexified group elements, which is most likely inerlaced with working out the extension of our calculations to the Lorentzian analogue of the Ponzano–Regge model \cite{Davids:2000kz,Freidel:2000uq,Freidel:2005bb}.


\section*{Acknowledgement}
The authors thank FQXi for the support of the workshop ``Holography in background independent approaches'' in Paris, October 2016, where part of this work was initiated. BD and AR thank
Laurent Freidel for discussions, CG acknowledges discussions with Clement Delcamp. This work
is supported by Perimeter Institute for Theoretical Physics. Research at Perimeter Institute is
supported by the Government of Canada through Industry Canada and by the Province of Ontario
through the Ministry of Research and Innovation.

\appendix

\section{Computation of the Hessians} \label{appHess1}

Here we will determine the Hessians for the action $S_{LS}$ as defined in (\ref{Hessiandef1}). 

\paragraph*{\bf $\phi\phi$-term}
This is the simplest term. 
\begin{align}
H_{\phi;\phi} & = 
\left.%
 -\frac{2L}{N_t} \sum_{t,x}  \left( \frac{\la +| G_{t+1,x}^{-1}  \E^{\frac{\varphi}{N_t}\tau_z} \tau_z^2 G_{t,x} | +\ra}{ \la +| G_{t+1,x}^{-1} \E^{\frac{\varphi}{N_t}\tau_z} G_{t,x} | +\ra} 
 -%
 \frac{\la +| G_{t+1,x}^{-1}  \E^{\frac{\varphi}{N_t}\tau_z} \tau_z G_{t,x} | +\ra^2}{ \la +| G_{t+1,x}^{-1} \E^{\frac{\varphi}{N_t}\tau_z} G_{t,x} | +\ra^2}\right)\right|_o \notag\\
 & =  \frac{L}{2N_t} \sum_{t,x} \left( 1  - (\hat z. G^o_{t,x}\triangleright \hat x)^2 \right) \notag\\
 & = \frac{L N_x}{2} \equiv F.
\end{align}
The topological $\varphi$-field has no kinetic term and a (positive) mass equal to one-half of the cylinder (spatial) circumference.\\

\paragraph*{\bf $\phi a$-term}
The mixed term is
\begin{align}
&  H^{k}_{t,x; \phi} = \notag\\
& = 
\left.%
 -\frac{2L}{N_t}\left(%
 \frac{\la +| G_{t+1,x}^{-1}  \E^{\frac{\varphi}{N_t}\tau_z} \tau_z G_{t,x} \tau^k | +\ra}{ \la +| G_{t+1,x}^{-1} \E^{\frac{\varphi}{N_t}\tau_z} G_{t,x} | +\ra} 
 -%
 \frac{\la +| G_{t+1,x}^{-1}  \E^{\frac{\varphi}{N_t}\tau_z} \tau_z G_{t,x} | +\ra\la +| G_{t+1,x}^{-1} \E^{\frac{\varphi}{N_t}\tau_z} G_{t,x} \tau^k| +\ra}{ \la +| G_{t+1,x}^{-1} \E^{\frac{\varphi}{N_t}\tau_z} G_{t,x} | +\ra^2} \right)\right|_o + \notag\\
 & \phantom{=} + \text{1 term from other link}\notag\\
 & =  - \frac{L}{2N_t}\Big( \la + | (G_{t,x}^o)^{-1} \sigma_z G^o_{t,x}\sigma^k | + \ra - %
 (\hat z. G_{t,x}^o\triangleright \hat x) \delta^k_1  - \la + | \sigma^k (G_{t,x}^o)^{-1} \sigma_z G^o_{t,x} | + \ra +  (\hat z. G_{t,x}^o\triangleright \hat x) \delta^k_1  \Big) \notag\\
 & = - \frac{L}{2N_t}\Big( \la + | \sigma_z \sigma^k | + \ra - %
 \la + | \sigma^k \sigma_z | + \ra \Big)  =  - \frac{iL}{N_t} \epsilon_{3kj} \la + |  \sigma^j | + \ra \notag\\
 & = \frac{iL}{N_t} \delta^k_2 \equiv D_k
\end{align}
Notice the independence from the position $(t,x)$.\\

\paragraph*{\bf $a a$-term}
Because of the near-neighbour interactions, this is the most complicated term.
We start from the ultralocal term (notice the symmetrization of the indices)
\begin{align}
H^{j;k}_{t,x;t,x} 
& = 2T \left.\left(- \frac{\la \up |\tau^{(j} \tau^{k)} G_{t,x}^{-1}G_{t,x-1} |\up\ra }{\la \up | G_{t,x}^{-1}G_{t,x-1} |\up\ra} + \frac{\la \up | \tau^k G_{t,x}^{-1}G_{t,x-1} |\up\ra \la \up | \tau^j G_{t,x}^{-1}G_{t,x-1} |\up\ra }{\la \up | G_{t,x}^{-1}G_{t,x-1} |\up\ra^2} \right)\right|_o+\notag\\ 
& \phantom{=} +  \text{3 other terms from other links}\notag\\
& = \frac{T}{2} \left( {\la \up |\sigma^{(j} \sigma^{k)} |\up\ra  -\la \up | \sigma^k  |\up\ra \la \up | \sigma^j |\up\ra } \right)+  \text{3 other terms from other links}\notag\\
& = \frac{T}{2} \left( \delta^{jk} - \delta^k_3\delta^j_3 \right) +  \text{3 other terms from other links}\notag\\
& =  T \left( \delta^{jk} - \delta^k_3\delta^j_3 \right) + L \left( \delta^{jk} - \delta^k_1\delta^j_1 \right) \equiv A^{jk}.
\end{align}
This term does not depend on the position $(t,x)$, either.

Then, we move on to the term which couples spatially separated cells.
In this case the two derivative commutes and one needs only to compute one of the two terms (we compute the second one). The result is 
\begin{align}
H^{j;k}_{t,x-1;t,x} 
& = 2T \left.\left( \frac{\la \up |\tau^{k} G_{t,x}^{-1}G_{t,x-1} \tau^j|\up\ra }{\la \up | G_{t,x}^{-1}G_{t,x-1} |\up\ra} - \frac{ \la \up | \tau^kG_{t,x}^{-1}G_{t,x-1} |\up\ra\la \up |  G_{t,x}^{-1}G_{t,x-1} \tau^j |\up\ra  }{\la \up | G_{t,x}^{-1}G_{t,x-1} |\up\ra^2} \right)\right|_o \notag\\ 
& = -\frac{T}{2}\left(  \la \up |\E^{ \psi^T_o \tau_z}\sigma^{k} \E^{- \psi^T_o \tau_z} \sigma^j |\up\ra -   \la \up |\sigma^{k} |\up\ra \la \up |\sigma^{j} |\up\ra    \right)\notag\\
& =  -\frac{T}{2}\left(  R_z(-\psi^T_o)^k{}_i \la \up |\sigma^{i}\sigma^j |\up\ra -   \la \up |\sigma^{k} |\up\ra \la \up |\sigma^{j} |\up\ra    \right)\notag\\
& =  -\frac{T}{2}\left(  R_z(-\psi^T_o)^k{}_j  + i R_z(-\psi^T_o)^k{}_i \epsilon^{ij3} -   \delta^k_3\delta^j_3    \right) \equiv B^{jk},
\end{align}
where a rapid computation shows
\be
B = - \frac{T}{2} \E^{-i\psi^T_o} \mat{ccc}{ 1 & -i & 0 \\ i & 1 & 0 \\ 0 & 0 & 0  }.
\ee
Similarly, one finds that
\be
H^{j;k}_{t,x+1;t,x} = B^{kj},
\ee
which is just the transpose of the above matrix. Notice that this could have been deduced from $H^{j;k}_{t,x-1;t,x} = H^{k;j}_{t,x; t,x-1}$ and the independence of these matrices from the position $(t,x)$.

Finally, for timely separated neighboring cells, one obtains
\begin{align}
H^{j;k}_{t-1,x;t,x} 
& = 2L \left.\left(  \frac{\la + | \tau^k G_{t,x}^{-1}\E^{\frac{\varphi}{N_t} \tau_z} G_{t-1,x} \tau^j |+\ra }{\la + | G_{t,x}^{-1}\E^{\frac{\varphi}{N_t} \tau_z} G_{t-1,x}  |+ \ra} - \frac{\la + | \tau^k G_{t,x}^{-1}\E^{\frac{\varphi}{N_t} \tau_z} G_{t-1,x}  |+\ra \la + |G_{t,x}^{-1}\E^{\frac{\varphi}{N_t} \tau_z} G_{t-1,x} \tau^j  |+\ra }{\la + | G_{t,x}^{-1}\E^{\frac{\varphi}{N_t} \tau_z} G_{t-1,x}  |+ \ra^2}   \right)\right|_o \notag\\ 
& = -\frac{L}{2} \Big(   \la + | \sigma^k \sigma^j |+\ra  - \la + | \sigma^k   |+\ra \la + |  \sigma^j  |+\ra   \Big) \notag\\ 
& = -\frac{L}{2} \Big(  \delta^{kj} + i\epsilon^{kj1} - \delta^k_1\delta^j_1 \Big) \equiv C^{jk},
\end{align}
where 
\be
C = - \frac{L}{2} \mat{ccc}{ 0 & 0 & 0 \\ 0  & 1 & -i \\ 0 & i & 1  }.
\ee
And similarly
\be
H^{j;k}_{t+1,x;t,x} = C^{kj} .
\ee

\section{Evaluation of Gaussian Integrals} \label{App_Gauss}

Here we evaluate the Gaussian integrals appearing in the amplitude (\ref{AmplF1}).

For  the integrals in the first parenthesis  in (\ref{AmplF1}) we have
\begin{align}
\mathcal N_{(0,0),\phi} &\equiv \left( \int_{\mathbb R^+}\d \phi \int_{\mathbb R^3} \d\hat a_{0,0} \;  \E^{\frac12 F \phi^2 + \phi {} D. \hat{{} a}_{0,0} + \frac12 \hat{ a}{}^*_{0,0} \hat H'_{0,0} \hat a_{0,0}   } \right) \notag\\
&=
\left(\frac{(2\pi)^3}{LT(1-\E^{-i\psi n})\left( \frac{L}{4 N_t^2} + \frac{TN_t}{2}(1- \E^{-i\psi n}) \right) }\right)^{1/2},
\end{align}
a result which is independent of $\gamma$ (recall that $\psi=\frac{2\pi}{N_x}n$).

 The $(E,p)$-th factor of the  following term  in  (\ref{AmplF1})  gives
\begin{align}
\mathcal N_{E,p}& \equiv \left( \int_{\mathbb C^3} \d \hat a_{E,p}  \,  \E^{ \frac12  \hat{{} a}{}^*_{E,p} \hat H_{E,p} \hat a_{E,p}   } \right)^{1/2} 
\,=\,
 \left( \frac{(2\pi)^3}{\det \hat H_{E,p} }\right)^{1/2} \notag\\
& = \left(\frac{(2\pi)^3}{LT \E^{-i \psi n} (2- 2\cos \chi_{E,p})\Big( (L+T) (\cos(n\psi) - \cos(p \psi) ) + i L \sin(n\psi)  \Big) }\right)^{1/2}.
\end{align}
As expected, the result is even in $(E,p)\mapsto(-E,-p)$. To multiply over the $(E,p)\neq(0,0)$, we reorganize the product as follows:
\be
\prod_{(E,p)\neq(0,0)} \mathcal N_{E,p} = \left(\prod_{E=1}^{N_t-1}\mathcal  N_{E,0}\right)\left(  \prod_{p=1}^{N_x-1}\prod_{E=0}^{N_t-1} \mathcal N_{E,p}\right).
\ee
It is then useful to recall the following identity (recall that $\chi_{E,p}=\frac{2\pi}{N_t} \left( E - \frac{ \gamma}{2\pi}p \right)$):
\be
\prod_{E=0}^{N_t-1} (2-2\cos\chi_{E,p}) = 2- 2\cos(\gamma p).
\ee
From this, one finds
\be
\prod_{E=1}^{N_t-1} (2-2\cos\chi_{E,0})
= \lim_{p\to 0}\frac{\prod_{E=0}^{N_t-1} (2-2\cos\chi_{E,p})}{2 - 2\cos\chi_{0,p}}
=\lim_{p\to 0}\frac{ 2- 2\cos(\gamma p)}{2 - 2\cos\chi_{0,p}}
= N_t^2
\ee
and hence, 
\begin{align}
\prod_{(E,p)\neq(0,0)} \mathcal N_{E,p} & =%
\left(\frac{(2\pi)^3}{LT \E^{-i \psi n}}\right)^{\frac{N_x N_t -1}{2}}%
\left(\frac{1}{(L+T)(\cos(n\psi)-1) + i L \sin(n\psi)}\right)^{\frac{N_t-1}{2}} \frac{1}{N_t} \times\notag\\%
& \phantom{=}\times%
\prod_{p=1}^{N_x-1} \left(   \frac{1}{\Big(2-2\cos(\gamma p)\Big)\Big((L+T)(\cos(n\psi)-\cos(p\psi)) + i L \sin(n\psi)\Big)^{N_t}}  %
\right)^\frac{1}{2}  .
\label{eq_prodNEp}
\end{align}

This formula holds whenever $(N_x/2 , N_\gamma/2)$ and $(N_x/2, (N_\gamma+N_t)/2)$ are {\it not} in $\mathbb N \times \mathbb N$, since in these cases we would be overcounting one real mode. In other words, these conditions ensure that the only real Fourier mode in the first Brillouin zone is $(E,p)=0$  (see equations \eqref{eq_oddeven}).   

We shall now consider two cases separately: ({\it i}) $N_x$ is odd, and ({\it ii}) $N_x$ and $N_t$ are even, while $N_\gamma$ is odd.\\

{\bf $N_x$ is odd.}\quad
In this case, a simple rearrangement of the terms above gives 
\begin{align}
\prod_{(E,p)\neq(0,0)} \mathcal N_{E,p} & \stackrel{N_x\text{ odd}}{=}%
\left(\frac{(2\pi)^3}{LT \E^{-i \psi n}}\right)^{\frac{N_x N_t -1}{2}}%
\left(\frac{1}{(L+T)(\cos(n\psi)-1) + i L \sin(n\psi)}\right)^{\frac{N_t-1}{2}} \frac{1}{N_t} \times\notag\\%
& \phantom{=}\qquad\times%
\prod_{p=1}^{\frac{N_x-1}{2}}  \frac{1}{2-2\cos(\gamma p)} \times
\prod_{p=1}^{\frac{N_x-1}{2}} \left(\frac{1}{(L+T)(\cos(n\psi)-\cos(p\psi)) + i L \sin(n\psi)}  %
\right)^{N_t} .
\end{align}

{\bf $N_x$ and $N_t$ are even, $N_\gamma$ is odd.}\quad
Conversely, in this case we obtain (for simplicity, we label this case just as ``$N_x$ even'')
\begin{align}
\prod_{(E,p)\neq(0,0)} \mathcal N_{E,p} & \stackrel{N_x\text{ even}}{=}%
\left(\frac{(2\pi)^3}{LT \E^{-i \psi n}}\right)^{\frac{N_x N_t -1}{2}}%
\left(\frac{1}{(L+T)(\cos(n\psi)-1) + i L \sin(n\psi)}\right)^{\frac{N_t-1}{2}} \frac{1}{N_t} \times\notag\\%
& \phantom{=}\qquad\times%
\frac12\times\left(\frac{1}{(L+T)(\cos(n\psi)+1) + i L \sin(n\psi)}  \right)^{\frac{N_t}{2}} \times\notag\\%
& \phantom{=}\qquad\times%
\prod_{p=1}^{\frac{N_x}{2}-1}  \frac{1}{2-2\cos(\gamma p)} \times
\prod_{p=1}^{\frac{N_x}{2}-1} \left(\frac{1}{(L+T)(\cos(n\psi)-\cos(p\psi)) + i L \sin(n\psi)}  \right)^{N_t}.
\end{align}
where the second line is the contribution coming from the $p=N_x/2$ on the rhs of equation \eqref{eq_prodNEp} (recall in particular that $N_\gamma$ is supposed to be odd, and therefore $2-2\cos(\gamma N_x/2)$ is necessarily equal to 4).\\

An important remark, studied in detail in the next appendix, is that the products above are ill-defined for $n=0$ and---in the even-$N_x$ case---$n=N_x/2$.

\section{The $n=0$ and $n=N_x/2$ critical points} \label{app_n=0}

\subsection{$n=0$}

At $n=0$, all the dihedral angles $\psi^{T,L}_{t,x}$ vanish as in the flat representation of the torus in terms of a rectangle with the opposite edges topologically identified.

As equation \eqref{eq_hatHEp} shows, the Hessian of the action evaluated at this solution is also highly degenerate, since it displays 1 extra degenerate modes for {\it any} value of $E$ provided $p=0$:
\be
 \hat{H}_{E,p}(n=0) =
 \mat{ccc}{ %
 T\left[1- \cos(\psi  p) \right] &- T\sin(\psi  p)   & 0 \\%
 T\sin(\psi p)   & T\left[1 - \cos(\psi p)\right] + L\left[1- \cos\left (\chi_{E,p}\right)\right] & -L\sin\left (\chi_{E,p}\right)\\%
 0  & L\sin\left ( \chi_{E,p}\right)  & L\left [1 -\cos\left ( \chi_{E,p}\right)\right] }.
\ee
(this is also readily seen by analyzing how $\mathcal A_\LS$ diverges for $n\to0$, which shows that there are exactly $N_t-1$ zeros developing in the determinant).

The null modes of $\hat H_{E,0}$ correspond to arbitrary rotations along the $\hat x$ axis. 
The question then is whether these are just null modes of the Hessian around the regular configuration, or whether they do generate a larger critical locus. 

To start with, recall that if the $\psi^T$ all vanish , so does $\varphi=0$, and this solution coincides with the only viable potential candidate set of solutions from the $X$-family. It is indeed easy to see that for any $\psi^L_{t,x} = \psi^L_t$ such that modulo $\varphi_o = \sum_{t'=1}^{N_t} \psi_{t'}^L = 0$ mod $2\pi$.
\be
\tl G_{t,x} = \pm_{t,x}\tl G_{0,0} \E^{\sum_{t'<t}\psi^L_{t'} \tau_x}
\ee
is a solution of the saddle point equations.
Now, the on-shell LS action of one of this solutions is
\be
S|_o = -\I L N_x\sum_{t'=1}^{N_t}\psi_{t'}^L = - 2 \I \pi L N_x q,
\ee
where we introduced the integer $q\in\mathbb Z$.
Thus, we found that the action is exactly constant throughout some continuum sectors of solutions of the equations of motion labeled by $q$.
Each sector, or critical locus, is of course compact. 

\begin{figure}
\begin{center}
\includegraphics[width=.6\textwidth]{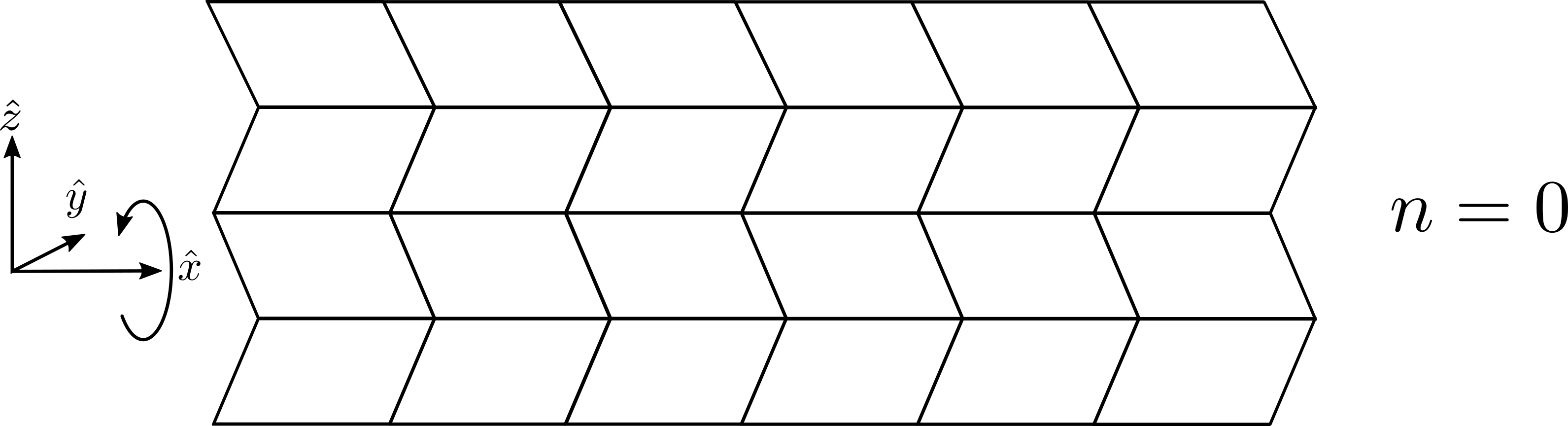}
\caption{A sketch of the bending along constant time slices which is at the origin of the continuous set of solutions of which the $n=0$ solution is part.}
\label{fig_n=0}
\end{center}
\end{figure}

Geometrically, these solutions correspond to flat rectangles that are homogeneous in the spacial direction and that can be freely bent along the equal time edges (modulo the fact that $\sum_{t'=1}^{N_t} \psi_{t'}^L$ = 0 mod $2\pi$), see figure \ref{fig_n=0}.

In this case, a saddle point analysis might not be an available tool (unless, for example, the extendedness of the critical locus is a consequence of an exact symmetry of the action, or at least of a symmetry of the linearized action around the critical locus), and we will not attempt to perform it. 

Nevertheless, we can argue as follows. 
Since the critical locus is compact, and all our integrals are indeed finite, we know that the contribution of the integral in the $\tl G_{t,x}$ in a neighborhood of the critical locus must be finite, and---evidently---independent of $n$. 
Considering now the integral over $\varphi$, we see that the only value of this variable compatible with the equations of motion is $\varphi_o=0$ mod $2\pi$, which corresponds precisely to conjugacy classes of vanishing volume.

In other words, the saddle-point contribution of this critical locus is ``killed'' by the measure factor in the $\varphi$ integral, and is hence sub-dominant.

\subsection{$n=N_x/2$}
This case, only allowed for even values of $N_x$, displays very similar degeneracies to the case discussed above.
The situation is in this case even more ``crampled'': all dihedral angles are equal to $\pi$ and the reconstructed geometry is maximally degenerate. 
Moreover, as in the case $n=0$, the reconstructed boundary geometry is again planar (albeit for the ``opposite'' reason).
This means that analogous null modes to those present in the $n=0$ case shall appear here too, except that their are now associated to a spacial frequency $p=N_x/2$ (rather than being constant in the horizontal direction). As in the previous case the symmetry is expected to appear at any time slice independently (modulo the global constraint $\sum_t \psi^L_t = 0 \,\text{mod}\,2\pi$).
This expectation is confirmed by the degeneracy of the Hessian in the $x$-direction precisely for $p=N_x/2$ and $E$ arbitrary.

Contrary to the $n=0$ case, however, the size of the conjugacy class of $\varphi$ at this saddle, $\varphi(n=N_x/2)=\pi$, is maximal rather than vanishing.
As a consequence, the contribution of this saddle is very hard to evaluate.
Moreover, for values of $n$ which are {\it close} to $N_x/2$, the determinant will be very small due to the emergence of an approximate symmetry, and the relative saddles will be enhanced. 

A similar mechanism is actually expected for the odd-$N_x$ case, a fact that can be easily confirmed by a numerical analysis.\\

Notice that in both the $n=0$ and $n=N_x/2$ cases the existence of these emerging symmetries is strictly bound to the regularity of the chosen discretization.

\section{Expansion of the dual boundary field theory around critical points \label{DBF}}

The form  (\ref{2ampl}) of the partition function defines a dual boundary field theory for the Ponzano--Regge model. The linearization of this theory around the critical points is described by the Hessian matrix given by
 \begin{align}
\mathbf{b}. \mathbf H \bfa
= & F \phi' \phi
+ \Big( \phi'  {} D.\hat{{} a}_{0,0} + \phi  \hat{{} b}_{0,0}. {} D \Big) + \sum_{E,p} \hat{{} b}{}_{-E,-p}. \hat{H}_{E, p}  \hat{{} a}_{E,p}
\end{align}
with
\be \label{Hess01}
 \hat{H}_{E,p} =
 \mat{ccc}{ %
 T\left[1- \E^{-\I n \psi }\cos(\psi  p) \right] &- T\E^{-\I n \psi }\sin(\psi  p)   & 0 \\%
 T\E^{-\I n \psi }\sin(\psi p)   & T\left[1 - \E^{-\I n  \psi }\cos(\psi p)\right] + L\left[1- \cos\left (\chi_{E,p}\right)\right] & -L\sin\left (\chi_{E,p}\right)\\%
 0  & L\sin\left ( \chi_{E,p}\right)  & L\left [1 -\cos\left ( \chi_{E,p}\right)\right] }.
\ee
using the notation
\be
\psi  = \frac{2\pi }{N_x},
\qquad
\chi_{E,p} = \frac{2\pi}{N_t} \left( E - \frac{ \gamma}{2\pi}p \right)
\qquad\text{and}\qquad
\gamma = \frac{2\pi N_\gamma}{N_x}.
\label{eq_psiANDchi}
\ee

We can get some insight into the dynamics of the $G$--field described by $\hat{H}_{E,p}$ by integrating out some components. If we integrate out the $z$--component only we are left with an effective Hessian
\be\label{Hess02}
\hat{H}^{(z)}_{p}=
\mat{cc}{%
T\left[1- \E^{-\I n \psi }\cos(\psi  p) \right]  &   - T\E^{-\I n \psi }\sin(\psi  p)    \\%
T\E^{-\I n \psi }\sin(\psi  p)       &  2L + T\left[1- \E^{-\I n \psi }\cos(\psi  p) \right]
}   .
\ee
This Hessian does not anymore involve the frequencies $\chi_{E,p}$ -- that is does not depend on $E$ nor on the twisting angle $\gamma$.

Integrating out only the $x$--field gives a slightly more complicated effective Hessian. This is due to the modification $(2-2\cos(\psi p)) \rightarrow (2-2 \exp(-\I n \psi ) \cos(\psi p))$ of the Laplacian in spatial direction, which gives the diagonal $xx$--term in the initial Hessian:
\be\label{Hess03}
\hat{H}^{(x)}_{E,p}=
\mat{cc}{%
2T+ L \left [1 -\cos\left ( \chi_{E,p}\right)\right] - \frac{2\I T \sin \left( n \psi  \right) }{ \cos\left( n \psi\right) -\cos\left( p \psi \right) + \I \sin \left( n \psi  \right)  }  &    -L\sin\left (\chi_{E,p}\right) \\
L\sin\left ( \chi_{E,p}\right)   &      L \left [1 -\cos\left ( \chi_{E,p}\right)\right]
}   .
\ee



\bibliographystyle{bibstyle_aldo}
\bibliography{PRholo2_biblio}

\end{document}